\documentclass[aps, prd, superscriptaddress, nofootinbib, preprintnumbers, onecolumn, 11pt]{revtex4-2}

\usepackage{style}

\begin{document}

\title{Effective Field Theory for Dark Matter Absorption on Single Phonons}

\author{Andrea Mitridate\,\orcidlink{0000-0003-2898-5844}}
\email{andrea.mitridate@desy.de}
\affiliation{Deutsches Elektronen-Synchrotron DESY, Notkestr. 85, 22607 Hamburg, Germany}

\author{Kris Pardo\,\orcidlink{0000-0002-9910-6782}}
\email{kmpardo@usc.edu}
\affiliation{Walter Burke Institute for Theoretical Physics, California Institute of Technology, Pasadena, CA 91125, USA}
\affiliation{Department of Physics \& Astronomy, University of Southern
California, Los Angeles, CA, 90089, USA}

\author{Tanner Trickle\,\orcidlink{0000-0003-1371-4988}}
\email{ttrickle@fnal.gov}
\affiliation{Theoretical Physics Division, Fermi National Accelerator Laboratory, Batavia, IL 60510, USA}

\author{Kathryn M. Zurek\,\orcidlink{0000-0002-2629-337X}}
\email{kzurek@caltech.edu}
\affiliation{Walter Burke Institute for Theoretical Physics, California Institute of Technology, Pasadena, CA 91125, USA}

\preprint{CALT-TH-2023-032}
\preprint{DESY-23-113}
\preprint{FERMILAB-PUB-23-417-T}

\begin{abstract}
    Single phonon excitations, with energies in the $1-100 \, \text{meV}$ range, are a powerful probe of light dark matter (DM). Utilizing effective field theory, we derive a framework to compute DM absorption rates into single phonons starting from general DM-electron, proton, and neutron interactions. We apply the framework to a variety of DM models: Yukawa coupled scalars, axionlike particles (ALPs) with derivative interactions, and vector DM coupling via gauge interactions or Standard Model electric and magnetic dipole moments. We find that \ce{GaAs} or \ce{Al2O3} targets can set powerful constraints on a $U(1)_{B-L}$ model, and targets with electronic spin ordering are similarly sensitive to DM coupling to the electron magnetic dipole moment. Lastly, we make the code, \textsf{PhonoDark-abs} (an extension of the existing \textsf{PhonoDark} code which computes general DM-single phonon scattering rates), publicly available.
\end{abstract}

\maketitle
\newpage
\tableofcontents
\clearpage 

\section{Introduction} 
\label{sec:introduction}

Recent years have seen rapid development in dark matter (DM) direct detection technology. As experimental constraints utilizing nuclear recoil, e.g., ANAIS~\cite{Amare:2019jul}, CRESST~\cite{Probst:2002qb,CRESST:2017cdd,CRESST:2015txj}, DAMA/LIBRA~\cite{Baum:2018ekm}, DAMIC~\cite{DAMIC:2015znm,DAMIC:2019dcn}, DarkSide~\cite{DarkSide:2018bpj}, DM-Ice~\cite{Jo:2016qql}, KIMS~\cite{Kim:2015prm}, LUX~\cite{LUX:2018zdm,LUX:2019npm,LUX:2018akb}, SABRE~\cite{Shields:2015wka}, SuperCDMS~\cite{SuperCDMS:2014cds,SuperCDMS:2016wui,SuperCDMS:2015eex,SuperCDMS:2017nns,SuperCDMS:2018gro,SuperCDMS:2018mne}, and XENON~\cite{XENON:2007uwm,XENON100:2016sjq,XENON:2019gfn,XENON:2018voc,XENONCollaboration:2023orw},  continue to increase sensitivity, others, such as, CDEX~\cite{CDEX:2022kcd}, DAMIC~\cite{DAMIC:2020cut,DAMIC:2016qck,Settimo:2020cbq,DAMIC:2015znm,DAMIC:2019dcn}, DarkSide~\cite{DarkSide:2018ppu,DarkSide:2022knj,DarkSide:2018bpja}, EDELWEISS~\cite{EDELWEISS:2018tde,EDELWEISS:2019vjv,EDELWEISS:2020fxc}, SENSEI~\cite{Crisler:2018gci,SENSEI:2020dpa,SENSEI:2019ibb}, SuperCDMS~\cite{SuperCDMS:2020ymb,SuperCDMS:2018mne,CDMS:2009fba}, and XENON~\cite{XENON:2022ltv,XENON:2019gfna,Bloch:2016sjj}, are utilizing electronic excitations to drive sensitivity to smaller DM masses. The lightest DM mass the experiments utilizing electronic excitations are sensitive to is set by the ionization energy in noble liquids, $\mathcal{O}(10 \, \text{eV})$, and the band gap in crystal targets, typically $\mathcal{O}(\text{eV})$. While a scattering DM particle needs to be heavier than an MeV to generate these electronic transitions, a DM particle being absorbed may be much lighter, since the energy deposited is approximately the DM mass. Therefore the lightest DM masses direct detection experiments are currently sensitive to is $\mathcal{O}(\text{eV})$.

The same production mechanisms for $\mathcal{O}(\text{eV})$ scale DM, e.g., inflationary production~\cite{Graham:2015rva}, parametric resonance~\cite{Dror:2018pdh,Adshead:2023qiw}, or misalignment mechanisms~\cite{Arias:2012az,Nelson:2011sf,Dimopoulos:2006ms} also allow for lighter, sub-eV scale DM candidates. Therefore the search for light DM should not end at $\mathcal{O}(\text{eV})$, and there have been a variety of proposals to explore this sub-eV mass region. Electronic excitations can be utilized in targets with small excitation gaps such as superconductors~\cite{Hochberg:2015fth,Hochberg:2015pha,Hochberg:2016ajh,Hochberg:2019cyy,Hochberg:2021ymx,Hochberg:2021yud,Gelmini:2020xir,Mitridate:2021ctr}, Dirac materials~\cite{Hochberg:2017wce,Coskuner:2019odd,Geilhufe:2019ndy}, doped semiconductors~\cite{Du:2022dxf}, graphene~\cite{Hochberg:2016ntt, Catena:2023qkj,Catena:2023awl}, narrow gap semiconductors~\cite{SuperCDMS:2022kse}, and spin-orbit coupled targets~\cite{Inzani:2020szg,Chen:2022pyd}.

Collective excitations, such as phonons~\cite{Schutz:2016tid,Knapen:2016cue,Knapen:2017ekk,Griffin:2018bjn,Trickle:2019nya} and magnons~\cite{Barbieri:1985cp,Trickle:2019ovy,Chigusa:2020gfs,Mitridate:2020kly,Trickle:2020oki}, have also been proposed as an avenue to detect light DM. These excitations have energies in the $\mathcal{O}(1 - 100 \, \text{meV})$ range and targets typically have $\mathcal{O}(10)$ modes, making them excellent prospects for direct detection of sub-GeV DM. In addition to being kinematically favorable, the experimental program for single phonon detection is being actively pursued. The TESSARACT experiment~\cite{Chang2020}, currently in development, will utilize single phonon excitations in \ce{GaAs} and \ce{Al2O3} (sapphire) targets. The combination of motivated DM models and a maturing experimental program compels us to quantitatively compute the reach of experiments to a broad range of theoretically consistent DM models.

The theory of DM-single phonon scattering has been well developed in the literature~\cite{Knapen:2017ekk,Griffin:2018bjn,Trickle:2019nya,Caputo:2019cyg,Cox:2019cod,Griffin:2019mvc,Trickle:2020oki,Lasenby:2021wsc}. Recently an effective field theory (EFT) approach was used to compute the general DM-single phonon scattering rate~\cite{Trickle:2020oki}, building on the EFT framework first developed for general DM-nucleus scattering~\cite{Fitzpatrick:2012ix,Cirelli:2013ufw,Anand:2013yka,Gresham:2014vja,Anand:2014kea,DelNobile:2018dfg}. However, in the absence of external electromagnetic fields, DM absorption on single phonons has only been computed for the kinetically mixed dark photon DM model~\cite{Knapen:2017ekk,Knapen:2021bwg}. The purpose of this work is to generalize the computation of DM absorption to {\it any} DM model which has a Yukawa-like interaction Lagrangian of the form,
\begin{align}
    \mathcal{L} \supset g \, \phi \, \bar{\Psi} \, \mathcal{O} \,  \Psi \, ,
    \label{eq:uv_lag}
\end{align}
where $g$ is a perturbatively small coupling constant, $\phi$ is the DM field, $\Psi \in \{ e, p, n \}$ is either an electron, proton, or neutron Standard Model (SM) field, and $\mathcal{O}$ is an operator. The simplest example of this interaction Lagrangian is when $\mathcal{O} = 1$, and $\phi$ is a scalar field coupling to, e.g., electrons; then Eq.~\eqref{eq:uv_lag} is simply $\mathcal{L} \supset g \, \phi \, \bar{e} e$. Eq.~\eqref{eq:uv_lag} can also apply to vector DM, $V_\mu$, when $\mathcal{O}$ has a matching Lorentz index, e.g., $\mathcal{L} \supset g \, V_\mu \, \bar{e} \gamma^\mu e$ and $\mathcal{O} = \gamma^\mu$. This can be further extended when the operator $\mathcal{O}$ is allowed to contain momentum (derivatives) acting on the DM and SM fields, such as for an axionlike particle (ALP), $a$, with derivative coupling to electrons: $\mathcal{L} \supset g \, \partial_\mu a \, \bar{e} \gamma^\mu \gamma^5 e$. In momentum space, this simply corresponds to $\mathcal{O} = - i \, q_\mu \, \gamma^\mu \gamma^5$, where $q^\mu$ is the four-momentum of the ALP field.

To compute the DM absorption rate for a general DM model we utilize the self-energy formalism developed in Refs.~\cite{Chen:2022pyd,Krnjaic:2023nxe,Mitridate:2021ctra,Hardy:2016kme} for electronic excitations. Using the optical theorem, the absorption rate can then be computed diagrammatically. Computing the single phonon absorption rate involves similar diagrams to electron absorption, although here the intermediate excitations are phonons instead of electrons. This framework has two main benefits: first, it automatically includes screening effects, which arise from DM-photon mixing. Second, setting up the calculation as a Feynman diagram calculation allows for straightforward generalizations to different DM models, by simply changing the Feynman rules at the vertex. The problem becomes finding Feynman rules of the DM-phonon vertex, ascertained from a Lagrangian coupling the DM to the ion displacement/phonon operator, $\mathbf{u}$. We develop a method to find this DM-phonon EFT Lagrangian starting from a UV Lagrangian of the form Eq.~\eqref{eq:uv_lag}.

In addition to the general operators encompassed by Eq.~\eqref{eq:uv_lag}, we consider targets with and without fermionic spin ordering, e.g., (anti) ferromagnets. We find that targets with different spin orderings can be sensitive to different DM models. For example, in the absence of spin ordering, ALP DM does not couple to phonons via the derivative coupling; however, in a spin ordered target phonons can be excited.

We make the code used to compute the absorption rates for all targets, \textsf{PhonoDark-abs}~\GithubLink, publicly available here~\GithubLink. This program complements \textsf{PhonoDark}~\cite{PhonoDark, Trickle:2020oki}, which was developed to compute general DM-single phonon scattering rates. 

The paper is organized as follows. In Sec.~\ref{sec:formalism} we provide a theoretical framework to compute the DM absorption rate into single phonons via a general, Yukawa-like interaction in the form of Eq.~\eqref{eq:uv_lag}. This derivation will proceed in three steps: first, in Sec.~\ref{subsec:abs_rate_self_energy} we write the absorption rate in terms of self-energy diagrams using the optical theorem. Second, in Sec.~\ref{subsec:self_energies} we compute the phonon contribution to these diagrams, whose imaginary part leads to single phonon absorption, in terms of vertex Feynman rules, which are then derived in detail in Sec.~\ref{subsec:EFT}. Before computing the DM absorption rate into single phonons, we compute the single phonon contribution to the dielectric function in Sec.~\ref{subsec:dielectric_calc}, which serves both as a cross check of the formalism, and verification of the first principles calculations of the target properties. In Secs.~\ref{subsec:scalar_DM} -~\ref{subsec:vector_DM_edm_mdm} we compute the DM absorption rate into single phonons for different DM models and show the projected constraints. Specifically, we consider scalar DM (Sec.~\ref{subsec:scalar_DM}), the derivative coupling of ALP DM (Sec.~\ref{subsec:alp_DM}), and two models of vector DM, one which couples to the SM vector currents (Sec.~\ref{subsec:vector_DM_gauge}), and another which couples to the SM particles electric and magnetic dipole moments (Sec.~\ref{subsec:vector_DM_edm_mdm}).

\section{General Single Phonon Absorption Rate} 
\label{sec:formalism}

The purpose of this section is to compute DM single phonon absorption rates due to Yukawa-like interactions in the form of Eq.~\eqref{eq:uv_lag}. The derivation proceeds in three steps. First, in Sec.~\ref{subsec:abs_rate_self_energy}, we use the optical theorem to write the DM absorption rate in terms of in-medium self-energies. Second, in Sec.~\ref{subsec:self_energies}, we write the phonon contribution to the self-energies in terms of crystal form factors, $\vec{\mathcal{F}}$, describing how the DM field couples to phonons. These form factors will depend on the properties of the ions at each lattice site, e.g., the number of protons or the electronic spin. Third, in Sec.~\ref{subsec:EFT}, we detail how the form factors are derived using non-relativistic effective field theory (NR EFT) starting from Eq.~\eqref{eq:uv_lag}. 

\subsection{Absorption Rate In Terms of Self-Energies}
\label{subsec:abs_rate_self_energy}

Following Refs.~\cite{Chen:2022pyd,Krnjaic:2023nxe,Mitridate:2021ctra,Hardy:2016kme}, we start by deriving the DM absorption rate in terms of in-medium self-energies, $\Pi$. While this formalism was originally developed to compute DM absorption on electrons, it is agnostic about the underlying crystal degrees of freedom and can be similarly applied to the case of phonon absorption. The optical theorem states that the absorption rate of the $\lambda^\text{th}$ polarization of the DM field, $\phi$, is,
\begin{align}
    \Gamma^\lambda = -\frac{1}{m_\phi} \text{Im} \left[ \Pi_{\phi \phi}^\lambda \right] \, ,
\end{align}
where $m_\phi$ is the DM mass, $\Pi_{\phi \phi}^\lambda$ is the self-energy between two $\phi$ particles of the $\lambda^\text{th}$ polarization. However, when $\phi$ can mix with the photon field, $A$, this introduces complications since one must include diagrams mixing $\phi$ and $A$. The sum of all the diagrams can be succinctly written in terms of 1PI diagrams by first going to the ``in-medium" basis, where the fields are no longer mixed. When $\phi$ and $A$ are perturbatively coupled, the leading order self-energy of the in-medium ``DM-like" state, $\hat{\phi}$, is given by,
\begin{align}
    \Pi_{\hat{\phi} \hat{\phi}}^\lambda \simeq \Pi_{\phi \phi}^\lambda + \sum_\eta \frac{\Pi_{\phi A}^{\lambda \eta} \Pi_{A \phi}^{\eta \lambda}}{m_\phi^2 - \Pi_{AA}^\eta} \, ,
    \label{eq:in_med_Pi}
\end{align}
where we have introduced the self-energies polarization components defined as, $\Pi^{\lambda\lambda'} \equiv - e_\mu^\lambda \, \Pi^{\mu\nu}  e_\nu^{\lambda'*}$, where $e_\mu^\lambda$ are the polarization vectors of the subscripted fields, e.g., $\Pi_{\phi A}^{\lambda \lambda'} = - e_{\phi, \mu}^\lambda \, \Pi^{\mu\nu}  e_{A, \nu}^{\lambda'*}$. The polarization vectors are defined to diagonalize the $\Pi_{\phi \phi}$ and $\Pi_{AA}$ self-energies, i.e., $\Pi_{AA}^{\lambda\lambda'}=\Pi_{AA}^\lambda\delta^{\lambda\lambda'}$, and in general will differ from the vacuum longitudinal and transverse polarization vectors. Given Eq.~\eqref{eq:in_med_Pi}, the total DM absorption rate is given by,
\begin{align}
    \Gamma^\lambda = -\frac{1}{m_\phi} \text{Im} \left[ \Pi_{\hat{\phi} \hat{\phi}}^\lambda \right] = - \frac{1}{m_\phi} \text{Im} \left[ \Pi_{\phi \phi}^\lambda + \sum_\eta \frac{\Pi_{\phi A}^{\lambda \eta} \Pi_{A \phi}^{\eta \lambda}}{m_\phi^2 - \Pi_{AA}^\eta} \right] \,,
\end{align}
and the DM-polarization averaged absorption rate, per target mass and exposure time, $R$, is then,
\begin{align}
    R = \frac{\rho_\phi}{\rho_{\scriptscriptstyle T} m_\phi} \frac{1}{n} \sum_{\lambda} \Gamma^\lambda  \, ,
    \label{eq:rate_in_self_energy}
\end{align}
where $\rho_\phi$ is the DM density, taken to be $0.4 \, \text{GeV} / \text{cm}^3$, $\rho_{\scriptscriptstyle T}$ is target density and $n$ is the number of polarizations of the DM field.

The absorption rate in Eq.~\eqref{eq:rate_in_self_energy} can be simplified further if we assume that the photon self-energy is independent of polarization, i.e., $\Pi_{AA}^\lambda \simeq \Pi_{AA}$, which is true in the isotropic limit. To simplify the following analysis we assume isotropy, and leave a study of anisotropic corrections to future work. In this case, the sum over $\eta$ can be performed exactly using the completeness relation $\sum_{\eta} e_{A, \eta}^\mu e_{A, \eta}^{\nu, *} = - g^{\mu \nu}$. Moreover, the Ward identity, $Q_\mu \Pi^{\mu \nu} = 0$, demands the time components to be $q$-suppressed relative to the spatial components. Therefore, $ - g_{\mu \nu} \Pi_{\phi A}^{\lambda \mu} \Pi_{A \phi}^{\nu \lambda} \approx \Pi_{\phi A}^{\lambda i} \Pi_{A \phi}^{i \lambda}$, and Eq.~\eqref{eq:rate_in_self_energy} simplifies to,
\begin{align}
    R = - \frac{\rho_\phi}{\rho_{\scriptscriptstyle T} m_\phi^2} \frac{1}{n} \sum_{\lambda} \text{Im} \left[ \Pi_{\phi \phi}^\lambda + \frac{\Pi_{\phi A}^{\lambda i} \Pi_{A \phi}^{i \lambda}}{m_\phi^2 - \Pi_{AA}} \right]  \, .
    \label{eq:rate_in_self_energy_2}
\end{align}
This leads to our final absorption rate expressions for (pseudo) scalar DM, $R_S$, and vector DM, $R_V$:\footnote{Any directional $\hat{\mathbf{q}}$ dependence within the rate is averaged over, i.e., we compute $\bar{R} = \frac{1}{4 \pi} \int d\Omega_{\hat{\mathbf{q}}} R(\hat{\mathbf{q}})$ for both scalar and vector DM models. Furthermore we take $q = \omega v_0$, where $v_0 = 230 \, \text{km}/\text{s}$.}
\begin{align}
    R_S = - \frac{\rho_\phi}{\rho_{\scriptscriptstyle T} m_\phi^2} \text{Im} \left[ \Pi_{\phi \phi} + \frac{\Pi_{\phi A}^{i} \Pi_{A \phi}^{i}}{m_\phi^2 - \Pi_{AA}} \right]\,, \qquad
    R_V =  - \frac{\rho_\phi}{3 \rho_{\scriptscriptstyle T} m_\phi^2} \text{Im} \left[ \Pi_{\phi \phi}^{ii} + \frac{\Pi_{\phi A}^{i j} \Pi_{A \phi}^{j i}}{m_\phi^2 - \Pi_{AA}} \right] \,.
    \label{eq:final_rate}
\end{align}
When deriving the absorption rate for vector DM we have once again used the completeness relation of the polarization vectors to perform the sum over the DM polarizations. Notice how in both the scalar and vector case we were able to remove the dependence on the photon and DM polarizations, such that the problem of computing the absorption rate has now shifted to deriving the spatial components of the in-medium self-energies.

The in-medium self-energies receive contributions from both electronic excitations, $\Pi_{\rm el}$, and phonon excitations, $\Pi_{\rm ph}$. For example, at one loop, the following graph topologies will contribute:
\vspace{1em}
\begin{align}
    \hspace{5em}
    \begin{fmffile}{Pi_PhiPhi_ph}
        \begin{fmfgraph*}(70,20)
            \fmfleft{i} \fmfright{f}
            \fmf{dashes}{i,m1}
            \fmf{double}{m1,m2}
            \fmf{dashes}{m2,f}
            \fmfblob{.05w}{m1,m2}
            \fmflabel{$\Pi = \Pi_\text{ph} + \Pi_\text{el} =\quad$}{i}
            \fmflabel{$\quad +$}{f}
        \end{fmfgraph*}
    \end{fmffile}
    \hspace{4em}
    \begin{fmffile}{Pi_PhiPhi_e1}
        \begin{fmfgraph*}(70,20)
            \fmfleft{i} \fmfright{f}
            \fmf{dashes}{i,m1}
            \fmf{fermion,right,tension=.4}{m1,m2}
            \fmf{fermion,right,tension=.4}{m2,m1}
            \fmf{dashes}{m2,f}
            \fmflabel{$\quad +$}{f}
        \end{fmfgraph*}
    \end{fmffile}
    \hspace{4em}
    \begin{fmffile}{Pi_PhiPhi_e2}
        \begin{fmfgraph*}(70,20)
            \fmfleft{i} \fmfright{f}
            \fmf{dashes}{i,m1}
            \fmf{fermion}{m1,m1} 
            \fmf{dashes}{m1,f}
            \fmflabel{$\quad\text{\,,}$}{f}
        \end{fmfgraph*}
    \end{fmffile}
\end{align}
\noindent where the first diagram represents the phonon contribution while the last two diagrams encode the contribution from electronic excitations. Here we will assume that the electronic band gap is much larger than the energy of phonon excitations, such that no electron excitation can go on shell at energies relevant for DM absorption into phonons. As a result, $\text{Im} \left[ \Pi_\text{el} \right] \simeq 0$ and
\begin{align}
    \Pi \simeq \Pi_\text{ph} + \text{Re} \left[ \Pi_\text{el} \right]\, .
    \label{eq:total_self_energy_approxed}
\end{align}
Therefore, in order to compute the absorption rates given in Eq.~\eqref{eq:final_rate}, in addition to the phonon contribution to the self-energies one also has to compute the real contribution from electronic excitations. The electron contribution to in-medium self-energies has been extensively studied in Refs.~\cite{Chen:2022pyd,Krnjaic:2023nxe,Mitridate:2021ctr}, therefore, in the following we will focus on the novel phonon contribution and use the values of $\text{Re}\left[\Pi_\text{el}\right]$ derived in these previous works. 

\subsection{Phonon Contribution To Self-Energies}
\label{subsec:self_energies}

The phonon contributions to the diagrams in $\Pi_{\phi \phi}, \Pi_{\phi A}$ and $\Pi_{AA}$ can all be understood from the same diagram:

\begin{minipage}{\textwidth}
    \centering 
    \begin{fmffile}{Pi_PhiPhi_ph2}
        \begin{fmfgraph*}(150,50)
            \fmfleft{i} \fmfright{f}
            \fmf{dashes, label.side=left, label=$\overset{Q}{\longrightarrow}$}{i,m1}
            \fmf{double}{m1,m2}
            \fmf{dashes}{m2,f}
            \fmfblob{.05w}{m1,m2}
            \fmflabel{$\Phi$}{i}
            \fmflabel{$\Phi'$}{f}
        \end{fmfgraph*}
    \end{fmffile}
\end{minipage}
where $Q^\mu = (\omega, \mathbf{q})$ is the incoming four-momentum, $\Phi$ and $\Phi'$ can be either $\phi$ or $A$ (the diagram inherits the Lorentz indices of the field, e.g., $\Pi_{A A} \rightarrow \Pi_{A A}^{\mu\nu}$), and the central double line is a phonon propagator. To compute the diagram the phonon propagator and vertex rule are needed. The phonon propagator, $D_{\nu \mathbf{k}}(\omega)$, is given by~\cite{mah00},
\begin{align}
    D_{\nu \mathbf{k}}(\omega; \gamma_{\nu\mathbf{k}}) = \frac{2 i \, \omega_{\nu \mathbf{k}}}{\omega^2 - \omega_{\nu\mathbf{k}}^2 + i\omega \gamma_{\nu\mathbf{k}} } \, ,
    \label{eq:phonon_propagator}
\end{align}
where $\nu, \mathbf{k}$ index the phonon branch and momentum within the first Brillouin zone (1BZ), respectively, $\omega_{\nu \mathbf{k}}$ is the phonon energy, and $\gamma_{\nu\mathbf{k}}$ is the phonon linewidth, or inverse of the phonon lifetime. Assuming, for now, that the left and right vertex rules are given by, $M_{\Phi, \nu\mathbf{k}}$, $M^*_{\Phi', \nu\mathbf{k}}$, respectively, the self-energy is,
\begin{align}
    i \Pi_{\Phi \Phi'}(Q) = \frac{1}{N \Omega} \sum_{\nu \mathbf{k}} M_{\Phi, \,\nu \mathbf{k}} \, D_{\nu\mathbf{k}} \,M_{\Phi',\, \nu\mathbf{k}}^* \, ,
    \label{eq:self_energy_1}
\end{align}
where $V = N \Omega$ is the volume of the target, $N$ is the number of unit cells, and $\Omega$ is the unit cell volume. Analogous to the self-energy, $M_{\Phi, \nu \mathbf{k}}$ will inherit the Lorentz indices of the field $\Phi$.

In order to separate the part of the vertex that depends on the structure of the UV Lagrangian from the part that is common among different UV interactions, we parameterize the vertices for scalar and vector fields as 
\begin{align}
\parbox[c][60pt][c]{90pt}{\centering
	\begin{fmffile}{se1-1loop}
	\begin{fmfgraph*}(70,40)
	\fmfleft{in}
	\fmfright{out}
	\fmf{dashes,tension=1,l.side=left,l.d=3pt}{in,v1}
    \fmf{double, tension=1}{v1,out}
	\fmfv{label={$S$\;\;},label.angle=-110,l.d=8pt}{in}
        \fmfv{decor.shape=circle,decor.filled=30,decor.size=3thick,label={},label.angle=-110,l.d=8pt}{v1}
        \fmfv{label={$\nu\mathbf{k}$\;\;},label.angle=-110,l.d=8pt}{out}
	\end{fmfgraph*}
	\end{fmffile}
}
& = i M_{S, \,\nu\mathbf{k}} \equiv - \sqrt{N} \, \delta_{\mathbf{q},\mathbf{k}} \, \sum_{j} \bm{\mathcal{F}}_{S,\, j} \cdot \bm{T}_{j \nu \mathbf{k}}  \label{eq:vertex_S} \\
\parbox[c][60pt][c]{90pt}{\centering
	\begin{fmffile}{se3-1loop}
	\begin{fmfgraph*}(70,40)
	\fmfleft{in}
	\fmfright{out}
	\fmf{photon,tension=1,l.side=left,l.d=3pt}{in,v1}
    \fmf{double, tension=1}{v1,out}
	\fmfv{label={$V$\;\;},label.angle=-110,l.d=8pt}{in}
        \fmfv{decor.shape=circle,decor.filled=30,decor.size=3thick,label={},label.angle=-110,l.d=8pt}{v1}
        \fmfv{label={$\nu\mathbf{k}$\;\;},label.angle=-110,l.d=8pt}{out}
	\end{fmfgraph*}
	\end{fmffile}
}
& = i M_{V, \, \nu\mathbf{k}}^\mu \equiv - \sqrt{N} \, \delta_{\mathbf{q},\mathbf{k}} \, \sum_{j} \bm{\mathcal{F}}_{V,\, j}^{\mu} \cdot \, \bm{T}_{j \nu \mathbf{k}} \, , \label{eq:vertex_V}
\end{align}
respectively, where $S$ is a scalar field, $V$ is a vector field, and the double line indicates a phonon. We will summarize the meaning of each term here, and provide a detailed derivation of $M_{\Phi, \nu \mathbf{k}}$ in Sec.~\ref{subsec:EFT}. A factor of $\sqrt{N}$ has been factored out to cancel the $1/N$ in Eq.~\eqref{eq:self_energy_1}, as well as the momentum conservation factor $\delta_{\mathbf{q}, \mathbf{k}}$. $j$ indexes the ions in each unit cell, and the sum indicates that all the ions in the unit cell contribute to the generation of a phonon. $\bm{T}_{j \nu \mathbf{k}}$ is defined as the phonon transition matrix element,
\begin{align}
    \bm{T}_{j \nu \mathbf{k}} = \sqrt{N} e^{i \mathbf{k} \cdot \mathbf{x}^0_{\ell j}} \langle \nu \mathbf{k} | \mathbf{u}_{\ell j} | 0 \rangle = \frac{1}{\sqrt{2 m_j \omega_{\nu \mathbf{k}}}} \bm{\epsilon}^*_{j \nu \mathbf{k}} \, ,
    \label{eq:phonon_transition_matrix_element}
\end{align}
$\mathbf{u}_{\ell j}$ is the displacement operator,
\begin{align}
    \mathbf{u}_{\ell j} = \frac{1}{\sqrt{2 N m_j}} \sum_{\nu \mathbf{k}} \frac{e^{i \mathbf{k} \cdot \mathbf{x}^0_{\ell j}}}{\sqrt{\omega_{\nu \mathbf{k}}}} \left( a_{\nu \mathbf{k}} + a^\dagger_{\nu \mathbf{k}} \right) \bm{\epsilon}_{j \nu \mathbf{k}}  \, ,
    \label{eq:displacement_operator}
\end{align}
$| \nu \mathbf{k} \rangle = a^\dagger_{\nu \mathbf{k}} |0 \rangle$ is a single phonon state at $\nu,\, \mathbf{k}$, $\mathbf{x}^0_{\ell j}$ is the equilibrium position of the $j^\text{th}$ ion in the $\ell^\text{th}$ unit cell, $m_j$ is the mass of the $j^\text{th}$ ion, and $\omega_{\nu\mathbf{k}}, \bm{\epsilon}_{j \nu \mathbf{k}}$ are the phonon energies and polarization vectors, respectively. The phonon transition matrix element, $\bm{T}_{j \nu \mathbf{k}}$, is coming from a $|0 \rangle \rightarrow | \nu \mathbf{k} \rangle$ transition in the vertex. Lastly, the form factors $\bm{\mathcal{F}}_{S,\, j}$ and $\bm{\mathcal{F}}_{V,\, j}^\mu$ are vectors that contain the information about how the scalar, $S$, or vector field, $V$ couple to the displacement operator, $\mathbf{u}_{\ell j}$, via macroscopic properties of the ion, e.g., total number of electrons. These form factors may contain contributions from each fermion type, $e, p, n$ at the lattice site $j$, and therefore can be further decomposed as $\bm{\mathcal{F}}_{j} = \sum_{\psi} \bm{\mathcal{F}}_{j\psi}$, where $\psi \in \{ e, p, n \}$. The detailed derivation of these form factors is given in Sec.~\ref{subsec:EFT}, and summarized in Table~\ref{tab:UV_to_form_factor} for the DM models of interest. 

Substituting Eqs.~\eqref{eq:phonon_propagator} and~\eqref{eq:phonon_transition_matrix_element} into Eq.~\eqref{eq:self_energy_1} gives the final expression for the phonon contribution to the self-energies,
\begin{align}
    \Pi_{\Phi \Phi'}(Q) = \frac{1}{\Omega} \sum_{\nu} \,  \left( \sum_{j} \bm{\mathcal{F}}_{j} \cdot \bm{T}_{j \nu \mathbf{q}} \right) \frac{2 \omega_\nu}{\omega^2 - \omega_\nu^2 + i \omega \gamma_\nu} \left( \sum_{j} \bm{\mathcal{F}}_{j} \cdot \bm{T}_{j \nu \mathbf{q}} \right)^* \, ,
    \label{eq:self_energy_phonon_contribution}
\end{align}
where the $S, V$ index on $\bm{\mathcal{F}}$ follows from directly from the $\Phi, \Phi'$ particle type, e.g., when computing $\Pi_{\phi \phi}$ for scalar $\phi$, $\bm{\mathcal{F}}_{S,\, j}$ should be used. We have also simplified the phonon propagator since for absorption kinematics, $q \ll \omega$, $\omega_{\nu \mathbf{q}} \approx \omega_{\nu}$. 

\subsection{Dark Matter-Phonon Interaction Form Factors}
\label{subsec:EFT}

\renewcommand{\arraystretch}{1.5}
\setlength{\tabcolsep}{12pt}
\begin{table*}
    \begin{center}
        \begin{tabular}{@{}ccc@{}} \toprule
            \bf{Model} & \multicolumn{2}{c}{\bf{(LO) Form Factors, }$\displaystyle \bm{\mathcal{F}}_{j}$}  \\
& No Spin Ordering & Spin Ordering\\\toprule
            \multicolumn{3}{c}{\bf{Spin-0 DM} ($\bm{\mathcal{F}}_{S,\, j}$)} \\[2pt]
            $g \, \phi \bar{\Psi} \Psi$ & $ g \, N_{j} \, \mathbf{q}$ & $g \, N_{j} \, \mathbf{q}$\\
            $\displaystyle \frac{g}{2 m_\Psi} \partial_\mu a \, \bar{\Psi} \gamma^\mu \gamma^5 \Psi$ & - & $\displaystyle - \frac{i \, g \, \omega^2}{m_\Psi} \, \mathbf{S}_{j}$ \\[5pt]\hline
            \multicolumn{3}{c}{\bf{Spin-1 DM} $\left( \bm{\mathcal{F}}^\mu_{V,\, j} = \mathcal{F}^{\mu i}_{V,\, j} = \left(  \mathcal{F}^{0 i}_{V,\, j}, \mathcal{F}_{V,\, j}^{k i} \right) \right)$} \\[2pt]
            $g \, V_\mu \bar{\Psi} \gamma^\mu \Psi$ & $g \, N_j \left(q^i, \; \omega \, \delta^{ki} \right)$ &$g \, N_j \left( q^i, \; \omega \, \delta^{ki} \right)$\\[2pt]
            $\displaystyle \frac{d_M}{2} V_{\mu\nu} \bar{\Psi} \sigma^{\mu \nu} \Psi$ & $\displaystyle d_M \frac{\omega^2}{2 m_\Psi} N_j \left( q^i \, , \, \omega \,  \delta^{ki} \right)$ & $\displaystyle  2 i \, d_M \omega \, \left( \epsilon^{i a b} \, q^a \, S^b_j \,,  -\epsilon^{b k i} \, \omega \, S_j^b \right)$  \\[2pt]
            $\displaystyle \frac{d_E}{2} V_{\mu\nu} \bar{\Psi} \sigma^{\mu \nu} i \gamma^5 \Psi$ & - & $\displaystyle 2 i \, d_E \, \left( \mathbf{q} \cdot \mathbf{S}_j \right) \left( q^i \, , \, \omega \, \delta^{ki} \right) $ \\
            \bottomrule
        \end{tabular}
    \end{center}
    \caption{DM-phonon interaction form factors, $\bm{\mathcal{F}}_{j}$, Eqs.~\eqref{eq:vertex_S} and~\eqref{eq:vertex_V}, for the DM Model (UV Lagrangian) shown in the left column. The leading order form factor is shown for targets with no spin ordering (middle column), and targets with spin ordering (right column). For spin-1 DM we write the $i$ index on the vector $\bm{\mathcal{F}}_{V,\, j}$ to avoid confusion with the $\mu$ index. Explicitly, $\mathcal{F}_{V, j}^{0 i}$ ($\bm{\mathcal{F}}^0_{V, j}$) is the left component inside the parentheses, and $\mathcal{F}_{V, j}^{k i}$, the $i^\text{th}$ component of the vector $\bm{\mathcal{F}}^k_{V, j}$, is the right component. Dashed lines indicate negligible, higher order responses. Note the $\psi$ index on the form factor has been dropped from $N, S$ for simplicity. The ``0" components of the spin-1 DM form factors are related by the Ward identity, $Q_\mu \bm{\mathcal{F}}^{\mu}_{V,\, j} = 0$, where $Q^\mu = (\omega, \mathbf{q})$ is the incoming DM four-momentum.} 
    \label{tab:UV_to_form_factor}
\end{table*}

In Sec.~\ref{subsec:abs_rate_self_energy} we wrote the single phonon absorption rate in terms of electron and phonon self-energies, and in Sec.~\ref{subsec:self_energies} we wrote the phonon self-energies, Eq.~\eqref{eq:self_energy_phonon_contribution}, in terms of some form factors, $\bm{\mathcal{F}}_{j}$. These results were independent of both the UV Lagrangian and target material, whose dependence manifests within the aforementioned form factors. In this section, we derive these form factors starting from the UV Lagrangian in Eq.~\eqref{eq:uv_lag}. Schematically, the derivation proceeds as follows,
\begin{align}
    \mathcal{L}_\text{UV}(\Psi) \overset{\text{NR} \, \text{EFT}}{\longrightarrow} \mathcal{L}_\text{NR}(\psi) \overset{\langle \rangle_{\ell j}}{\longrightarrow} \mathcal{L}(\mathbf{u}_{\ell j}) \overset{|0 \rangle \, \rightarrow \, |\nu \mathbf{k} \rangle}{\longrightarrow} \bm{\mathcal{F}}_{j} \, .
    \label{eq:eft_game_plan}
\end{align}

The first step ``NR EFT" (Sec.~\ref{subsubsec:nr_eft}) (non-relativistic effective field theory) reduces the UV Lagrangian, $\mathcal{L}_\text{UV}$, written in terms of the four-component $\Psi$ fields, to the NR Lagrangian, $\mathcal{L}_\text{NR}$, written in terms of two-component fields, $\psi$, which describe the electron, proton, and neutrons in the target. The NR expansion is appropriate when the energy and momentum transfers are much smaller than the fermion masses, which is certainly the case for absorption processes. 

While the NR Lagrangian is written directly in terms of the particles constituting the target, it does not contain any information about the target itself. In the case of a crystal, the target state is simply a lattice of ions at positions $\mathbf{x}_{\ell j} = \mathbf{x}^0_{\ell j} + \mathbf{u}_{\ell j}$, where $\mathbf{x}^0_{\ell j}$ is the equilibrium position of the ion, $\mathbf{u}_{\ell j}$ are the displacement operators, and each site indexed by $\ell j$, where $\ell$ indexes the unit cell, and $j$ indexes the ion inside the unit cell. This information is added in the second step, labeled ``$\langle \rangle_{\ell j}$" in Eq.~\eqref{eq:eft_game_plan}, or ``Target Expectation Value" (Sec.~\ref{subsubsec:target_expectation}), which transforms the DM-$\psi$ interaction Lagrangian to DM coupling to the lattice properties at each site, e.g., $\phi \, \psi^\dagger \psi \rightarrow \sum_{\ell j} \phi \, n_{j\psi}(\mathbf{x} - \mathbf{x}_{\ell j})$, where $n_{j\psi}$ is the number density of $\psi$ particles on the $j^\text{th}$ site, by summing the expectation values at each lattice site. 

In the last step, ``$|0 \rangle \, \rightarrow \, |\nu \mathbf{k} \rangle$" in Eq.~\eqref{eq:eft_game_plan}, or ``Form Factor Calculation" (Sec.~\ref{subsubsec:form_factor}), the form factors, or vertex rules in Eqs.~\eqref{eq:vertex_S} and~\eqref{eq:vertex_V}, are derived from the interaction Lagrangian, $\mathcal{L}(\mathbf{u}_{\ell j})$, which is written in terms of the displacement operators. This step is simply computing quantum mechanical matrix elements of the transition between an initial state with no phonon and incoming DM particle, to a final state containing no DM particle and a single phonon indexed by $\nu \mathbf{k}$.

These three steps are described in the following subsections. While the procedure to connect the UV Lagrangian to the form factors is the same for each operator $\mathcal{O}$ in Eq.~\eqref{eq:uv_lag}, the details can differ. Therefore to avoid repetition, at each step in the derivation we begin with a general discussion, and then provide an example calculation for vector DM, $V$, coupling the (spatial part) of a vector current, $\mathcal{L}_\text{UV}(\Psi) = g V_\mu \bar{\Psi} \gamma^\mu \Psi \supset - g \, V^i \, \bar{\Psi} \gamma^i \Psi$. The form factors for all of the DM models considered in Sec.~\ref{sec:results} can be found in Table~\ref{tab:UV_to_form_factor}.

\subsubsection{NR EFT}
\label{subsubsec:nr_eft}

The purpose of finding the NR limit of a UV Lagrangian is to isolate the dynamics of the two-component field, $\psi$, which satisfies the Sch\"odigner equation, within the Lagrangian containing two two-component fields inside $\Psi$. Our starting point is the Dirac Lagrangian,
\begin{align}
    \mathcal{L} = \bar{\Psi} \left( i \gamma^\mu D_\mu - m_\Psi \right) \Psi \, ,
    \label{eq:dirac_L}
\end{align}
where $D_\mu$ is the gauge covariant derivative, and $m_\Psi$ is the mass of the fermion. The problem becomes more clear after a change of variables, $\Psi \rightarrow e^{-i m_\Psi t} \Psi$ which transforms Eq.~\eqref{eq:dirac_L} to,
\begin{align}
    \mathcal{L} = \bar{\Psi} \left( i \gamma^\mu D_\mu + 2 m_\Psi P_- \right) \Psi \, ,
    \label{eq:dirac_L_2}
\end{align}
where $P_\pm = (1 \pm \gamma^0) / 2$ are projection operators. Eq.~\eqref{eq:dirac_L_2} describes the dynamics of two two-component fields, $P_\pm \Psi$, where $P_+ \Psi$ is massless, and $P_- \Psi$ is massive. For this reason, we will refer to $P_+ \Psi$ as the ``light" field and $P_- \Psi$ as the ``heavy" field. If there were no terms in Eq.~\eqref{eq:dirac_L_2} which mixed the heavy and light fields then there would be no problem; the dynamics of the two fields are decoupled.

However, the $\gamma^i D_i$ term mixes the heavy and light fields, and therefore to isolate the dynamics of the light field, a procedure to remove the heavy field needs to be performed. This is the fundamental problem of NRQED/QCD~\cite{PhysRevA.82.052520,Balk:1993eva,Gardestig:2007mka}, and there are many different approaches. We will give a summary of two methods that have been utilized in the context of DM direct detection, Refs.~\cite{Mitridate:2021ctr} and~\cite{Krnjaic:2023nxe}, and refer the reader to these references for more details.

Ref.~\cite{Mitridate:2021ctr} used the ``equation of motion" (EOM) method, which is the most physically intuitive. One simply solves for the EOM of the heavy field in terms of the light field, and then substitutes the heavy field back into the Lagrangian. This generates a Lagrangian which only depends on the light field. The NR limit of the interaction Lagrangian in Eq.~\eqref{eq:uv_lag} is also readily found; to first order in the DM coupling, one can simply substitute the heavy field which satisfies the EOM when no DM field is present. Including the $\mathcal{O}(g)$ dependence in the heavy field EOM only introduces extra $\mathcal{O}(g^2)$ terms.

While physically straightforward, when integrating out the heavy field extra time derivatives enter, which require careful field redefinitions to keep canonically normalized fields. Another approach, used in Ref.~\cite{Krnjaic:2023nxe} and known more generally as a Foldy-Wouthuysen (FW) transformation~\cite{bjorken,Gardestig:2007mk,Foldy:1949wa,Foldy_1952,Balk:1993ev,Smith:2023htu}, avoids this by removing the mixing with consecutive field redefinitions at each order in $1/m_\Psi$. That is, one finds $n$ Hermitian operators, $\{ X_0, X_1, ..., X_{n - 1}\}$ such that,
\begin{align}
    \Psi \rightarrow e^{-i m_\Psi t} \left[ \exp\left( -i \frac{X_0}{m_\Psi} \right)  \ldots \exp\left( -i \frac{X_{n - 1}}{m_\Psi^n} \right) \right] \Psi \, ,
    \label{eq:FW_field_redef}
\end{align}
removes all heavy/light field mixing to $\mathcal{O}(m_\Psi^{-n})$. One can show that the operators,
\begin{align}
    X_0 = \frac{1}{2} \gamma^i D^i ~~~,~~~
    X_1 = \frac{i}{4} \gamma^0 \gamma^i \left[ D^0, D^i \right] \, , 
    \label{eq:X0_X1}
\end{align}
remove the heavy/light field mixing to $\mathcal{O}(1/m_\Psi^2)$ when substituted into Eq.~\eqref{eq:FW_field_redef} and then Eq.~\eqref{eq:dirac_L}. These operators can then be used to simplify any Yukawa-like DM interaction in Eq.~\eqref{eq:uv_lag}, to $\mathcal{O}(g / m_\Psi^2)$, by simply substituting Eq.~\eqref{eq:FW_field_redef} into Eq.~\eqref{eq:uv_lag},
\begin{align}
    \bar{\Psi} \mathcal{O} \Psi \approx \; \psi^\dagger \, \text{Tr} \left[ P_+ \left[ \gamma^0 \mathcal{O} + \frac{i}{m_\Psi} \left[ X_0, \gamma^0 \mathcal{O} \right] - \frac{1}{m_\Psi^2} \left[ X_0, \left[ X_0, \gamma^0 \mathcal{O} \right] \right] + \frac{i}{m_\Psi^2} \left[ X_1, \gamma^0 \mathcal{O} \right]  \right] \right] \psi \, ,
    \label{eq:expanded_general_op}
\end{align}
where the $\text{Tr}$ is performed over the $2 \times 2$ block diagonal matrix, and $\psi$, in the Dirac basis, is the upper two components of $\Psi$ on the right-hand side of Eq.~\eqref{eq:FW_field_redef}. Eq.~\eqref{eq:expanded_general_op} gives the general form of the first step in Eq.~\eqref{eq:eft_game_plan}. Applying Eq.~\eqref{eq:expanded_general_op} to the example UV DM Lagrangian, and keeping terms leading order in both $1/m_\Psi$ and absorption kinematics ($q \ll \omega$), yields,
\begin{align}
    \mathcal{L}_\text{UV}(\Psi) = - g \, V^i \, \bar{\Psi} \gamma^i \Psi \longrightarrow \mathcal{L}_\text{NR}(\psi) \approx - g \, V^i \, \psi^\dagger \left( \frac{i D^i}{m_\Psi} \right) \psi \, ,
    \label{eq:example_EFT_step}
\end{align}
which comes solely from the second term in Eq.~\eqref{eq:expanded_general_op} using $[ \gamma^i, \gamma^0 \gamma^j ] = 2 \delta^{ij} $.

\subsubsection{Target Expectation Value}
\label{subsubsec:target_expectation}

Given the NR Lagrangian in terms of the electron, proton, and neutron fields, $\psi$, the DM-phonon interaction Lagrangian is simply a sum over the expectation value at each lattice site,
\begin{align}
    \mathcal{L}(\mathbf{u}_{\ell j}) = \sum_{\ell j} \langle \, \mathcal{L}_\text{NR}(\psi) \, \rangle_{\ell j} \, .
    \label{eq:target_expectation_value_equation}
\end{align}
These expectation values will then be written in terms of the target properties at each site, e.g.,
\begin{align}
    \langle \psi^\dagger \psi \rangle_{\ell j} & = n_{j \psi}(\mathbf{x} - \mathbf{x}_{\ell j}) \, ,
    \label{eq:num_density}
\end{align}
where $n_{j \psi}$ is the number density of the $\psi$ field. This step is analogous to the EFT calculation performed in Ref.~\cite{Trickle:2020oki} for DM-single phonon scattering. However, in Ref.~\cite{Trickle:2020oki} it was the scattering potential, $\mathcal{V}$, which was written as a sum of the scattering potential at each lattice site, $\mathcal{V} = \sum_{\ell j} \langle \, \mathcal{V} \, \rangle_{\ell j}$. Further simplifications were made assuming that the scattering potential only depends on $\mathbf{x}_{\ell j}$ in the same way as Eq.~\eqref{eq:num_density}, i.e., $\mathcal{V} = \sum_{\ell j} \mathcal{V}_{\ell j}(\mathbf{x} - \mathbf{x}_{\ell j})$. This simplified calculations by allowing the $\mathbf{x}_{\ell j}$ dependence to be factored out in the Fourier transform of the scattering potential, $\widetilde{V}(-\mathbf{q}) = \sum_{\ell j} e^{i \mathbf{q} \cdot \mathbf{x}_{\ell j}} \widetilde{V}_{\ell j}(-\mathbf{q})$. 

Since we are only concerned with single phonon excitations in the $q \ll \omega$ limit here, we perform a different simplification of these expectation values than in Ref.~\cite{Trickle:2020oki} by focusing on the terms that are linear in $\mathbf{u}_{\ell j}$. The other terms will not enter the matrix element calculations of the form factors performed in Sec.~\ref{subsubsec:form_factor}, and avoids the exponential dependence on $\mathbf{x}_{\ell j}$. Additionally, the derivation performed here will keep expectation values that are $\mathcal{O}(\omega)$ which were subdominant in for the scattering EFT and implicitly dropped when assuming $\mathcal{V} = \sum_{\ell j} \mathcal{V}_{\ell j}(\mathbf{x} - \mathbf{x}_{\ell j})$.

As an example, the linear order in $\mathbf{u}_{\ell j}$ term in Eq.~\eqref{eq:num_density} is,
\begin{align}
    \langle \psi^\dagger \psi \rangle_{\ell j} \rightarrow - u^i_{\ell j} \nabla^i n_{j \psi}(\mathbf{x} - \mathbf{x}^0_{\ell j}) \, .
\end{align}
An additional simplification can be made when we consider that these expectation values multiply the DM field inside the Lagrangian $\mathcal{L}_{\text{NR}}$. Therefore we can integrate by parts and move the derivative acting on the number density to the DM field, and convert to momentum space with $q^\mu = i \partial^\mu$,
\begin{align}
    \langle \psi^\dagger \psi \rangle_{\ell j} & \rightarrow i \, q^i \, u^i_{\ell j} \, n_{j \psi}(\mathbf{x} - \mathbf{x}^0_{\ell j}) \, .
\end{align}
Similar simplifications can be performed for the spin density, $s_{j \psi}^i$, 
\begin{align}
    \langle \psi^\dagger \sigma^i \psi \rangle_{\ell j} & \rightarrow 2 i \, q^k \, u^k_{\ell j} \, s^i_{j \psi}(\mathbf{x} - \mathbf{x}_{\ell j}^0) \, ,
    \label{eq:spin_density_coupling}
\end{align}
where the factor of two enters from the definition of spin, $\mathbf{S} = \bm{\sigma} / 2$.

More complicated operators can be simplified using Ehrenfest's theorem. For example, consider $\langle \psi^\dagger k^i \psi \rangle_{\ell j}$, Ehrenfest's theorem states that,
\begin{align}
    k^i = i m_\Psi \left[ H_0, x^i \right] \, ,
    \label{eq:ehrenfest_k}
\end{align}
and therefore,
\begin{align}
    \langle \psi^\dagger k^i \psi \rangle_{\ell j} & = i m_\Psi \langle \psi^\dagger \left[ H_0, x^i \right] \psi \rangle_{\ell j} = m_\Psi \left( \langle \partial_0 \psi^\dagger x^i \psi \rangle_{\ell j} + \langle \psi^\dagger x^i \partial_0 \psi \rangle_{\ell j}\right) = i \, m_{\Psi} \omega \, \langle \psi^\dagger x^i \psi \rangle_{\ell j} \nonumber \\ 
    & \rightarrow i \, m_{\Psi} \omega \, u^i_{\ell j} \, n_{j\psi}(\mathbf{x} - \mathbf{x}^0_{\ell j}) \label{eq:k_target_expectation}\, ,
\end{align}
where we have used the Schr\"odinger equation, $H_0 \psi = i \partial_0 \psi$, and, similar to $\mathbf{q}$ previously, $\omega$ represents a time derivative acting on the DM field. With one exception that will be discussed later, the operators, $1, \, \sigma^i$ and $k^i$ are the only operators needed to compute the form factors for all the models discussed here. Furthermore, we assume that there are no background vector gauge fields, and therefore $\langle \psi^\dagger i D^i \psi \rangle_{\ell j} = \langle \psi^\dagger k^i \psi \rangle_{\ell j}$.

With these target expectation values computing the example DM-phonon interaction Lagrangian from Eq.~\eqref{eq:example_EFT_step} is trivial,
\begin{align}
    \mathcal{L}_\text{NR}(\psi) \approx - g \, V^i \psi^\dagger \left( \frac{i D^i}{m_\Psi} \right) \psi \longrightarrow \mathcal{L}(\mathbf{u}_{\ell j}) \approx - i \, g \, \omega \, V^i \sum_{\ell j}  u_{\ell j}^i \, n_{j \psi}(\mathbf{x} - \mathbf{x}^0_{\ell j})  \, .
    \label{eq:example_expectation_value_step}
\end{align}

\subsubsection{Form Factor Calculation}
\label{subsubsec:form_factor}

The last step in the derivation is to identify the form factor from the DM-phonon interaction Lagrangian. This is done by computing the matrix elements from Eqs.~\eqref{eq:vertex_S} and~\eqref{eq:vertex_V},
\begin{align}
    i M_{S,\, \nu\mathbf{k}} & = i \int d^3 \mathbf{x} \, e^{i \mathbf{q} \cdot \mathbf{x}} \, \left\langle \nu\mathbf{k} \left| \, \frac{\delta \mathcal{L}(\mathbf{u}_{\ell j}) }{\delta \phi}  \, \right| 0 \right\rangle = -\sqrt{N} \delta_{\mathbf{q}, \mathbf{k}} \sum_{j} \, \bm{\mathcal{F}}_{S, j} \cdot \bm{T}_{j \nu \mathbf{k} } \\ 
    i M_{V,\, \nu\mathbf{k}}^\mu & = i \int d^3 \mathbf{x} \, e^{i \mathbf{q} \cdot \mathbf{x}} \, \left\langle \nu\mathbf{k} \left| \, \frac{\delta \mathcal{L}(\mathbf{u}_{\ell j}) }{\delta V_\mu}  \, \right| 0 \right\rangle = -\sqrt{N} \delta_{\mathbf{q}, \mathbf{k}} \sum_{j} \, \bm{\mathcal{F}}^\mu_{V,\, j} \cdot \bm{T}_{j \nu \mathbf{k} } \, ,
    \label{eq:matrix_element_form_factor_derivation}
\end{align}
respectively, where the $\delta / \delta \phi$ ($\delta / \delta V_\mu$) simply removes the scalar field, $\phi$ (vector field, $V$) from the interaction vertex, leaving only a function of the phonon operators, $\mathbf{u}_{\ell j}$. It is easiest to understand this formula in practice, and similar simplifications hold for all DM-phonon interaction Lagrangians. Consider the example $\mathcal{L}(\mathbf{u}_{\ell j})$ in Eq.~\eqref{eq:example_expectation_value_step}, in this case we have 
\begin{align}
    i M_{V, \nu\mathbf{k}}^k = - g \omega \frac{1}{\sqrt{N}}  \sum_{\ell j} T^k_{j\nu\mathbf{k}} \, e^{- i \mathbf{k} \cdot \mathbf{x}^0_{\ell j}} \int d^3\mathbf{x} \, e^{i \mathbf{q} \cdot \mathbf{x}} n_{j \psi}(\mathbf{x} - \mathbf{x}^0_{\ell j}) \, ,
    \label{eq:example_mat}
\end{align}
where we have written the displacement operator matrix element in terms of the phonon transition matrix element with Eq.~\eqref{eq:phonon_transition_matrix_element}. The integral in Eq.~\eqref{eq:example_mat} can be related to the total particle number, $N_{j\psi}$,
\begin{align}
    \int d^3\mathbf{x} \, e^{i \mathbf{q} \cdot \mathbf{x}} \, n_{j \psi}(\mathbf{x} - \mathbf{x}^0_{\ell j}) = e^{i \mathbf{q} \cdot \mathbf{x}^0_{\ell j}} \, \widetilde{n}_{j \psi}(q) \approx e^{i \mathbf{q} \cdot \mathbf{x}^0_{\ell j}} \, N_{j \psi} 
\end{align}
and the sum over $\ell$ enforces momentum conservation within the crystal, $\sum_{\ell} e^{i ( \mathbf{q} - \mathbf{k} ) \cdot \mathbf{x}_\ell} = N \delta_{\mathbf{q}, \mathbf{k}}$. After these simplifications, one can isolate the form factor,
\begin{align}
    \mathcal{F}^{k i}_{j\psi} = g \, \omega \, \delta^{ki} N_{j\psi} \, ,
    \label{eq:example_form_factor}
\end{align}
where the $k$ index corresponds to the spatial part of $\mu$, and $i$ corresponds to the index of the vector, $\bm{\mathcal{F}}_{V, \, j }^\mu$, i.e., what gets contracted with $\bm{T}_{j\nu\mathbf{k}}$ in Eq.~\eqref{eq:vertex_S}.

\section{Applications}
\label{sec:results}

In this section, we apply the formalism developed in Sec.~\ref{sec:formalism} to compute DM absorption rates into single phonon excitations for a variety of targets and DM models. Our focus will be on four classes of DM models with Yukawa-like interactions: scalar DM with couplings of the form, $\phi \bar{\Psi} \Psi$ (Sec.~\ref{subsec:scalar_DM}), pseudoscalar DM with axionlike particle (ALP) derivative couplings, $\partial_\mu a \bar{\Psi} \gamma^\mu \gamma^5 \Psi$ (Sec.~\ref{subsec:alp_DM}), vector DM from spontaneously broken gauge theories coupling to the vector current, $V_\mu \bar{\Psi} \gamma^\mu \Psi$ (Sec.~\ref{subsec:vector_DM_gauge}), and lastly vector DM coupling to the SM electric, $V_{\mu \nu} \bar{\Psi} \sigma^{\mu \nu} i \gamma^5 \Psi$, and magnetic, $V_{\mu \nu} \bar{\Psi} \sigma^{\mu \nu} \Psi$ dipole (Sec.~\ref{subsec:vector_DM_edm_mdm}). The absorption rate for each of these models can be easily computed with the help of the form factors in Table~\ref{tab:UV_to_form_factor}. Specifically, for each DM model, we derive the phonon contribution to the in-medium self-energy by substituting the form factors given in Table~\ref{tab:UV_to_form_factor} into Eq.~\eqref{eq:self_energy_phonon_contribution}. These self-energies are then substituted into Eq.~\eqref{eq:final_rate} to compute the total absorption rates. 

The phonon transition matrix elements, $\bm{T}_{j \nu \mathbf{k}}$, are computed from first principles in two steps. First, using first principles density functional theory (DFT)~\cite{Martin_2004} calculations within \textsf{VASP}~\cite{VASP_1,VASP_2,VASP_3,VASP_4,VASP_5}, the lattice is relaxed to its equilibrium position and the equilibrium positions, $\mathbf{x}_{\ell j}^0$, are found. Each ion is then displaced from its equilibrium position and the forces on the ion are computed, which generates the spring constants between the ions. \textsf{VASP} is also used to compute the high-frequency dielectric constant, $\varepsilon_\infty$. These three pieces of information are contained in the \textsf{POSCAR}, \textsf{FORCE\_SETS}, and \textsf{BORN} files output from \text{VASP} (or similar DFT software). Second, these files are then input to the \textsf{phonopy} program~\cite{phonopy} which diagonalizes the system and calculates the phonon energies, $\omega_{\nu \mathbf{k}}$, and eigenvectors, $\bm{\epsilon}_{j\nu\mathbf{k}}$. 

\textsf{PhonoDark-abs}~\GithubLink~is used to compute all absorption rates shown here. \textsf{PhonoDark-abs} performs the second step (calling \textsf{phonopy}) internally, and therefore one simply needs to supply the DFT input files (\textsf{POSCAR}, \textsf{FORCE\_SETS}, and \textsf{BORN}), for any target material, to compute the absorption rate. \textsf{PhonoDark-abs} is publicly available here~\GithubLink.

Before computing DM absorption rates, in Sec.~\ref{subsec:dielectric_calc}, we apply our formalism to compute the long-wavelength dielectric function, $\varepsilon(\omega)$. This calculation serves a variety of purposes.  First, it is needed to compute the screened contribution to the DM absorption rates in Eqs.~\eqref{eq:final_rate}, as the dielectric function is related to the self-energy of the photon via $\Pi_{AA} = \omega^2 \, ( 1 - \varepsilon(\omega) )$. Second, it provides a cross-check of our formalism since it allows us to compare our results with previous calculations of the dielectric function in terms of the low-energy electronic and phononic responses~\cite{Gonze_1997}. Finally, by comparing our results with experimentally measured values of the dielectric we can tune the phonon widths, $\gamma_\nu$, in the phonon propagator in Eq.~\eqref{eq:phonon_propagator}.

In Secs.~\ref{subsec:scalar_DM} -~\ref{subsec:vector_DM_edm_mdm} we compute the DM absorption rates into \ce{GaAs}, \ce{Al2O3} (sapphire), and \ce{SiO2} (quartz) targets. \ce{GaAs} was the first studied due to the simple structure of its unit cell~\cite{Knapen:2017ekk}, while \ce{Al2O3} is desirable for its large number of resonances and directionality~\cite{Griffin:2018bjn} and ready availability in terms of fabrication of ultra-pure single crystals. Both of these targets will be used in the TESSERACT experiment~\cite{Chang2020}. \ce{SiO2} has been previously identified as an optimal target in terms of reach to light DM scattering off phonons~\cite{Trickle:2019nya}. The DFT input files for \ce{GaAs}, \ce{Al2O3}, and \ce{SiO2} are identical to those used in previous works~\cite{Coskuner:2021qxo,Griffin:2019mvc,Trickle:2019nya,Mitridate:2020kly,Trickle:2020oki}. These targets have no spin ordering, i.e., the fermion spins are not periodically aligned, $\mathbf{S}_{\ell j \, \psi} \neq \mathbf{S}_{j \, \psi}$. Note that spin ordering includes both ferromagnetic and anti-ferromagnetic ordering. The lack of spin ordering limits the DM models that can be reached. Specifically, without spin ordering only scalar DM, vector DM with gauge interactions, and vector DM with magnetic dipole interactions (at a detrimentally suppressed rate) can be targeted. Targets with no spin ordering have no sensitivity to ALP DM, one of the most theoretically motivated DM candidates, although this can be alleviated if the sample is placed in an external $B$-field~\cite{Mitridate:2020kly,Berlin:2023ppd}. 

Therefore we also consider a magnetically ordered target, \ce{FeBr2}. This magnetic target was chosen because the first-principles calculations of its phonon properties are publicly available~\cite{Jain_2013,Ong_2013,Ong_2015,phonondb}; its purpose is to serve as an example calculation, not promote this specific target as a detector concept. The $\text{Fe}^{2+}$ ion has a magnetic moment of $\mu \approx 3.9 \, \mu_B$~\cite{FeBr2_data}, where $\mu_B$ is the Bohr magneton. Assuming that the 3$d$ electrons are orbitally quenched, as is common for transition metal electrons due to crystal field effects, the spin quantum number is $S \approx 1.8$. While \ce{FeBr2} is ferromagnetically ordered within the unit cell~\cite{Ong_2015}, it is anti-ferromagnetically ordered in adjacent cells~\cite{FeBr2_structure}. Since this target is only meant to serve as an example, we will treat it as a ferromagnet and take $\mathbf{S}_{e} \approx [ 0, 0, 1.8 ]$ on the $\text{Fe}^{2+}$ lattice site.
\begin{figure}[ht]
    \centering
    \includegraphics[width=\textwidth]{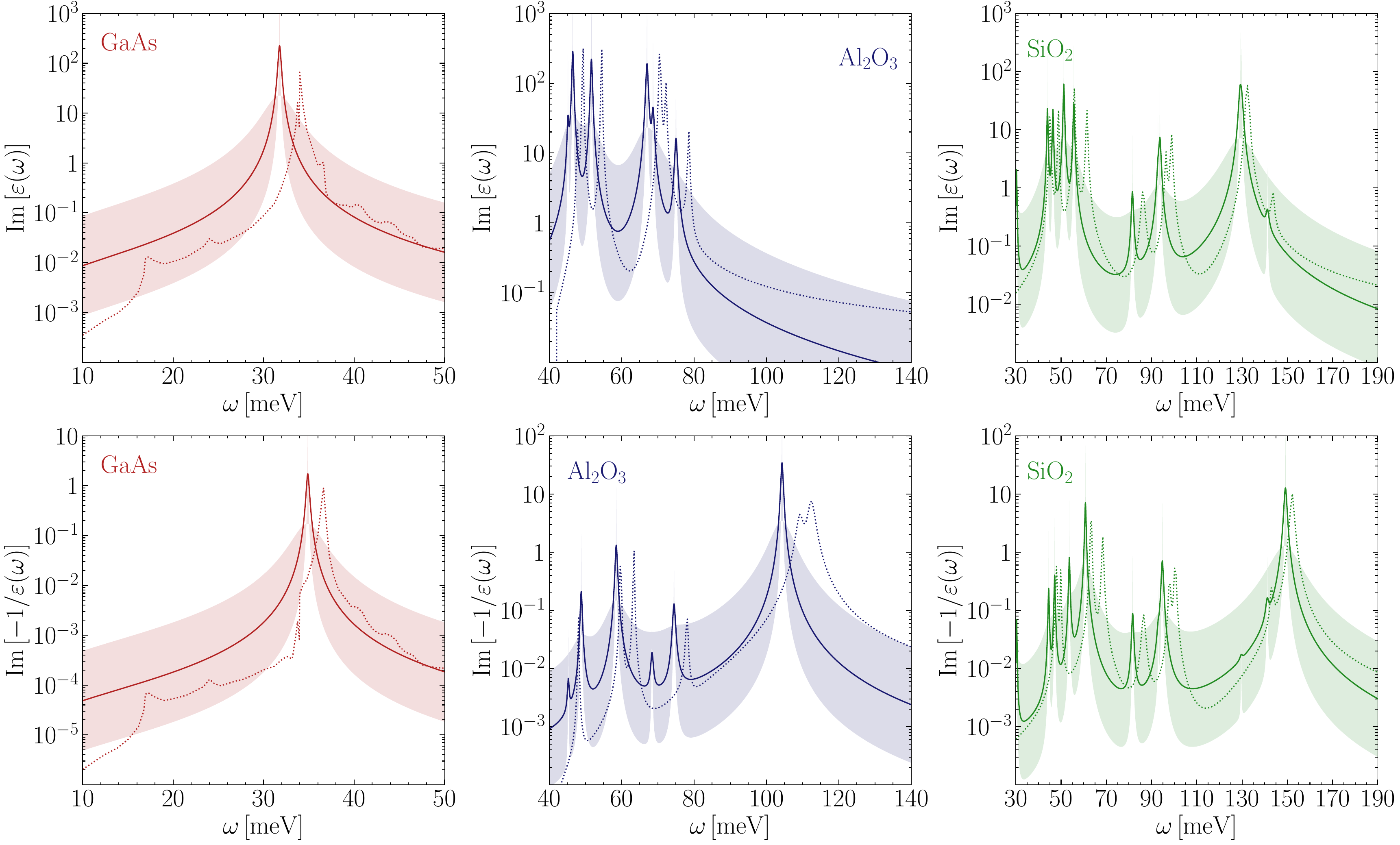}
    \caption{Comparison of the imaginary part of the dielectric function, $\text{Im} \left[ \varepsilon(\omega) \right]$ (top row), and energy loss function, $\text{Im}\left[ -1/\varepsilon(\omega) \right]$ (bottom row), from first principles calculation using Eq.~\eqref{eq:total_dielectric} (solid lines) and measurements from Ref.~\cite{Knapen:2021bwg} (dotted lines) for \ce{GaAs}, \ce{Al2O3} (sapphire), and \ce{SiO2} (quartz). Solid lines correspond to $\gamma_\nu = 10^{-2} \, \omega_\nu$, and the boundaries of the shaded regions assume $\gamma_\nu = 10^{-1} \, \omega_\nu$ and $\gamma_\nu = 10^{-3} \, \omega_\nu$.}
    \label{fig:dielectric_comp_rate}
\end{figure}
\subsection{Dielectric Function}
\label{subsec:dielectric_calc}

The long-wavelength ($q \approx 0$) dielectric function, $\varepsilon(\omega)$, receives contributions from both the electron and phonon degrees of freedom in a crystal. This can be understood simply in terms of two contributions to the photon self-energy, $\Pi_{AA} = \Pi_{AA}^\text{el} + \Pi_{AA}^\text{ph}$, where $\Pi_{AA}^\text{el}$ is the electronic response, and $\Pi_{AA}^\text{ph}$ is the phononic response. Well below the band gap, the electronic contribution is directly related to the ``high-frequency" dielectric constant, $\Pi_{AA}^\text{el} = \omega^2 (1 - \varepsilon_\infty)$, encoding the response of the electrons if the lattice ions were not allowed to move, or ``clamped". Using this, and the definition of the total dielectric function defined in terms of the total photon self-energy, $\Pi_{AA} = \omega^2 (1 - \varepsilon(\omega))$, we can write the dielectric function as 
\begin{align}
    \varepsilon(\omega) & = \varepsilon_\infty - \frac{1}{\omega^2} \Pi_{AA}^\text{ph} \, .
\end{align}
With the help of Table~\ref{tab:UV_to_form_factor} and Eq.~\eqref{eq:self_energy_phonon_contribution}, we can then compute $\Pi_{AA}^\text{ph}$ to obtain
\begin{align}
    \varepsilon(\omega) = \varepsilon_\infty + \frac{e^2}{3\Omega} \sum_{\nu} \frac{2 \omega_\nu}{\omega_\nu^2 - \omega^2 - i \omega \gamma_\nu} \,  \left( \sum_{j} Q_{j} \, T^i_{j \nu \mathbf{q}} \right) \left( \sum_{j} Q_{j} \, T^i_{j \nu \mathbf{q}} \right)^* \, ,
    \label{eq:total_dielectric}
\end{align}
where the sums over $\psi$ have been simplified in terms of the total electric charge of the ion, $\sum_{\psi} g_\psi N_{j\psi} = e \, N_{j, p} - e\, N_{j, e} \equiv e \, Q_j$, since $g_p = - g_e = e$ in the QED Lagrangian, $\mathcal{L}_\text{QED} \supset - e A_\mu \, \bar{e} \gamma^\mu e + e A_\mu \, \bar{p} \gamma^\mu p$, following the convention in Ref.~\cite{Peskin:1995ev} where $e = - |e|$. The factor of $3$ comes from taking the isotropic limit and averaging over the spatial components, $\Pi_{AA, \, \text{ph}} = \Pi^{ii}_{AA, \, \text{ph}} / 3$. This agrees with the standard result in, e.g., Ref.~\cite{Gonze_1997}, providing a validation of the formalism.\footnote{
Eq.~\eqref{eq:total_dielectric} is derived assuming that the electronic wave functions do not distort under ionic motion. These effects can be incorporated by loosening the assumption, $\langle \psi^\dagger \gamma^0 \psi \rangle_{\ell j} \approx n_{\ell j, e}(\mathbf{x} - \mathbf{x}_{\ell j}) + \delta n_{\ell j, e}$, where $\delta n_{\ell j, e} \equiv i q^i \, \delta Z_j^{ik} \, u^k_{\ell j}$, and $\delta Z^{ik}_j \equiv Z^{ik}_j - Q_j$, where $Z_j^{ik}$ are the ``Born effective charges". The spatial component, $\langle \psi^\dagger \gamma^i \psi \rangle_{\ell j}$, follows a similar simplification by the Ward identity. This adds a form factor to the photon-electron coupling, $\delta \mathcal{F}_{j e}^{\mu i} = (q^m \, \delta Z^{mi}_{j}, \omega \, \delta Z^{ki}_{j})$, replacing $Q_{j} \, T^i_{j \nu \mathbf{q}} \rightarrow Z^{ik}_{j} \, T^k_{j \nu \mathbf{q}}$ in Eq.~\eqref{eq:total_dielectric}. See Ref.~\cite{Trickle:2019nya} for more details.}

In Fig.~\ref{fig:dielectric_comp_rate} we compare the imaginary part of the dielectric function (top row) and energy loss function (ELF), $\text{Im}\left[ -1/\varepsilon(\omega) \right]$ (bottom row), computed from first principles with Eq.~\eqref{eq:total_dielectric}, to measured data from Ref.~\cite{Knapen:2021bwg} for the non-spin ordered targets \ce{GaAs}, \ce{Al2O3}, and \ce{SiO2}. The computed dielectric function is shown for different assumptions about the phonon widths, $\gamma_\nu \in \{10^{-3} \,\omega_\nu,\, 10^{-2} \, \omega_\nu,\, 10^{-1}\, \omega_\nu \}$. Smaller widths correspond to a larger resonance peak and smaller off-resonance behavior, and vice versa for larger widths. We find that the measured data can be well reproduced with phonon widths in this range, with slight shifts to the exact locations of the resonances. More sophisticated models of the widths as a function of energy could further improve these fits. For the results shown in Secs.~\ref{subsec:scalar_DM} -~\ref{subsec:vector_DM_edm_mdm} we use $\gamma_\nu = 10^{-2}\, \omega_\nu$.   

It is known that the absorption rate on phonons of some DM models, e.g., the kinetically mixed dark photon~\cite{Knapen:2017ekk,Knapen:2021bwg} and ALPs (in an external magnetic field~\cite{Berlin:2023ppd}) can be related to the measured ELF shown in Fig.~\ref{fig:dielectric_comp_rate}. That first principles calculation can reproduce the measured ELF further validates the first principles approach of computing single phonon absorption rates in these models as studied in, e.g., Ref.~\cite{Mitridate:2020kly}. In Fig.~\ref{fig:vector_kineticmixing} we explicitly compare the constraints on the kinetically mixed dark photon model from the measured and calculated ELFs, whose differences are due to the differences shown in Fig.~\ref{fig:dielectric_comp_rate}.
\begin{figure}[ht]
    \centering
    \includegraphics[width=0.6\textwidth]{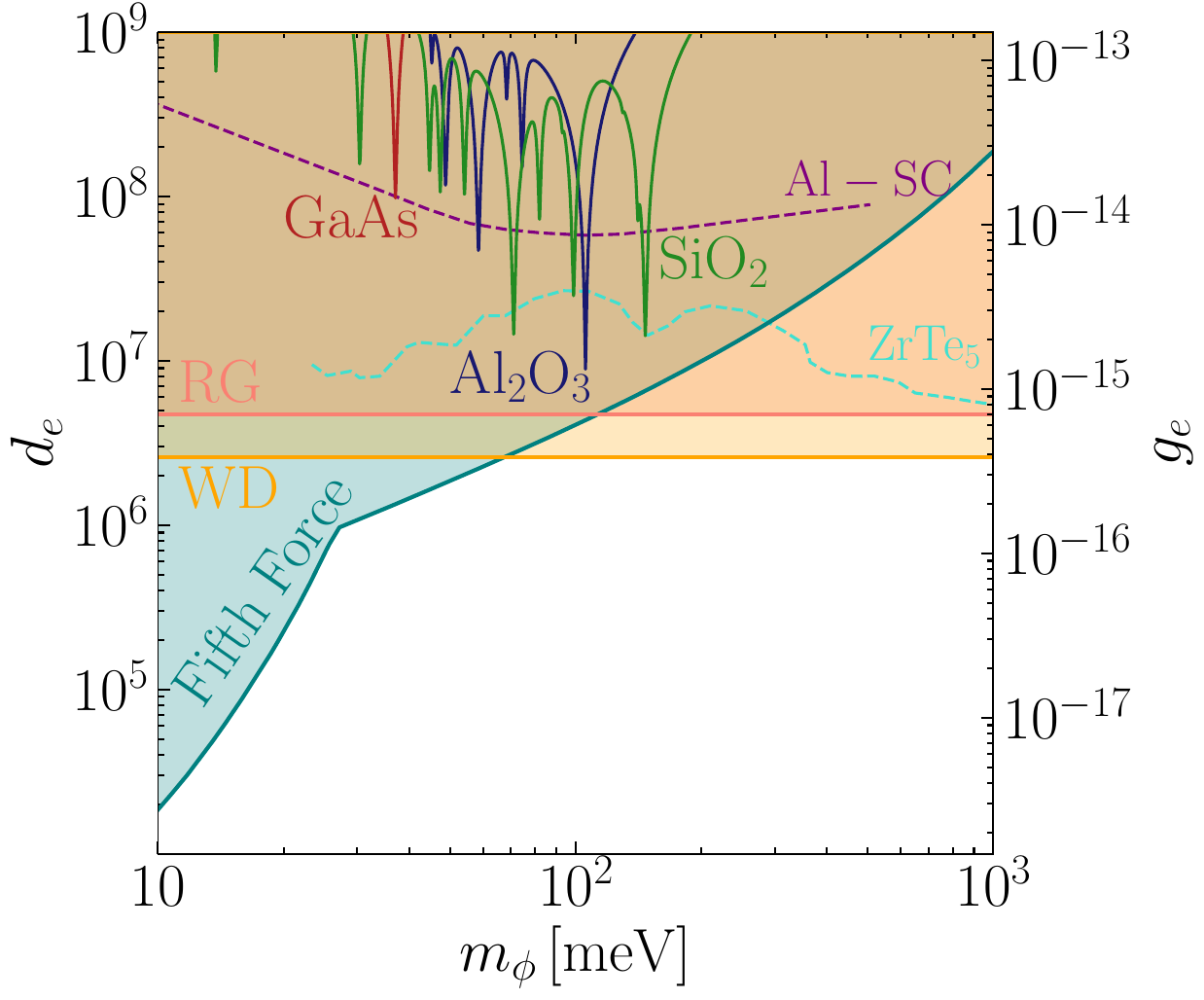}
    \caption{Projected 95\% C.L. constraints (3 events) on $d_e = g_e \Lambda / m_e$ (Eq.~\eqref{eq:L_UV_scalar}, $\Lambda = M_\text{Pl} / \sqrt{4 \pi}$), utilizing single phonon excitations in \ce{GaAs} (solid red), \ce{Al2O3} (solid blue), and \ce{SiO2} (solid green) targets, assuming a $\text{kg} \cdot \text{yr}$ exposure and no backgrounds. Dashed lines correspond to projected constraints from absorption on electrons in small band-gap targets, i.e., Al superconductors (``Al-SC", purple)~\cite{Hochberg:2016ajh, Mitridate:2021ctr}, and spin-orbit coupled targets \ce{ZrTe5} (turquoise)~\cite{Chen:2022pyd}. Shaded regions correspond to constraints from fifth force experiments (teal)~\cite{Adelberger:2003zx,KONOPLIV2011401} and stellar cooling bounds from red giants (``RG", pink)~\cite{Hardy:2016kme} and white dwarfs (``WD", orange)~\cite{Bottaro:2023gep}.}
    \label{fig:scalar_de}
\end{figure}
\subsection{Scalar DM}
\label{subsec:scalar_DM}

The first model we consider is scalar DM, $\phi$, whose couplings to electrons and nucleons are given by the Lagrangian,
\begin{align}
    \mathcal{L} \supset \sum_{\Psi \in \{ e, p, n \} } \, g_\Psi \, \phi \, \bar{\Psi} \Psi \quad\longrightarrow\quad\mathcal{L}_\text{NR} = \sum_{\psi} g_\Psi \, \psi^\dagger \psi + \mathcal{O}(1/m_\Psi^2)\,.
    \label{eq:L_UV_scalar}
\end{align}
Using Table~\ref{tab:UV_to_form_factor} the relevant self-energies are easily computed, 
\begin{align}
    \Pi_{\phi \phi} & = -i \sum_{\nu} \frac{D_{\nu}(\omega)}{\Omega} \,  \left( \sum_{j \psi} g_{\Psi} N_{j\psi} \, q^i \, T^i_{j \nu \mathbf{q}} \right) \left( \sum_{j \psi} g_{\Psi} N_{j\psi} \, q^i \, T^i_{j \nu \mathbf{q}} \right)^* \label{eq:pi_phi_phi_ph_scalar}\,, \\ 
    \Pi_{\phi A}^i & = -i e \omega \sum_{\nu} \frac{D_{\nu}(\omega)}{\Omega} \,  \left( \sum_{j \psi} g_{\Psi} N_{j\psi} \, q^k \, T^k_{j \nu \mathbf{q}} \right) \left( \sum_{j \psi} Q_{j\psi} \, T^i_{j \nu \mathbf{q}} \right)^* - \frac{g_e}{e} \, \omega q^i \, (1 - \varepsilon_\infty) \, ,\label{eq:pi_phi_A_scalar}
\end{align}
and $\Pi_{\phi A}^\text{el} = \Pi_{A \phi}^\text{el}$. Since only the imaginary component of $\Pi_{\phi \phi}$ enters in the absorption rate given in Eq.~\eqref{eq:final_rate}, we have ignored the electron contribution to $\Pi_{\phi \phi}$ as it is purely real. This will also apply to the self-energies discussed in Secs.~\ref{subsec:alp_DM}~-~\ref{subsec:vector_DM_edm_mdm}. The electron contribution to $\Pi_{\phi A}$ was derived in Refs.~\cite{Mitridate:2021ctr,Chen:2022pyd}, and is given by the last term in Eq.~\eqref{eq:pi_phi_A_scalar}.\footnote{The $\mathcal{O}(v_e^2)$ term, where $v_e$ is the velocity of the electron, in the NR Lagrangian, Eq.~\eqref{eq:L_UV_scalar} was important for absorption into electrons in Ref.~\cite{Mitridate:2021ctr}. Here its contribution to $\Pi_{\phi A}$ is suppressed relative to the term in Eq.~\eqref{eq:L_UV_scalar}, since both are $q$ suppressed in targets respecting parity. See Ref.~\cite{Chen:2022pyd} for more details.} 

While the general absorption rate is given by substituting Eqs.~\eqref{eq:pi_phi_phi_ph_scalar} and~\eqref{eq:pi_phi_A_scalar} into Eq.~\eqref{eq:final_rate}, it is illuminating to study specific combinations of the coupling constants. For example, if the $g_\Psi$ coefficients are ``photon-like", i.e., $g_p = - g_e = g, g_n = 0$, then all the self-energies are proportional to $\Pi_{AA}$ (assuming an isotropic target), indicating that the total absorption rate can be written in terms of the ELF:
\begin{align}
    \qquad R\approx \frac{g^2}{e^2} \frac{q^2}{\omega^2} \frac{\rho_\phi}{\rho_{\scriptscriptstyle T}} \, \text{Im} \left[ \frac{-1}{\varepsilon(\omega)} \right] \qquad\qquad \text{(photon-like }\phi\text{)}\,.
\end{align}
Since the absorption rate can be written in terms of the ELF, it can also be related to the dark photon absorption rate, $R_\text{dp} = (\rho_\phi / \rho_{\scriptscriptstyle T}) \, \kappa^2 \, \text{Im} \left[ - 1 / \varepsilon(\omega) \right]$~\cite{Knapen:2021bwg}. Therefore the constraints on $g$ can be related to the constraints on the mixing parameter, $\kappa$, of the dark photon model:
\begin{align}
    g  \sim 4 \times 10^{-14} \, \left( \frac{\kappa}{10^{-16}} \right) \label{eq:g_scalar_DP}\,.
\end{align}

When the couplings are proportional to the particle masses, $g_\Psi = g \, m_\Psi$, the ions are shaken in-phase, and therefore optical, or out-of-phase, oscillations are not excited. Therefore, as optical phonons are the only ones that can match DM absorption kinematics, this leads to a vanishing absorption rate. This effect, sometimes referred to as the ``coupling to mass" effect~\cite{Cox:2019cod,Griffin:2018bjn,Knapen:2017ekk,Mitridate:2020kly}, mathematically corresponds to the statement that
\begin{align}
    \sum_{j} m_j T^i_{j \nu \mathbf{q}} \approx 0 \, ,
    \label{eq:coupling_to_mass}
\end{align}
in the absorption kinematics limit. While the cancellation is exact for couplings $g_\Psi \propto m_\Psi$, it is also important even when the couplings are approximately proportional to the masses. For example, consider only coupling to the electron number on each site in \ce{GaAs}, $N_{\text{Ga},\, e} = 28$, $N_{\text{As},\, e} = 36$. Parameterically, one would expect $\mathcal{F}_j \propto N_{j\, e}$, however because the masses are $m_\text{Ga} = 69.7 \, \text{u}, m_\text{As} = 74.9 \, \text{u}$, when one subtracts off the contribution which vanishes due to the coupling to mass effect, $\mathcal{F}_j \rightarrow \mathcal{F}_j - \hat{m}_j \sum_j \mathcal{F}_j \hat{m}_j$, where $\hat{m}_j = m_j / \sqrt{\sum_j m_j^2}$, the form factors are roughly a factor of $10$ smaller than $N_{j\,e}$.

The scalar DM models affected by the coupling to mass effect are fairly generic. This is because both the proton and neutron masses are dominantly dependent on the same quantity, the QCD scale. Therefore scalar DM models which couple to the QCD field strength kinetic term~\cite{Damour:2010rp}, or the benchmark hadrophilic DM model~\cite{Knapen:2017xzo} satisfy $g_{p, n} = g \, m_{p, n}$, and single phonon excitations will have limited reach due to the coupling to mass effect. Because of this, and since constraints on DM models with photon-like couplings can be trivially related to dark photon constraints, we focus on a DM model with only coupling to electrons, which suffers less from the coupling to mass effect, as discussed previously.

In Fig.~\ref{fig:scalar_de} we compare the constraints on the electron coupling derived in this work to stellar cooling~\cite{Hardy:2016kme,Bottaro:2023gep} and fifth force~\cite{Adelberger:2003zx,KONOPLIV2011401} constraints assuming no backgrounds and a $\text{kg} \cdot \text{yr}$ exposure. To facilitate the comparison with other conventions for the coupling constant, we show the constraints both on $g_e$, defined in Eq.~\eqref{eq:L_UV_scalar}, and the commonly adopted parameterization $d_e = g_e \Lambda / m_e$, where $\Lambda = M_\text{Pl} / \sqrt{4 \pi}$, and $M_\text{Pl}$ is the Planck mass. Note that, due to the mixing term in Eq.~\eqref{eq:final_rate}, the resonance structure of the constraints in Fig.~\ref{fig:scalar_de} does not necessarily match the resonance structure of the self-energies, which have resonances at $\omega_\nu$. This is analogous to the difference in the resonance structure between the top and bottom rows of Fig.~\ref{fig:dielectric_comp_rate}. The resonances of $\text{Im}\left[ \varepsilon(\omega) \right]$ at $\omega_\nu$ are inherited from the propagator in Eq.~\eqref{eq:phonon_propagator}, which differ from the resonances in the ELF.
\begin{figure}[ht]
    \centering
    \includegraphics[width=\textwidth]{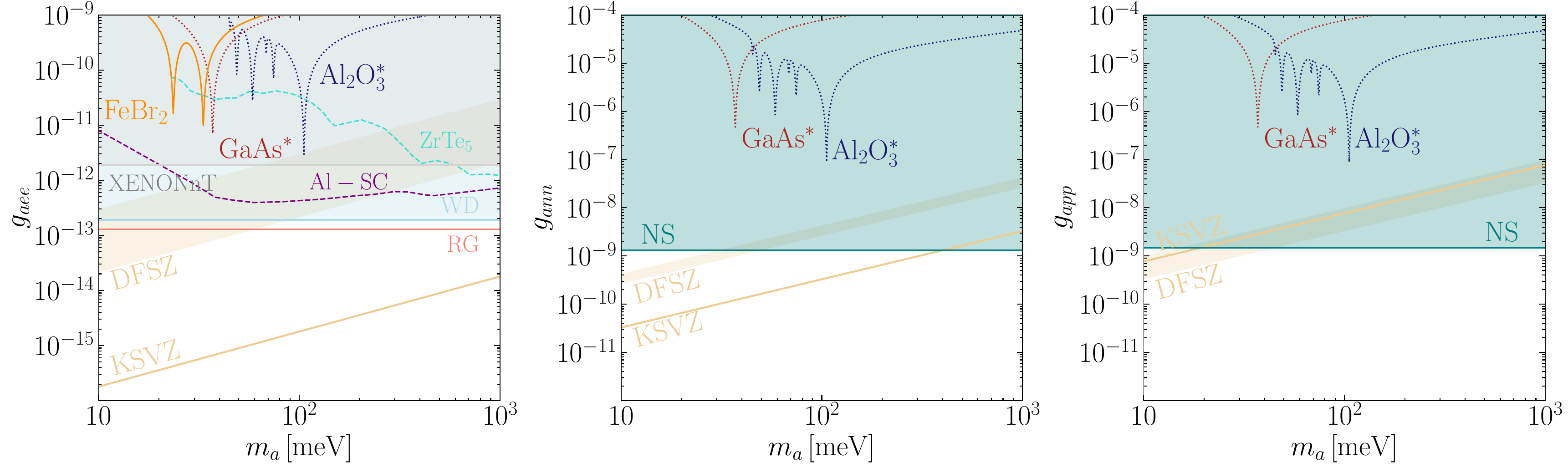}
    \caption{Projected 95\% C.L. constraints (3 events) on the ALP couplings $g_{aee}, g_{ann}, g_{app}$, Eq.~\eqref{eq:L_UV_axion} shown in the left, middle, and right panels, respectively, utilizing single phonon excitations in a variety of targets assuming a $\text{kg} \cdot \text{yr}$ exposure and no backgrounds. \ce{FeBr2} (solid orange) is a ferromagnetic target with polarized electronic spins on the \ce{Fe} site. In each panel the dotted curves labeled \ce{GaAs}$^*$ (blue) and \ce{Al2O3}$^*$ (red) correspond to a \ce{GaAs} and \ce{Al2O3} target whose total fermionic spin at each lattice site has been set to $\mathbf{S}_{j, e} = [ 0, 0, 0.5 ],\, \mathbf{S}_{j, p/n} = N_{j, p/n} \, [ 0, 0, 0.5 ]$, where $\{ e, p, n\}$ correspond to the left, middle, and right panels, respectively. Since these are not real targets, their purpose is to give an estimate of a target that does have non-zero spin ordering. As in Fig.~\ref{fig:scalar_de}, dashed lines correspond to projected constraints from absorption on electrons in small band-gap targets, i.e., Al superconductors (``Al-SC", purple)~\cite{Hochberg:2016ajh, Mitridate:2021ctr}, and spin-orbit coupled targets \ce{ZrTe5} (turquoise)~\cite{Chen:2022pyd}. In the left panel, the shaded gray region corresponds to constraints from solar axion searches with the XENONnT experiment~\cite{XENON:2022ltv}, and the shaded light blue region corresponds to white dwarf (``WD") cooling constraints~\cite{Bottaro:2023gep}. The red line corresponds to constraints from red giant (``RG") cooling~\cite{Capozzi:2020cbu,Giannotti:2017hny}, which have come under recent scrutiny~\cite{Dennis:2023kfe}. In the middle and right panel, the shaded teal region corresponds to neutron star (``NS") cooling~\cite{Buschmann:2021juv}. Tan lines correspond to the prototypical KSVZ and DFSZ QCD axion models~\cite{ParticleDataGroup:2022pth}, assuming $0.28 \leq \tan{\beta} \leq 140$ in the DFSZ model.}
    \label{fig:alp_gaff}
\end{figure}
\subsection{Axionlike Particle DM}
\label{subsec:alp_DM}

The QCD axion~\cite{Peccei:1977ur,Peccei:1977hh,Wilczek:1977pj,Weinberg:1975ui} is one of the most theoretically motivated DM candidates since it also provides a solution to the strong-CP problem. The canonical QCD axion DM candidate, with an abundance set by the post-inflationary misalignment mechanism~\cite{Abbott:1982af,Turner:1983he,Turner:1985si,Preskill:1982cy,Dine:1982ah}, is predicted to have a mass in the $10^{-6} \, \text{eV} \lesssim m_a \lesssim 10^{-5} \, \text{eV}$ range, well below the scale of gapped phonon excitations in crystal targets. However, non-standard production mechanisms, as well as ALPs that do not necessarily solve the strong-CP problem, can prefer larger values of the axion mass~\cite{Chang:2019tvx,Co:2019jts,Co:2019wyp,Co:2017mop,Harigaya:2019qnl,Co:2020dya,Daido:2017wwb,Daido:2017tbr,Battye:1994au,Hagmann:1998me,Hiramatsu:2010yu,Klaer:2017qhr,Klaer:2017ond,Gorghetto:2018myk,Vaquero:2018tib,Buschmann:2019icd,Hindmarsh:2019csc,Gorghetto:2020qws,Dine:2020pds,Buschmann:2021sdq,Hindmarsh:2021zkt}.

Our focus will be on the derivative ALP couplings,\footnote{The Lagrangian in Eq.~\eqref{eq:L_UV_axion} is equivalent to $\mathcal{L} \supset - \sum_\Psi g_{a\Psi\Psi} \, a \, \bar{\Psi} i \gamma^5 \Psi$. Both forms of the Lagrangians give the same form factor in Table~\ref{tab:UV_to_form_factor}, but one needs to expand to $\mathcal{O}(1/m_\Psi^2)$ when taking the NR limit of the $a \bar{\Psi} i \gamma^5 \Psi$ Lagrangian. The equivalence can be shown explicitly using the relationship in Eq.~\eqref{eq:comm_relationship}. See Sec.~\ref{subsec:vector_DM_edm_mdm} for more details.}
\begin{align}
    \mathcal{L} \supset \sum_{\Psi \in \{ e, p, n \} } \frac{g_{a\Psi\Psi}}{2 m_\Psi} \, \partial_\mu a \bar{\Psi} \gamma^\mu \gamma^5 \Psi \quad\longrightarrow\quad \mathcal{L}_\text{NR} \approx -i  \sum_\psi \frac{g_{a\Psi\Psi}}{2 m_\Psi^2} \omega \, a\, \psi^\dagger \, \bm{\sigma} \cdot (i \mathbf{D}) \, \psi\,.
    \label{eq:L_UV_axion}
\end{align}
Na\"ively, the leading order term in the NR Lagrangian seems to be higher order than the ``axion wind" term, $\propto g_{a\Psi \Psi} \,a \, \mathbf{q} \cdot \bm{\sigma} / m_\Psi$. However, when evaluating the target expectation value of the ``wind" term, an additional factor of $q$ enters the form factor via Eq.~\eqref{eq:spin_density_coupling}. Therefore, the form factor for the ``wind" term is order $q^2 / m_\Psi$, which is much smaller than the form factor for the term in Eq.~\eqref{eq:L_UV_axion} (see  Table~\ref{tab:UV_to_form_factor}). More generally, when the leading order term in the Lagrangian is $\mathcal{O}(q / m_\Psi)$, it is important to check if at next-to-leading order in the $\mathcal{O}(1/m_\Psi^2)$ NR expansion there are terms which dominate. For example, a term of the form $\omega k / m_\Psi^2$, where $k$ is the fermion momentum, is dominant compared to the $q / m_\Psi$ term. This feature will also appear when discussing the dipole DM models in Sec.~\ref{subsec:vector_DM_edm_mdm}.

While the phonon contribution to the self-energies is straightforward to derive using the form factor in Table~\ref{tab:UV_to_form_factor}, simplifying the electron contribution to $\Pi_{aA}$ in a spin ordered target requires an additional approximation. This can be understood physically: the high-frequency dielectric corresponds to the electronic response to an electric field, which couples identically to all electrons in the target. Therefore for any effect to be related to the high-frequency dielectric, it must affect all the electrons in the same way, i.e., all the electrons have the same spin such that the spin density is proportional to the number density, $s_e^i = \hat{s}_e^i n_e$. However, while all the electrons must be spin polarized for an exact correspondence, $\Pi_{aA}$ may be approximately written in terms of $\varepsilon_\infty$ as long as the electrons which give the dominant contribution to $\varepsilon_\infty$ are spin polarized. Under these approximations the relevant self-energies are,
\begin{align}
    \Pi_{aa} & = -i \omega^4 \sum_{\nu} \frac{D_{\nu}(\omega)}{\Omega} \,  \left( \sum_{j \psi}  \, \frac{g_{a \Psi \Psi}}{m_\psi} \, S_{j\psi}^i \, T^i_{j \nu \mathbf{q}} \right) \left( \sum_{j \psi} \frac{g_{a \Psi \Psi}}{m_\psi} \, S_{j\psi}^i \, T^i_{j \nu \mathbf{q}} \right)^* \label{eq:pi_a_a_ph_alp} \\ 
    \Pi_{a A}^i & = -e \omega^3 \sum_{\nu} \frac{D_{\nu}(\omega)}{\Omega} \,  \left( \sum_{j \psi}  \, \frac{g_{a \Psi \Psi}}{m_\psi} \, S_{j\psi}^k \, T^k_{j \nu \mathbf{q}} \right) \left( \sum_{j \psi} Q_{j\Psi} \, T^i_{j \nu \mathbf{q}} \right)^* + \frac{i g_{aee} \omega^3}{e m_e} \hat{s}_e^i (1 - \varepsilon_\infty) \, ,\label{eq:pi_a_A_alp}
\end{align}
and $\Pi_{a A}^\text{el} = -\Pi_{A a}^\text{el}$.

Similar to Sec.~\ref{subsec:scalar_DM} the absorption rate can be simplified with specific combinations of the coupling constants. For example, if the target is a ferromagnet, and $g_{app} = - m_p g_{aee} / m_e, g_{ann} = 0$, then the absorption rate can be expressed in terms of the ELF (and dark photon absorption rate) as,
\begin{align}
    \qquad R \approx \frac{1}{4} \frac{ g_{aee} }{ e } \frac{\omega}{m_e} \frac{\rho_\phi}{\rho_{\scriptscriptstyle T}} \text{Im} \left[ \frac{-1}{\varepsilon(\omega)} \right] \qquad\qquad \text{(photon-like }a\text{)}\,.
\end{align}
In this case, the constraints on $g_{aee}$ can be related to the constraints on the dark photon coupling $\kappa$:
\begin{align}
    g_{aee} \sim 10^{-9} \left( \frac{100 \, \text{meV}}{\omega} \right) \left( \frac{\kappa}{10^{-16}} \right) 
\end{align}
Additionally, if $\sum_{\psi} g_{a\Psi\Psi} S_{j \psi} / m_\psi \propto m_j$, then the coupling to mass selection rule in Eq.~\eqref{eq:coupling_to_mass} applies and the absorption rate vanishes. While the coupling combinations are more contrived than for the scalar DM models, these two scenarios serve as benchmark points to understand different limits of the theory.

In Fig.~\ref{fig:alp_gaff} we compute the projections (assuming a $\text{kg} \cdot \text{yr}$ exposure and no backgrounds) for three models, where the only non-zero coupling is the one plotted. We compare these to stellar cooling bounds~\cite{Capozzi:2020cbu,Giannotti:2017hny,Bottaro:2023gep,Buschmann:2021juv}, the canonical DFSZ and KSVZ QCD axion model~\cite{ParticleDataGroup:2022pth} predictions. The solid line in the left panel corresponds to the ferromagnetic \ce{FeBr2} target. The dashed lines, labeled \ce{GaAs}$^*$, \ce{Al2O3}$^*$ do not correspond to real targets; \ce{GaAs} and \ce{Al2O3} do not have electron, neutron, or proton spin ordering. We show these curves to illustrate what the projections might be in similar targets with proton or neutron spin ordering, which can be achieved in the presence of a strong magnetic field~\cite{Berlin:2022mia}. 

\begin{figure}[ht]
    \centering
    \includegraphics[width=\textwidth]{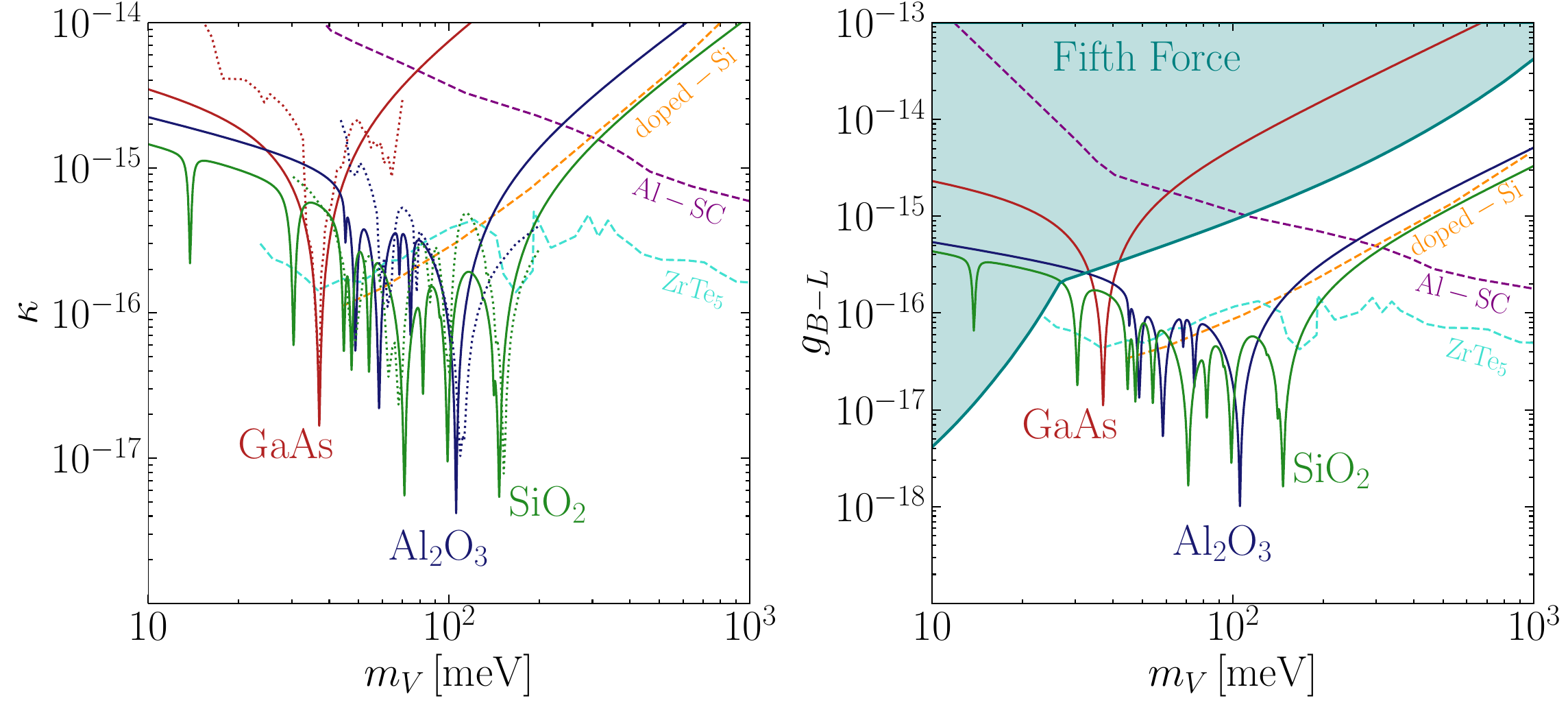}
    \caption{Projected 95\% C.L. constraints (3 events) on $\kappa = -g_e / e = g_p / e$ (left panel) and $g_{B-L}$, Eq.~\eqref{eq:vec_lag}, in \ce{GaAs} (solid red), \ce{Al2O3} (solid blue), and \ce{SiO2} (solid green) targets utilizing single phonon excitations assuming a $\text{kg} \cdot \text{yr}$ exposure and no backgrounds. Projected constraints on the kinetically mixed dark photon model have also been shown in Ref.~\cite{Knapen:2021bwg}; the purpose of the comparison here is to illustrate the good agreement between the first principles calculation performed here, and the data-driven approach (dotted lines) utilizing the ELF~\cite{Knapen:2021bwg}, also compared in Fig.~\ref{fig:dielectric_comp_rate}. Dashed lines are projected constraints from targets utilizing electronic absorption: doped Si (orange)~\cite{Du:2022dxf}, Al superconductors (``Al-SC", purple)~\cite{Hochberg:2016ajh,Mitridate:2021ctr}, and the spin-orbit coupled target \ce{ZrTe5} (turquoise)~\cite{Chen:2022pyd} Fifth force constraints are from Ref.~\cite{ADELBERGER2009102}.}
    \label{fig:vector_kineticmixing}
\end{figure}

We find that single phonon absorption is weaker than stellar cooling bounds for all couplings, especially $g_{app}$ and $g_{ann}$. We note that the reach on $g_{app}$ and $g_{ann}$ is severely affected by the coupling to mass effect. If the proton or neutron spins could anti-align on different sites, to avoid the coupling to mass selection rule, the reach would improve. However this seems experimentally unfeasible. While the $g_{aee}$ constraint from \ce{Al2O3}$^*$ is competitive with the XENONnT bounds~\cite{XENON:2022ltv}, and nearly reaches the DFSZ band, the white dwarf~\cite{Bottaro:2023gep} and red giant~\cite{Capozzi:2020cbu,Giannotti:2017hny} cooling bounds are stronger by roughly an order of magnitude on resonance. However, recently there has been some uncertainty surrounding the stellar cooling bounds on $g_{aee}$~\cite{Dennis:2023kfe}, which may re-open the parameter space. Absorption into magnons via the wind coupling~\cite{Mitridate:2020kly} is still the dominant process in electron spin ordered targets. This is because the magnon response from the wind coupling does not suffer the extra $q$ suppression that the phonon response does, as discussed previously. However, the strict selection rules governing that process~\cite{Mitridate:2020kly} severely limits the number of useful modes in simple targets, especially in the absence of an external magnetic field, and single magnon readout is still a developing technology.

\subsection{Vector DM: Gauge Coupling}
\label{subsec:vector_DM_gauge}

Next, we consider vector DM, $V$, coupled to the SM fermion vector currents, 
\begin{align}\label{eq:vec_lag}
    \mathcal{L} \supset \sum_{\Psi \in \{ e, p, n \}} g_{\Psi} \, V_\mu \bar{\Psi} \gamma^\mu \Psi\qquad\longrightarrow\qquad \mathcal{L}_\text{NR}(\psi) \approx - g \, V^i \, \psi^\dagger \left( \frac{i D^i}{m_\Psi} \right) \psi  \, ,
\end{align}
generally arising from $U(1)$ gauge theories. The self-energies are straightforwardly computed,
\begin{align}
    \Pi_{VV}^{ii} & = -i \omega^2 \sum_{\nu} \frac{D_{\nu}(\omega)}{\Omega} \,  \left( \sum_{j \psi} g_\Psi \, N_{j\psi} \, T^i_{j \nu \mathbf{q}} \right) \left( \sum_{j \psi} g_\Psi \, N_{j\psi} \, T^i_{j \nu \mathbf{q}} \right)^* \label{eq:pi_phi_phi_ph_vec_gauge} \\ 
    \Pi_{V A}^{ik} & = -i e \omega^2 \sum_{\nu} \frac{D_{\nu}(\omega)}{\Omega} \,  \left( \sum_{j \psi} g_\Psi \, N_{j\psi} \, T^i_{j \nu \mathbf{q}} \right) \left( \sum_{j \psi} Q_{j\psi} \, T^k_{j \nu \mathbf{q}} \right)^* - \frac{g_e}{e} \omega^2 \delta^{ik} (1 - \varepsilon_\infty) \, , \label{eq:phi_phi_A_vec_gauge}
\end{align}
and $\Pi_{V A}^\text{el} = \Pi_{A V}^\text{el}$.

For $g_p = - g_e = \kappa e$, and $g_n = 0$, we recover the kinetically mixed dark photon model~\cite{Fabbrichesi:2020wbt}, where $\kappa$ is the kinetic mixing parameter. As shown in Refs.~\cite{Knapen:2017ekk,Knapen:2021bwg}, the absorption rate for this model is directly related to ELF shown in Fig.~\ref{fig:dielectric_comp_rate}. Interactions of the form given in Eq.~\eqref{eq:vec_lag}, also arise in models where global symmetries of the SM are gauged at some high energy scale, and then subsequently broken to introduce a mass to the DM. Two common examples of interest here are $U(1)_B$ and $U(1)_{B - L}$, where $B$ is baryon number. However, due to the coupling to mass effect, the $U(1)_B$ gauge field cannot be absorbed into single phonon excitations. Therefore we focus on the $U(1)_{B - L}$ model, which behaves identically to a $U(1)_L$ model.

In Fig.~\ref{fig:vector_kineticmixing} we compute projected constraints on the kinetically mixed dark photon model (left panel) and $U(1)_{B-L}$ model (right panel) assuming a $\text{kg} \cdot \text{yr}$ exposure and no backgrounds. In the $U(1)_{B - L}$ we also compare our results to the constraints from fifth force experiments~\cite{ADELBERGER2009102}. Since the constraints on the kinetically mixed dark photon model have been computed previously, the main purpose of this figure is to compare our results with the ones derived by using the ELF. We see that absorption into single phonons in any of the \ce{GaAs}, \ce{Al2O3} or \ce{SiO2} targets can be far superior not only to fifth force constraints in the $30 \, \text{meV} \lesssim m_V \lesssim 100 \, \text{meV}$ mass window, but also absorption into small gap electronic excitations.
\begin{figure}[ht]
    \centering
    \includegraphics[width=\textwidth]{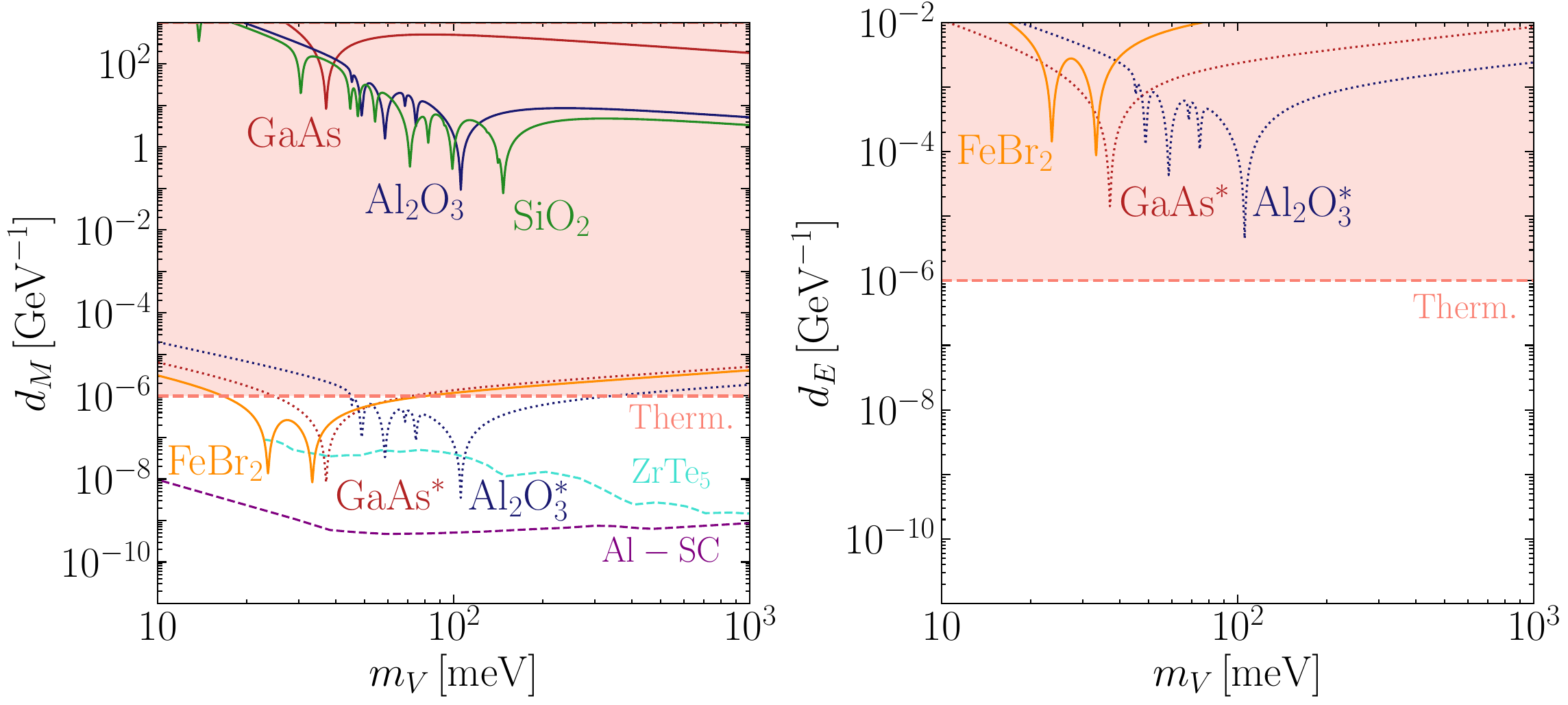}
    \caption{
    Projected 95\% C.L. constraints (3 events) utilizing single phonon excitations on the MDM coupling to electrons, $d_M$, and the EDM coupling to electrons, $d_E$, in the left and right panels, respectively, assuming a $\text{kg} \cdot \text{yr}$ exposure and no backgrounds. Constraints from \ce{GaAs} (solid red), \ce{Al2O3} (solid blue), and \ce{SiO2} (solid green) targets are weak due to the response in a non-spin ordered target coming at higher order. Similar to Fig.~\ref{fig:alp_gaff} we also consider a target with ferromagnetic ordering, \ce{FeBr2} (solid orange), and use \ce{GaAs}$^*$ (dotted red) and \ce{Al2O3}$^*$ (dotted blue) as an example for other targets with ferromagnetic ordering. Projected constraints from small band-gap electronic absorption in Al superconductors (``Al-SC", purple)~\cite{Hochberg:2016ajh, Mitridate:2021ctr}, and spin-orbit coupled target, \ce{ZrTe5},~\cite{Chen:2022pyd}, shown as dashed lines, have been rescaled according to Ref.~\cite{Krnjaic:2023nxe}. The shaded red region (``Therm.") is excluded cosmologically; couplings in this region would overproduce DM via freeze-in, even at a reheat temperature of $\sim \text{MeV}$~\cite{Krnjaic:2022wor,Krnjaic:2023nxe}.}
    \label{fig:vector_multipole_current}
\end{figure}
\subsection{Vector DM: Electric and Magnetic Dipole}
\label{subsec:vector_DM_edm_mdm}

The last DM models we consider are again vector DM models, but this time with a magnetic or electric dipole coupling to SM fermions. We will refer to these models as MDM and EDM models, respectively. These models were studied in Refs.~\cite{Dobrescu:2004wz,Krnjaic:2022wor,Krnjaic:2023nxe} in the context of DM with mass $m_V \gtrsim\text{keV}$, where it was shown that production could occur via the standard freeze-in mechanism~\cite{Krnjaic:2022wor} but is dominated during reheating due to the dimension five nature of the operators. Other production mechanisms could produce the DM nonthermally with the MDM/EDM couplings at much smaller masses~\cite{Graham:2015rva,Dror:2018pdh,Adshead:2023qiw,Arias:2012az,Nelson:2011sf,Dimopoulos:2006ms}, and therefore it is interesting to study the constraints on these models below the $\text{eV}$ scale where constraints from electronic excitations begin~\cite{Krnjaic:2023nxe}.

The UV Lagrangians of the MDM and EDM models are,
\begin{align}
    \mathcal{L}_M & \supset \sum_{\Psi \in \{ e, p, n \}} \frac{d_M^\Psi}{2} \, V_{\mu\nu} \bar{\Psi} \sigma^{\mu \nu} \Psi \\
    \mathcal{L}_E & \supset \sum_{\Psi \in \{e, p, n \}} \frac{d_E^\Psi}{2} \, V_{\mu\nu} \bar{\Psi} \sigma^{\mu \nu} i \gamma^5 \Psi \, ,
\end{align}
respectively, where $V_{\mu \nu} \equiv \partial_\mu V_\nu - \partial_\nu V_\mu$. The coupling to the spatial part of the vector DM can be further simplified as,
\begin{align}
    \frac{1}{2} V_{\mu \nu} \bar{\Psi} \sigma^{\mu \nu} \Psi & \supset V^i \left( i \omega \, \bar{\Psi} \sigma^{0i} \Psi + i q^j \, \bar{\Psi} \sigma^{ij} \Psi \right) \label{eq:expanded_MDM}\\
    \frac{1}{2} V_{\mu \nu} \bar{\Psi} \sigma^{\mu \nu} i \gamma^5 \Psi & \supset V^i \left( i \omega \, \bar{\Psi} \sigma^{0i} i \gamma^5 \Psi + i q^j \, \bar{\Psi} \sigma^{ij} i \gamma^5 \Psi \right) \, , \label{eq:expanded_EDM}
\end{align}
and the NR Lagrangians are,
\begin{align}
    \mathcal{L}_{M, \text{NR}} & \approx \sum_{\psi} d_M^\Psi \, \frac{i \omega}{m_\Psi} V^i \, \psi^\dagger \left[ \left( \bm{\sigma} \times (i \mathbf{D}) \right)^i + \frac{i}{2 m_\Psi} \left[ D^0, D^i \right] \right] \psi \\
    \mathcal{L}_{E, \text{NR}} & \approx \sum_{\psi} d_E^\Psi \, V^i \, \psi^\dagger \left[ - i \omega \sigma^i + \frac{i}{m_\Psi} \left( \sigma^i (\mathbf{q} \cdot (i \mathbf{D}))  - (i D^i) (\mathbf{q} \cdot \bm{\sigma}) \right) \right] \psi \label{eq:L_NR_EDM}\, .
\end{align}
Two terms are kept in the NR limit of the MDM model, the first is the leading order response when the target is spin ordered, and the second is the leading order response when the target is not spin ordered. The NR limit of the EDM has multiple terms due to contributions of similar order from the NR limit of $\bar{\Psi} \sigma^{0i} i \gamma^5 \Psi$ and $\bar{\Psi} \sigma^{ij} i \gamma^5 \Psi$ in Eq.~\eqref{eq:expanded_EDM}. Note that all terms in Eq.~\eqref{eq:L_NR_EDM} will contribute at the same order in the form factors.

Before continuing to the self-energies we comment on the target expectation value of $\left[ D^0, D^i \right]$, which is a bit subtle. We assume that the SM fermions are bound at the lattice site by the temporal component of the gauge fields. For example, the electrons are bound by the potential $e A^0$, which is simply the electrostatic potential generated by the ion. Assuming that the protons and neutrons are bound by similar strong forces, then $D^0 = \partial^0 + i \mathcal{V}$, where $\mathcal{V}$ is the binding potential, e.g., the electromagnetic part of $\mathcal{V}$ is simply $e A^0$. The NR Lagrangian of the SM fermions is then simply the Schr\"odinger equation with this potential, $H = p^2 / 2 m_\Psi + \mathcal{V}$. Furthermore, assuming that the target expectation value of the spatial part of the gauge fields vanishes (i.e., the spatial part of the gauge fields does not significantly impact binding) then $\left[ D^0, D^i \right] \rightarrow \left[ H, k^i \right]$, and therefore,
\begin{align}
    \langle \psi^\dagger \left[ D^0, D^i \right] \psi \rangle_{\ell j} = \omega \, \langle \psi^\dagger k^i \psi \rangle_{\ell j} \, ,
    \label{eq:comm_relationship}
\end{align}
which can then be straightforwardly written in terms of the displacement operator with Eq.~\eqref{eq:k_target_expectation}. 

In the case of no spin ordering, only the MDM self-energies are non-zero,
\begin{align}
    \Pi_{M, \, VV}^{ii} & = -i \frac{\omega^6}{4} \sum_{\nu} \frac{D_{\nu}(\omega)}{\Omega} \,  \left( \sum_{j \psi} \frac{d_M^\Psi}{m_\psi} \, N_{j\psi} \, T^i_{j \nu \mathbf{q}} \right) \left( \sum_{j \psi} \frac{d_M^\Psi}{m_\psi} \, N_{j\psi} \, T^i_{j \nu \mathbf{q}} \right)^* \label{eq:pi_phi_phi_ph_vec_mdm_N} \\ 
    \Pi_{M, \, V A}^{ik} & = -ie \frac{\omega^4}{2} \sum_{\nu} \frac{D_{\nu}(\omega)}{\Omega} \,  \left( \sum_{j \psi} \frac{d_M^\Psi}{m_\psi} \, N_{j\psi} \, T^i_{j \nu \mathbf{q}} \right) \left( \sum_{j \psi} Q_{j\psi} \, T^k_{j \nu \mathbf{q}} \right)^* - \frac{d_M^e}{e} \frac{\omega^4}{2 m_e} \, \delta^{ik} (1 - \varepsilon_\infty) \, ,
 \label{eq:phi_phi_A_vec_mdm_N}
\end{align}
where the $M$ subscript denotes the MDM model (the $E$ will subscript denotes the EDM model) and $\Pi_{M,\, V A}^\text{el} = \Pi_{M,\, A V}^\text{el}$. For targets with spin ordering both MDM and EDM have self-energies,
\begin{align}
    \Pi_{M, \, VV}^{ii} & = -4i \omega^4 \sum_{\nu} \frac{D_{\nu}(\omega)}{\Omega} \,  \left( \sum_{j \psi} d_M^\Psi \, \epsilon^{imk} S_{j \psi}^{m} \, T^k_{j \nu \mathbf{q}} \right) \left( \sum_{j \psi} d_M^\Psi \, \epsilon^{imk} S_{j \psi}^{m} \, T^k_{j \nu \mathbf{q}} \right)^* 
 \label{eq:pi_phi_phi_ph_vec_mdm} \\ 
    \Pi_{M, \, V A}^{ik} & = -2 e\omega^3 \sum_{\nu} \frac{D_{\nu}(\omega)}{\Omega} \,  \left( \sum_{j \psi} d_M^\Psi \, \epsilon^{imn} S_{j \psi}^{m} \, T^n_{j \nu \mathbf{q}} \right) \left( \sum_{j \psi} Q_{j\psi} \, T^k_{j \nu \mathbf{q}} \right)^* + 2 i \frac{d_M^e}{e} \omega^3 \epsilon^{ik m} \hat{s}_e^m (1 - \varepsilon_\infty) \label{eq:phi_phi_A_vec_mdm} \\
    \Pi_{E, \, VV}^{ii} & = -4i \omega^2 \sum_{\nu} \frac{D_{\nu}(\omega)}{\Omega} \,  \left( \sum_{j \psi} \, d_E^\Psi \, q^m S^m_{j \psi} \, T^i_{j \nu \mathbf{q}} \right) \left( \sum_{j \psi} \, d_E^\Psi \, q^m S^m_{j \psi} \, T^i_{j \nu \mathbf{q}} \right)^*  \label{eq:pi_phi_phi_ph_vec_edm} \\ 
    \Pi_{E, \, V A}^{ik} & = -2 e \omega^2 \sum_{\nu} \frac{D_{\nu}(\omega)}{\Omega} \,  \left( \sum_{j \psi} \, d_E^\Psi \, q^m S^m_{j \psi} \, T^i_{j \nu \mathbf{q}} \right) \left( \sum_{j \psi} \, Q_{j \psi} \, T^k_{j \nu \mathbf{q}} \right)^*  \label{eq:phi_phi_A_vec_edm} + 2i \frac{d_E^e}{e} \omega^2 q^m \hat{s}_e^m \delta^{ik} (1 - \varepsilon_\infty) \, ,
\end{align}
and $\Pi_{M/E,\, V A}^\text{el} = -\Pi_{M/E,\, A V}^\text{el}$.

In Fig.~\ref{fig:vector_multipole_current} we focus on models with only an electron coupling which were the focus of Ref.~\cite{Krnjaic:2022wor,Krnjaic:2023nxe}. Projections are computed assuming a $\text{kg} \cdot \text{yr}$ exposure with no backgrounds, and the line labeled, ``Therm." corresponds to the minimum coupling needed to not thermally produce the vector DM in a universe that reheats right before BBN. That is, we require $\Gamma \lesssim H$ at $T = \text{MeV}$, where $H$ is the Hubble constant, $\Gamma = \bar{n}_V \sigma_{E/M}$ is the interaction rate, $\bar{n}_V$ is the equilibrium number density of $V$ particles, and $\sigma_{E/M} \sim d_{E/M}^2$~\cite{Krnjaic:2022wor}. In addition to the thermalization bound we also show projected constraints from an Aluminum superconductor, and \ce{ZrTe5} target. Ref.~\cite{Krnjaic:2023nxe} showed that the electronic absorption rate of the MDM model could be related to $\text{Im}\left[ \varepsilon(\omega) \right]$, and therefore the constraints on $g_{aee}$ can simply be re-scaled accordingly. We see that spin ordering is crucial to be able to probe either of these models, since targets without spin ordering have no EDM response, and an MDM $N$ response only at higher in $1/m_\Psi$. In these spin ordered targets single phonon excitations are able to probe new parameter space for both the MDM model.

\section{Conclusions}
\label{sec:conclusions}

Single phonon excitations are an exciting avenue for direct detection of light DM with sub-$\text{eV}$ thresholds. In Sec.~\ref{sec:formalism}, using effective field theory (EFT) techniques we provided a framework for computing the DM absorption rate into single phonons starting from a fairly general UV Lagrangian (Eq.~\eqref{eq:uv_lag}). This complements previous work which computed general DM-single phonon scattering rates~\cite{Caputo:2019cyg,Cox:2019cod,Griffin:2019mvc,Griffin:2018bjn,Knapen:2017ekk,Trickle:2019nya,Trickle:2020oki,Lasenby:2021wsc} and further illustrates the variety of DM models that can excite single phonon excitations. Then in Sec.~\ref{sec:results} we applied this formalism to compute the DM absorption rate of five DM models (Secs.~\ref{subsec:scalar_DM} -~\ref{subsec:vector_DM_edm_mdm}) on spin ordered, e.g., \ce{FeBr2}, and non-spin ordered, \ce{GaAs}, \ce{Al2O3}, \ce{SiO2} targets. Additionally, in Sec.~\ref{subsec:dielectric_calc}, we used the formalism to compute the dielectric function which allows for a direct comparison between first principles calculation and experimental data. In Fig.~\ref{fig:dielectric_comp_rate} we find good agreement between both the dielectric function, $\varepsilon(\omega)$, and the ELF, $\text{Im} \left[ -1/\varepsilon(\omega) \right]$, in \ce{GaAs}, \ce{Al2O3}, and \ce{SiO2} targets, indicating the reliability of the first principles calculations. Moreover, this comparison allows for a data-driven approach to set the only (theoretically) free parameter, the phonon mode widths, $\gamma_\nu$. 

In addition to providing a theoretical framework to compute general DM absorption rates into single phonon excitations, we developed \textsf{PhonoDark-abs}~\GithubLink. \textsf{PhonoDark-abs} is an extension of \textsf{PhonoDark}~\cite{PhonoDark, Trickle:2020oki}, which computes general DM-single phonon scattering rates (see Refs.~\cite{Coskuner:2021qxo, Trickle:2019nya, Griffin:2019mvc, Trickle:2020oki,Taufertshofer:2023rgq,Romao:2023zqf} for examples), and numerically computes the DM absorption rate for any target material, given the input DFT files discussed in Sec.~\ref{sec:results}. Currently, \textsf{PhonoDark-abs} can reproduce all the results shown here, and future work will further extend its capabilities and integrate it with \textsf{PhonoDark} completely. \textsf{PhonoDark-abs} is publicly available here~\GithubLink.

Using \textsf{PhonoDark-abs}, we find that, assuming a $\text{kg} \cdot \text{yr}$ exposure and no backgrounds, single phonon excitations in \ce{GaAs}, \ce{Al2O3}, and \ce{SiO2} can probe new parameter space when DM is the gauge boson of a broken $U(1)_{B - L}$ symmetry (Fig.~\ref{fig:vector_kineticmixing}), and when DM couples to the electron magnetic dipole moment in spin ordered targets, e.g., the ferromagnetic \ce{FeBr2} (Fig.~\ref{fig:vector_multipole_current}).\footnote{In addition to the previously well studied kinetically mixed dark photon model~\cite{Knapen:2017ekk,Knapen:2021bwg}.} For the latter projected constraints the spin ordering is crucial; without spin ordering the target response is much higher order and therefore normal targets, e.g., \ce{GaAs} and \ce{Al2O3}, project rather weak constraints. While the projected constraints for the scalar, Fig.~\ref{fig:scalar_de}, and ALP, Fig.~\ref{fig:alp_gaff}, DM models coupling to electrons are competitive with targets utilizing small band-gap electronic transitions, e.g., Al superconductors~\cite{Mitridate:2021ctr,Hochberg:2016ajh,Gelmini:2020xir}, \ce{ZrTe5}~\cite{Chen:2022pyd}, and doped Si~\cite{Du:2022dxf}, strong stellar cooling and fifth force constraints are still superior in this parameter space. Furthermore the projected constraints for ALP DM coupling to protons and neutron spin with hyperpolarized targets are much weaker than the neutron star cooling. 

The theoretical framework here may be useful for other other collective excitations, e.g., magnons~\cite{Mitridate:2020kly,Trickle:2020oki,Barbieri:1985cp,Esposito:2022bnu,Trickle:2019ovy,Chigusa:2020gfs}. Formulating the absorption rate in terms of self-energies and using NR EFT, as done in Secs.~\ref{subsec:abs_rate_self_energy} and~\ref{subsubsec:nr_eft}, has the advantage of being independent of the internal excitations, and therefore may be used to understand general DM absorption into magnon excitations. Additionally, while only targets with particle number and spin ordering were considered, more novel targets may have, e.g., angular momentum ordering, $\langle \mathbf{L} \rangle_{\ell j} = \langle \mathbf{L} \rangle_{j}$, as considered in Ref.~\cite{Mitridate:2020kly}, or anisotropic responses, which could have interesting consequences for DM absorption rates.

\begin{acknowledgments}
We would like to thank Asher Berlin, Alex Millar, and Kevin Zhou for helpful conversations.
This work was supported by the Quantum Information Science Enabled Discovery (QuantISED) for High Energy Physics (KA2401032), and the Deutsche Forschungsgemeinschaft under Germany’s Excellence Strategy - EXC 2121 Quantum Universe - 390833306.
TT is supported by the Fermi Research Alliance, LLC under Contract No. DE-AC02-07CH11359 with the U.S. Department of Energy, Office of Science, Office of High Energy Physics.
KZ is supported by the U.S. Department of Energy, Office of Science, Office of High Energy Physics, under Award Number DE-SC0011632, and by the Walter Burke Institute for Theoretical Physics.

\end{acknowledgments}

\bibliographystyle{apsrev4-1}
\bibliography{ref}

\begin{thebibliography}{168}%
\makeatletter
\providecommand \@ifxundefined [1]{%
 \@ifx{#1\undefined}
}%
\providecommand \@ifnum [1]{%
 \ifnum #1\expandafter \@firstoftwo
 \else \expandafter \@secondoftwo
 \fi
}%
\providecommand \@ifx [1]{%
 \ifx #1\expandafter \@firstoftwo
 \else \expandafter \@secondoftwo
 \fi
}%
\providecommand \natexlab [1]{#1}%
\providecommand \enquote  [1]{``#1''}%
\providecommand \bibnamefont  [1]{#1}%
\providecommand \bibfnamefont [1]{#1}%
\providecommand \citenamefont [1]{#1}%
\providecommand \href@noop [0]{\@secondoftwo}%
\providecommand \href [0]{\begingroup \@sanitize@url \@href}%
\providecommand \@href[1]{\@@startlink{#1}\@@href}%
\providecommand \@@href[1]{\endgroup#1\@@endlink}%
\providecommand \@sanitize@url [0]{\catcode `\\12\catcode `\$12\catcode
  `\&12\catcode `\#12\catcode `\^12\catcode `\_12\catcode `\%12\relax}%
\providecommand \@@startlink[1]{}%
\providecommand \@@endlink[0]{}%
\providecommand \url  [0]{\begingroup\@sanitize@url \@url }%
\providecommand \@url [1]{\endgroup\@href {#1}{\urlprefix }}%
\providecommand \urlprefix  [0]{URL }%
\providecommand \Eprint [0]{\href }%
\providecommand \doibase [0]{http://dx.doi.org/}%
\providecommand \selectlanguage [0]{\@gobble}%
\providecommand \bibinfo  [0]{\@secondoftwo}%
\providecommand \bibfield  [0]{\@secondoftwo}%
\providecommand \translation [1]{[#1]}%
\providecommand \BibitemOpen [0]{}%
\providecommand \bibitemStop [0]{}%
\providecommand \bibitemNoStop [0]{.\EOS\space}%
\providecommand \EOS [0]{\spacefactor3000\relax}%
\providecommand \BibitemShut  [1]{\csname bibitem#1\endcsname}%
\let\auto@bib@innerbib\@empty
\bibitem [{\citenamefont {Amar\'e}\ \emph {et~al.}(2019)\citenamefont {Amar\'e}
  \emph {et~al.}}]{Amare:2019jul}%
  \BibitemOpen
  \bibfield  {author} {\bibinfo {author} {\bibfnamefont {J.}~\bibnamefont
  {Amar\'e}} \emph {et~al.},\ }\href {\doibase 10.1103/PhysRevLett.123.031301}
  {\bibfield  {journal} {\bibinfo  {journal} {Phys. Rev. Lett.}\ }\textbf
  {\bibinfo {volume} {123}},\ \bibinfo {pages} {031301} (\bibinfo {year}
  {2019})},\ \Eprint {http://arxiv.org/abs/1903.03973} {arXiv:1903.03973
  [astro-ph.IM]} \BibitemShut {NoStop}%
\bibitem [{\citenamefont {Pr\"obst}\ \emph {et~al.}(2002)\citenamefont
  {Pr\"obst} \emph {et~al.}}]{Probst:2002qb}%
  \BibitemOpen
  \bibfield  {author} {\bibinfo {author} {\bibfnamefont {F.}~\bibnamefont
  {Pr\"obst}} \emph {et~al.},\ }\href {\doibase 10.1016/S0920-5632(02)01453-6}
  {\bibfield  {journal} {\bibinfo  {journal} {Nucl. Phys. B Proc. Suppl.}\
  }\textbf {\bibinfo {volume} {110}},\ \bibinfo {pages} {67} (\bibinfo {year}
  {2002})}\BibitemShut {NoStop}%
\bibitem [{\citenamefont {Petricca}\ \emph {et~al.}(2020)\citenamefont
  {Petricca} \emph {et~al.}}]{CRESST:2017cdd}%
  \BibitemOpen
  \bibfield  {author} {\bibinfo {author} {\bibfnamefont {F.}~\bibnamefont
  {Petricca}} \emph {et~al.} (\bibinfo {collaboration} {CRESST}),\ }\href
  {\doibase 10.1088/1742-6596/1342/1/012076} {\bibfield  {journal} {\bibinfo
  {journal} {J. Phys. Conf. Ser.}\ }\textbf {\bibinfo {volume} {1342}},\
  \bibinfo {pages} {012076} (\bibinfo {year} {2020})},\ \Eprint
  {http://arxiv.org/abs/1711.07692} {arXiv:1711.07692 [astro-ph.CO]}
  \BibitemShut {NoStop}%
\bibitem [{\citenamefont {Angloher}\ \emph {et~al.}(2016)\citenamefont
  {Angloher} \emph {et~al.}}]{CRESST:2015txj}%
  \BibitemOpen
  \bibfield  {author} {\bibinfo {author} {\bibfnamefont {G.}~\bibnamefont
  {Angloher}} \emph {et~al.} (\bibinfo {collaboration} {CRESST}),\ }\href
  {\doibase 10.1140/epjc/s10052-016-3877-3} {\bibfield  {journal} {\bibinfo
  {journal} {Eur. Phys. J. C}\ }\textbf {\bibinfo {volume} {76}},\ \bibinfo
  {pages} {25} (\bibinfo {year} {2016})},\ \Eprint
  {http://arxiv.org/abs/1509.01515} {arXiv:1509.01515 [astro-ph.CO]}
  \BibitemShut {NoStop}%
\bibitem [{\citenamefont {Baum}\ \emph {et~al.}(2019)\citenamefont {Baum},
  \citenamefont {Freese},\ and\ \citenamefont {Kelso}}]{Baum:2018ekm}%
  \BibitemOpen
  \bibfield  {author} {\bibinfo {author} {\bibfnamefont {S.}~\bibnamefont
  {Baum}}, \bibinfo {author} {\bibfnamefont {K.}~\bibnamefont {Freese}}, \ and\
  \bibinfo {author} {\bibfnamefont {C.}~\bibnamefont {Kelso}},\ }\href
  {\doibase 10.1016/j.physletb.2018.12.036} {\bibfield  {journal} {\bibinfo
  {journal} {Phys. Lett. B}\ }\textbf {\bibinfo {volume} {789}},\ \bibinfo
  {pages} {262} (\bibinfo {year} {2019})},\ \Eprint
  {http://arxiv.org/abs/1804.01231} {arXiv:1804.01231 [astro-ph.CO]}
  \BibitemShut {NoStop}%
\bibitem [{\citenamefont {de~Mello~Neto}\ \emph {et~al.}(2016)\citenamefont
  {de~Mello~Neto} \emph {et~al.}}]{DAMIC:2015znm}%
  \BibitemOpen
  \bibfield  {author} {\bibinfo {author} {\bibfnamefont {J.~R.~T.}\
  \bibnamefont {de~Mello~Neto}} \emph {et~al.} (\bibinfo {collaboration}
  {DAMIC}),\ }\href {\doibase 10.22323/1.236.1221} {\bibfield  {journal}
  {\bibinfo  {journal} {PoS}\ }\textbf {\bibinfo {volume} {ICRC2015}},\
  \bibinfo {pages} {1221} (\bibinfo {year} {2016})},\ \Eprint
  {http://arxiv.org/abs/1510.02126} {arXiv:1510.02126 [physics.ins-det]}
  \BibitemShut {NoStop}%
\bibitem [{\citenamefont {Aguilar-Arevalo}\ \emph {et~al.}(2019)\citenamefont
  {Aguilar-Arevalo} \emph {et~al.}}]{DAMIC:2019dcn}%
  \BibitemOpen
  \bibfield  {author} {\bibinfo {author} {\bibfnamefont {A.}~\bibnamefont
  {Aguilar-Arevalo}} \emph {et~al.} (\bibinfo {collaboration} {DAMIC}),\ }\href
  {\doibase 10.1103/PhysRevLett.123.181802} {\bibfield  {journal} {\bibinfo
  {journal} {Phys. Rev. Lett.}\ }\textbf {\bibinfo {volume} {123}},\ \bibinfo
  {pages} {181802} (\bibinfo {year} {2019})},\ \Eprint
  {http://arxiv.org/abs/1907.12628} {arXiv:1907.12628 [astro-ph.CO]}
  \BibitemShut {NoStop}%
\bibitem [{\citenamefont {Agnes}\ \emph
  {et~al.}(2018{\natexlab{a}})\citenamefont {Agnes} \emph
  {et~al.}}]{DarkSide:2018bpj}%
  \BibitemOpen
  \bibfield  {author} {\bibinfo {author} {\bibfnamefont {P.}~\bibnamefont
  {Agnes}} \emph {et~al.} (\bibinfo {collaboration} {DarkSide}),\ }\href
  {\doibase 10.1103/PhysRevLett.121.081307} {\bibfield  {journal} {\bibinfo
  {journal} {Phys. Rev. Lett.}\ }\textbf {\bibinfo {volume} {121}},\ \bibinfo
  {pages} {081307} (\bibinfo {year} {2018}{\natexlab{a}})},\ \Eprint
  {http://arxiv.org/abs/1802.06994} {arXiv:1802.06994 [astro-ph.HE]}
  \BibitemShut {NoStop}%
\bibitem [{\citenamefont {Jo}(2017)}]{Jo:2016qql}%
  \BibitemOpen
  \bibfield  {author} {\bibinfo {author} {\bibfnamefont {J.~H.}\ \bibnamefont
  {Jo}} (\bibinfo {collaboration} {DM-Ice}),\ }\href {\doibase
  10.22323/1.282.1223} {\bibfield  {journal} {\bibinfo  {journal} {PoS}\
  }\textbf {\bibinfo {volume} {ICHEP2016}},\ \bibinfo {pages} {1223} (\bibinfo
  {year} {2017})},\ \Eprint {http://arxiv.org/abs/1612.07426} {arXiv:1612.07426
  [physics.ins-det]} \BibitemShut {NoStop}%
\bibitem [{\citenamefont {Kim}(2015)}]{Kim:2015prm}%
  \BibitemOpen
  \bibfield  {author} {\bibinfo {author} {\bibfnamefont {K.}~\bibnamefont
  {Kim}} (\bibinfo {collaboration} {KIMS}),\ }in\ \href@noop {} {\emph
  {\bibinfo {booktitle} {{Meeting of the APS Division of Particles and
  Fields}}}}\ (\bibinfo {year} {2015})\ \Eprint
  {http://arxiv.org/abs/1511.00023} {arXiv:1511.00023 [physics.ins-det]}
  \BibitemShut {NoStop}%
\bibitem [{\citenamefont {Akerib}\ \emph {et~al.}(2018)\citenamefont {Akerib}
  \emph {et~al.}}]{LUX:2018zdm}%
  \BibitemOpen
  \bibfield  {author} {\bibinfo {author} {\bibfnamefont {D.~S.}\ \bibnamefont
  {Akerib}} \emph {et~al.} (\bibinfo {collaboration} {LUX}),\ }\href {\doibase
  10.1103/PhysRevD.97.112002} {\bibfield  {journal} {\bibinfo  {journal} {Phys.
  Rev. D}\ }\textbf {\bibinfo {volume} {97}},\ \bibinfo {pages} {112002}
  (\bibinfo {year} {2018})},\ \Eprint {http://arxiv.org/abs/1802.06162}
  {arXiv:1802.06162 [physics.ins-det]} \BibitemShut {NoStop}%
\bibitem [{\citenamefont {Akerib}\ \emph {et~al.}(2020)\citenamefont {Akerib}
  \emph {et~al.}}]{LUX:2019npm}%
  \BibitemOpen
  \bibfield  {author} {\bibinfo {author} {\bibfnamefont {D.~S.}\ \bibnamefont
  {Akerib}} \emph {et~al.} (\bibinfo {collaboration} {LUX}),\ }\href {\doibase
  10.1103/PhysRevD.101.042001} {\bibfield  {journal} {\bibinfo  {journal}
  {Phys. Rev. D}\ }\textbf {\bibinfo {volume} {101}},\ \bibinfo {pages}
  {042001} (\bibinfo {year} {2020})},\ \Eprint
  {http://arxiv.org/abs/1907.06272} {arXiv:1907.06272 [astro-ph.CO]}
  \BibitemShut {NoStop}%
\bibitem [{\citenamefont {Akerib}\ \emph {et~al.}(2019)\citenamefont {Akerib}
  \emph {et~al.}}]{LUX:2018akb}%
  \BibitemOpen
  \bibfield  {author} {\bibinfo {author} {\bibfnamefont {D.~S.}\ \bibnamefont
  {Akerib}} \emph {et~al.} (\bibinfo {collaboration} {LUX}),\ }\href {\doibase
  10.1103/PhysRevLett.122.131301} {\bibfield  {journal} {\bibinfo  {journal}
  {Phys. Rev. Lett.}\ }\textbf {\bibinfo {volume} {122}},\ \bibinfo {pages}
  {131301} (\bibinfo {year} {2019})},\ \Eprint
  {http://arxiv.org/abs/1811.11241} {arXiv:1811.11241 [astro-ph.CO]}
  \BibitemShut {NoStop}%
\bibitem [{\citenamefont {Shields}\ \emph {et~al.}(2015)\citenamefont
  {Shields}, \citenamefont {Xu},\ and\ \citenamefont
  {Calaprice}}]{Shields:2015wka}%
  \BibitemOpen
  \bibfield  {author} {\bibinfo {author} {\bibfnamefont {E.}~\bibnamefont
  {Shields}}, \bibinfo {author} {\bibfnamefont {J.}~\bibnamefont {Xu}}, \ and\
  \bibinfo {author} {\bibfnamefont {F.}~\bibnamefont {Calaprice}},\ }\href
  {\doibase 10.1016/j.phpro.2014.12.028} {\bibfield  {journal} {\bibinfo
  {journal} {Phys. Procedia}\ }\textbf {\bibinfo {volume} {61}},\ \bibinfo
  {pages} {169} (\bibinfo {year} {2015})}\BibitemShut {NoStop}%
\bibitem [{\citenamefont {Agnese}\ \emph {et~al.}(2014)\citenamefont {Agnese}
  \emph {et~al.}}]{SuperCDMS:2014cds}%
  \BibitemOpen
  \bibfield  {author} {\bibinfo {author} {\bibfnamefont {R.}~\bibnamefont
  {Agnese}} \emph {et~al.} (\bibinfo {collaboration} {SuperCDMS}),\ }\href
  {\doibase 10.1103/PhysRevLett.112.241302} {\bibfield  {journal} {\bibinfo
  {journal} {Phys. Rev. Lett.}\ }\textbf {\bibinfo {volume} {112}},\ \bibinfo
  {pages} {241302} (\bibinfo {year} {2014})},\ \Eprint
  {http://arxiv.org/abs/1402.7137} {arXiv:1402.7137 [hep-ex]} \BibitemShut
  {NoStop}%
\bibitem [{\citenamefont {Agnese}\ \emph {et~al.}(2017)\citenamefont {Agnese}
  \emph {et~al.}}]{SuperCDMS:2016wui}%
  \BibitemOpen
  \bibfield  {author} {\bibinfo {author} {\bibfnamefont {R.}~\bibnamefont
  {Agnese}} \emph {et~al.} (\bibinfo {collaboration} {SuperCDMS}),\ }\href
  {\doibase 10.1103/PhysRevD.95.082002} {\bibfield  {journal} {\bibinfo
  {journal} {Phys. Rev. D}\ }\textbf {\bibinfo {volume} {95}},\ \bibinfo
  {pages} {082002} (\bibinfo {year} {2017})},\ \Eprint
  {http://arxiv.org/abs/1610.00006} {arXiv:1610.00006 [physics.ins-det]}
  \BibitemShut {NoStop}%
\bibitem [{\citenamefont {Agnese}\ \emph {et~al.}(2016)\citenamefont {Agnese}
  \emph {et~al.}}]{SuperCDMS:2015eex}%
  \BibitemOpen
  \bibfield  {author} {\bibinfo {author} {\bibfnamefont {R.}~\bibnamefont
  {Agnese}} \emph {et~al.} (\bibinfo {collaboration} {SuperCDMS}),\ }\href
  {\doibase 10.1103/PhysRevLett.116.071301} {\bibfield  {journal} {\bibinfo
  {journal} {Phys. Rev. Lett.}\ }\textbf {\bibinfo {volume} {116}},\ \bibinfo
  {pages} {071301} (\bibinfo {year} {2016})},\ \Eprint
  {http://arxiv.org/abs/1509.02448} {arXiv:1509.02448 [astro-ph.CO]}
  \BibitemShut {NoStop}%
\bibitem [{\citenamefont {Agnese}\ \emph
  {et~al.}(2018{\natexlab{a}})\citenamefont {Agnese} \emph
  {et~al.}}]{SuperCDMS:2017nns}%
  \BibitemOpen
  \bibfield  {author} {\bibinfo {author} {\bibfnamefont {R.}~\bibnamefont
  {Agnese}} \emph {et~al.} (\bibinfo {collaboration} {SuperCDMS}),\ }\href
  {\doibase 10.1103/PhysRevD.97.022002} {\bibfield  {journal} {\bibinfo
  {journal} {Phys. Rev. D}\ }\textbf {\bibinfo {volume} {97}},\ \bibinfo
  {pages} {022002} (\bibinfo {year} {2018}{\natexlab{a}})},\ \Eprint
  {http://arxiv.org/abs/1707.01632} {arXiv:1707.01632 [astro-ph.CO]}
  \BibitemShut {NoStop}%
\bibitem [{\citenamefont {Agnese}\ \emph {et~al.}(2019)\citenamefont {Agnese}
  \emph {et~al.}}]{SuperCDMS:2018gro}%
  \BibitemOpen
  \bibfield  {author} {\bibinfo {author} {\bibfnamefont {R.}~\bibnamefont
  {Agnese}} \emph {et~al.} (\bibinfo {collaboration} {SuperCDMS}),\ }\href
  {\doibase 10.1103/PhysRevD.99.062001} {\bibfield  {journal} {\bibinfo
  {journal} {Phys. Rev. D}\ }\textbf {\bibinfo {volume} {99}},\ \bibinfo
  {pages} {062001} (\bibinfo {year} {2019})},\ \Eprint
  {http://arxiv.org/abs/1808.09098} {arXiv:1808.09098 [astro-ph.CO]}
  \BibitemShut {NoStop}%
\bibitem [{\citenamefont {Agnese}\ \emph
  {et~al.}(2018{\natexlab{b}})\citenamefont {Agnese} \emph
  {et~al.}}]{SuperCDMS:2018mne}%
  \BibitemOpen
  \bibfield  {author} {\bibinfo {author} {\bibfnamefont {R.}~\bibnamefont
  {Agnese}} \emph {et~al.} (\bibinfo {collaboration} {SuperCDMS}),\ }\href
  {\doibase 10.1103/PhysRevLett.121.051301} {\bibfield  {journal} {\bibinfo
  {journal} {Phys. Rev. Lett.}\ }\textbf {\bibinfo {volume} {121}},\ \bibinfo
  {pages} {051301} (\bibinfo {year} {2018}{\natexlab{b}})},\ \bibinfo {note}
  {[Erratum: Phys.Rev.Lett. 122, 069901 (2019)]},\ \Eprint
  {http://arxiv.org/abs/1804.10697} {arXiv:1804.10697 [hep-ex]} \BibitemShut
  {NoStop}%
\bibitem [{\citenamefont {Angle}\ \emph {et~al.}(2008)\citenamefont {Angle}
  \emph {et~al.}}]{XENON:2007uwm}%
  \BibitemOpen
  \bibfield  {author} {\bibinfo {author} {\bibfnamefont {J.}~\bibnamefont
  {Angle}} \emph {et~al.} (\bibinfo {collaboration} {XENON}),\ }\href {\doibase
  10.1103/PhysRevLett.100.021303} {\bibfield  {journal} {\bibinfo  {journal}
  {Phys. Rev. Lett.}\ }\textbf {\bibinfo {volume} {100}},\ \bibinfo {pages}
  {021303} (\bibinfo {year} {2008})},\ \Eprint {http://arxiv.org/abs/0706.0039}
  {arXiv:0706.0039 [astro-ph]} \BibitemShut {NoStop}%
\bibitem [{\citenamefont {Aprile}\ \emph {et~al.}(2016)\citenamefont {Aprile}
  \emph {et~al.}}]{XENON100:2016sjq}%
  \BibitemOpen
  \bibfield  {author} {\bibinfo {author} {\bibfnamefont {E.}~\bibnamefont
  {Aprile}} \emph {et~al.} (\bibinfo {collaboration} {XENON100}),\ }\href
  {\doibase 10.1103/PhysRevD.94.122001} {\bibfield  {journal} {\bibinfo
  {journal} {Phys. Rev. D}\ }\textbf {\bibinfo {volume} {94}},\ \bibinfo
  {pages} {122001} (\bibinfo {year} {2016})},\ \Eprint
  {http://arxiv.org/abs/1609.06154} {arXiv:1609.06154 [astro-ph.CO]}
  \BibitemShut {NoStop}%
\bibitem [{\citenamefont {Aprile}\ \emph
  {et~al.}(2019{\natexlab{a}})\citenamefont {Aprile} \emph
  {et~al.}}]{XENON:2019gfn}%
  \BibitemOpen
  \bibfield  {author} {\bibinfo {author} {\bibfnamefont {E.}~\bibnamefont
  {Aprile}} \emph {et~al.} (\bibinfo {collaboration} {XENON}),\ }\href
  {\doibase 10.1103/PhysRevLett.123.251801} {\bibfield  {journal} {\bibinfo
  {journal} {Phys. Rev. Lett.}\ }\textbf {\bibinfo {volume} {123}},\ \bibinfo
  {pages} {251801} (\bibinfo {year} {2019}{\natexlab{a}})},\ \Eprint
  {http://arxiv.org/abs/1907.11485} {arXiv:1907.11485 [hep-ex]} \BibitemShut
  {NoStop}%
\bibitem [{\citenamefont {Aprile}\ \emph {et~al.}(2018)\citenamefont {Aprile}
  \emph {et~al.}}]{XENON:2018voc}%
  \BibitemOpen
  \bibfield  {author} {\bibinfo {author} {\bibfnamefont {E.}~\bibnamefont
  {Aprile}} \emph {et~al.} (\bibinfo {collaboration} {XENON}),\ }\href
  {\doibase 10.1103/PhysRevLett.121.111302} {\bibfield  {journal} {\bibinfo
  {journal} {Phys. Rev. Lett.}\ }\textbf {\bibinfo {volume} {121}},\ \bibinfo
  {pages} {111302} (\bibinfo {year} {2018})},\ \Eprint
  {http://arxiv.org/abs/1805.12562} {arXiv:1805.12562 [astro-ph.CO]}
  \BibitemShut {NoStop}%
\bibitem [{\citenamefont {Aprile}\ \emph {et~al.}(2023)\citenamefont {Aprile}
  \emph {et~al.}}]{XENONCollaboration:2023orw}%
  \BibitemOpen
  \bibfield  {author} {\bibinfo {author} {\bibfnamefont {E.}~\bibnamefont
  {Aprile}} \emph {et~al.} (\bibinfo {collaboration} {(XENON
  Collaboration)\textdagger{}\textdagger{}, XENON}),\ }\href {\doibase
  10.1103/PhysRevLett.131.041003} {\bibfield  {journal} {\bibinfo  {journal}
  {Phys. Rev. Lett.}\ }\textbf {\bibinfo {volume} {131}},\ \bibinfo {pages}
  {041003} (\bibinfo {year} {2023})},\ \Eprint
  {http://arxiv.org/abs/2303.14729} {arXiv:2303.14729 [hep-ex]} \BibitemShut
  {NoStop}%
\bibitem [{\citenamefont {Zhang}\ \emph {et~al.}(2022)\citenamefont {Zhang}
  \emph {et~al.}}]{CDEX:2022kcd}%
  \BibitemOpen
  \bibfield  {author} {\bibinfo {author} {\bibfnamefont {Z.~Y.}\ \bibnamefont
  {Zhang}} \emph {et~al.} (\bibinfo {collaboration} {CDEX}),\ }\href {\doibase
  10.1103/PhysRevLett.129.221301} {\bibfield  {journal} {\bibinfo  {journal}
  {Phys. Rev. Lett.}\ }\textbf {\bibinfo {volume} {129}},\ \bibinfo {pages}
  {221301} (\bibinfo {year} {2022})},\ \Eprint
  {http://arxiv.org/abs/2206.04128} {arXiv:2206.04128 [hep-ex]} \BibitemShut
  {NoStop}%
\bibitem [{\citenamefont {Aguilar-Arevalo}\ \emph {et~al.}(2020)\citenamefont
  {Aguilar-Arevalo} \emph {et~al.}}]{DAMIC:2020cut}%
  \BibitemOpen
  \bibfield  {author} {\bibinfo {author} {\bibfnamefont {A.}~\bibnamefont
  {Aguilar-Arevalo}} \emph {et~al.} (\bibinfo {collaboration} {DAMIC}),\ }\href
  {\doibase 10.1103/PhysRevLett.125.241803} {\bibfield  {journal} {\bibinfo
  {journal} {Phys. Rev. Lett.}\ }\textbf {\bibinfo {volume} {125}},\ \bibinfo
  {pages} {241803} (\bibinfo {year} {2020})},\ \Eprint
  {http://arxiv.org/abs/2007.15622} {arXiv:2007.15622 [astro-ph.CO]}
  \BibitemShut {NoStop}%
\bibitem [{\citenamefont {Aguilar-Arevalo}\ \emph {et~al.}(2017)\citenamefont
  {Aguilar-Arevalo} \emph {et~al.}}]{DAMIC:2016qck}%
  \BibitemOpen
  \bibfield  {author} {\bibinfo {author} {\bibfnamefont {A.}~\bibnamefont
  {Aguilar-Arevalo}} \emph {et~al.} (\bibinfo {collaboration} {DAMIC}),\ }\href
  {\doibase 10.1103/PhysRevLett.118.141803} {\bibfield  {journal} {\bibinfo
  {journal} {Phys. Rev. Lett.}\ }\textbf {\bibinfo {volume} {118}},\ \bibinfo
  {pages} {141803} (\bibinfo {year} {2017})},\ \Eprint
  {http://arxiv.org/abs/1611.03066} {arXiv:1611.03066 [astro-ph.CO]}
  \BibitemShut {NoStop}%
\bibitem [{\citenamefont {Settimo}(2020)}]{Settimo:2020cbq}%
  \BibitemOpen
  \bibfield  {author} {\bibinfo {author} {\bibfnamefont {M.}~\bibnamefont
  {Settimo}} (\bibinfo {collaboration} {DAMIC, DAMIC-M}),\ }in\ \href@noop {}
  {\emph {\bibinfo {booktitle} {{16th Rencontres du Vietnam}: {Theory meeting
  experiment: Particle Astrophysics and Cosmology}}}}\ (\bibinfo {year}
  {2020})\ \Eprint {http://arxiv.org/abs/2003.09497} {arXiv:2003.09497
  [hep-ex]} \BibitemShut {NoStop}%
\bibitem [{\citenamefont {Agnes}\ \emph
  {et~al.}(2018{\natexlab{b}})\citenamefont {Agnes} \emph
  {et~al.}}]{DarkSide:2018ppu}%
  \BibitemOpen
  \bibfield  {author} {\bibinfo {author} {\bibfnamefont {P.}~\bibnamefont
  {Agnes}} \emph {et~al.} (\bibinfo {collaboration} {DarkSide}),\ }\href
  {\doibase 10.1103/PhysRevLett.121.111303} {\bibfield  {journal} {\bibinfo
  {journal} {Phys. Rev. Lett.}\ }\textbf {\bibinfo {volume} {121}},\ \bibinfo
  {pages} {111303} (\bibinfo {year} {2018}{\natexlab{b}})},\ \Eprint
  {http://arxiv.org/abs/1802.06998} {arXiv:1802.06998 [astro-ph.CO]}
  \BibitemShut {NoStop}%
\bibitem [{\citenamefont {Agnes}\ \emph {et~al.}(2023)\citenamefont {Agnes}
  \emph {et~al.}}]{DarkSide:2022knj}%
  \BibitemOpen
  \bibfield  {author} {\bibinfo {author} {\bibfnamefont {P.}~\bibnamefont
  {Agnes}} \emph {et~al.} (\bibinfo {collaboration} {DarkSide}),\ }\href
  {\doibase 10.1103/PhysRevLett.130.101002} {\bibfield  {journal} {\bibinfo
  {journal} {Phys. Rev. Lett.}\ }\textbf {\bibinfo {volume} {130}},\ \bibinfo
  {pages} {101002} (\bibinfo {year} {2023})},\ \Eprint
  {http://arxiv.org/abs/2207.11968} {arXiv:2207.11968 [hep-ex]} \BibitemShut
  {NoStop}%
\bibitem [{\citenamefont {Agnes}\ \emph
  {et~al.}(2018{\natexlab{c}})\citenamefont {Agnes} \emph
  {et~al.}}]{DarkSide:2018bpja}%
  \BibitemOpen
  \bibfield  {author} {\bibinfo {author} {\bibfnamefont {P.}~\bibnamefont
  {Agnes}} \emph {et~al.} (\bibinfo {collaboration} {DarkSide}),\ }\href
  {\doibase 10.1103/PhysRevLett.121.081307} {\bibfield  {journal} {\bibinfo
  {journal} {Phys. Rev. Lett.}\ }\textbf {\bibinfo {volume} {121}},\ \bibinfo
  {pages} {081307} (\bibinfo {year} {2018}{\natexlab{c}})},\ \Eprint
  {http://arxiv.org/abs/1802.06994} {arXiv:1802.06994 [astro-ph.HE]}
  \BibitemShut {NoStop}%
\bibitem [{\citenamefont {Armengaud}\ \emph {et~al.}(2018)\citenamefont
  {Armengaud} \emph {et~al.}}]{EDELWEISS:2018tde}%
  \BibitemOpen
  \bibfield  {author} {\bibinfo {author} {\bibfnamefont {E.}~\bibnamefont
  {Armengaud}} \emph {et~al.} (\bibinfo {collaboration} {EDELWEISS}),\ }\href
  {\doibase 10.1103/PhysRevD.98.082004} {\bibfield  {journal} {\bibinfo
  {journal} {Phys. Rev. D}\ }\textbf {\bibinfo {volume} {98}},\ \bibinfo
  {pages} {082004} (\bibinfo {year} {2018})},\ \Eprint
  {http://arxiv.org/abs/1808.02340} {arXiv:1808.02340 [hep-ex]} \BibitemShut
  {NoStop}%
\bibitem [{\citenamefont {Armengaud}\ \emph {et~al.}(2019)\citenamefont
  {Armengaud} \emph {et~al.}}]{EDELWEISS:2019vjv}%
  \BibitemOpen
  \bibfield  {author} {\bibinfo {author} {\bibfnamefont {E.}~\bibnamefont
  {Armengaud}} \emph {et~al.} (\bibinfo {collaboration} {EDELWEISS}),\ }\href
  {\doibase 10.1103/PhysRevD.99.082003} {\bibfield  {journal} {\bibinfo
  {journal} {Phys. Rev. D}\ }\textbf {\bibinfo {volume} {99}},\ \bibinfo
  {pages} {082003} (\bibinfo {year} {2019})},\ \Eprint
  {http://arxiv.org/abs/1901.03588} {arXiv:1901.03588 [astro-ph.GA]}
  \BibitemShut {NoStop}%
\bibitem [{\citenamefont {Arnaud}\ \emph {et~al.}(2020)\citenamefont {Arnaud}
  \emph {et~al.}}]{EDELWEISS:2020fxc}%
  \BibitemOpen
  \bibfield  {author} {\bibinfo {author} {\bibfnamefont {Q.}~\bibnamefont
  {Arnaud}} \emph {et~al.} (\bibinfo {collaboration} {EDELWEISS}),\ }\href
  {\doibase 10.1103/PhysRevLett.125.141301} {\bibfield  {journal} {\bibinfo
  {journal} {Phys. Rev. Lett.}\ }\textbf {\bibinfo {volume} {125}},\ \bibinfo
  {pages} {141301} (\bibinfo {year} {2020})},\ \Eprint
  {http://arxiv.org/abs/2003.01046} {arXiv:2003.01046 [astro-ph.GA]}
  \BibitemShut {NoStop}%
\bibitem [{\citenamefont {Crisler}\ \emph {et~al.}(2018)\citenamefont
  {Crisler}, \citenamefont {Essig}, \citenamefont {Estrada}, \citenamefont
  {Fernandez}, \citenamefont {Tiffenberg}, \citenamefont {Sofo~haro},
  \citenamefont {Volansky},\ and\ \citenamefont {Yu}}]{Crisler:2018gci}%
  \BibitemOpen
  \bibfield  {author} {\bibinfo {author} {\bibfnamefont {M.}~\bibnamefont
  {Crisler}}, \bibinfo {author} {\bibfnamefont {R.}~\bibnamefont {Essig}},
  \bibinfo {author} {\bibfnamefont {J.}~\bibnamefont {Estrada}}, \bibinfo
  {author} {\bibfnamefont {G.}~\bibnamefont {Fernandez}}, \bibinfo {author}
  {\bibfnamefont {J.}~\bibnamefont {Tiffenberg}}, \bibinfo {author}
  {\bibfnamefont {M.}~\bibnamefont {Sofo~haro}}, \bibinfo {author}
  {\bibfnamefont {T.}~\bibnamefont {Volansky}}, \ and\ \bibinfo {author}
  {\bibfnamefont {T.-T.}\ \bibnamefont {Yu}} (\bibinfo {collaboration}
  {SENSEI}),\ }\href {\doibase 10.1103/PhysRevLett.121.061803} {\bibfield
  {journal} {\bibinfo  {journal} {Phys. Rev. Lett.}\ }\textbf {\bibinfo
  {volume} {121}},\ \bibinfo {pages} {061803} (\bibinfo {year} {2018})},\
  \Eprint {http://arxiv.org/abs/1804.00088} {arXiv:1804.00088 [hep-ex]}
  \BibitemShut {NoStop}%
\bibitem [{\citenamefont {Barak}\ \emph {et~al.}(2020)\citenamefont {Barak}
  \emph {et~al.}}]{SENSEI:2020dpa}%
  \BibitemOpen
  \bibfield  {author} {\bibinfo {author} {\bibfnamefont {L.}~\bibnamefont
  {Barak}} \emph {et~al.} (\bibinfo {collaboration} {SENSEI}),\ }\href
  {\doibase 10.1103/PhysRevLett.125.171802} {\bibfield  {journal} {\bibinfo
  {journal} {Phys. Rev. Lett.}\ }\textbf {\bibinfo {volume} {125}},\ \bibinfo
  {pages} {171802} (\bibinfo {year} {2020})},\ \Eprint
  {http://arxiv.org/abs/2004.11378} {arXiv:2004.11378 [astro-ph.CO]}
  \BibitemShut {NoStop}%
\bibitem [{\citenamefont {Abramoff}\ \emph {et~al.}(2019)\citenamefont
  {Abramoff} \emph {et~al.}}]{SENSEI:2019ibb}%
  \BibitemOpen
  \bibfield  {author} {\bibinfo {author} {\bibfnamefont {O.}~\bibnamefont
  {Abramoff}} \emph {et~al.} (\bibinfo {collaboration} {SENSEI}),\ }\href
  {\doibase 10.1103/PhysRevLett.122.161801} {\bibfield  {journal} {\bibinfo
  {journal} {Phys. Rev. Lett.}\ }\textbf {\bibinfo {volume} {122}},\ \bibinfo
  {pages} {161801} (\bibinfo {year} {2019})},\ \Eprint
  {http://arxiv.org/abs/1901.10478} {arXiv:1901.10478 [hep-ex]} \BibitemShut
  {NoStop}%
\bibitem [{\citenamefont {Amaral}\ \emph {et~al.}(2020)\citenamefont {Amaral}
  \emph {et~al.}}]{SuperCDMS:2020ymb}%
  \BibitemOpen
  \bibfield  {author} {\bibinfo {author} {\bibfnamefont {D.~W.}\ \bibnamefont
  {Amaral}} \emph {et~al.} (\bibinfo {collaboration} {SuperCDMS}),\ }\href
  {\doibase 10.1103/PhysRevD.102.091101} {\bibfield  {journal} {\bibinfo
  {journal} {Phys. Rev. D}\ }\textbf {\bibinfo {volume} {102}},\ \bibinfo
  {pages} {091101} (\bibinfo {year} {2020})},\ \Eprint
  {http://arxiv.org/abs/2005.14067} {arXiv:2005.14067 [hep-ex]} \BibitemShut
  {NoStop}%
\bibitem [{\citenamefont {Ahmed}\ \emph {et~al.}(2009)\citenamefont {Ahmed}
  \emph {et~al.}}]{CDMS:2009fba}%
  \BibitemOpen
  \bibfield  {author} {\bibinfo {author} {\bibfnamefont {Z.}~\bibnamefont
  {Ahmed}} \emph {et~al.} (\bibinfo {collaboration} {CDMS}),\ }\href {\doibase
  10.1103/PhysRevLett.103.141802} {\bibfield  {journal} {\bibinfo  {journal}
  {Phys. Rev. Lett.}\ }\textbf {\bibinfo {volume} {103}},\ \bibinfo {pages}
  {141802} (\bibinfo {year} {2009})},\ \Eprint {http://arxiv.org/abs/0902.4693}
  {arXiv:0902.4693 [hep-ex]} \BibitemShut {NoStop}%
\bibitem [{\citenamefont {Aprile}\ \emph {et~al.}(2022)\citenamefont {Aprile}
  \emph {et~al.}}]{XENON:2022ltv}%
  \BibitemOpen
  \bibfield  {author} {\bibinfo {author} {\bibfnamefont {E.}~\bibnamefont
  {Aprile}} \emph {et~al.} (\bibinfo {collaboration} {XENON}),\ }\href
  {\doibase 10.1103/PhysRevLett.129.161805} {\bibfield  {journal} {\bibinfo
  {journal} {Phys. Rev. Lett.}\ }\textbf {\bibinfo {volume} {129}},\ \bibinfo
  {pages} {161805} (\bibinfo {year} {2022})},\ \Eprint
  {http://arxiv.org/abs/2207.11330} {arXiv:2207.11330 [hep-ex]} \BibitemShut
  {NoStop}%
\bibitem [{\citenamefont {Aprile}\ \emph
  {et~al.}(2019{\natexlab{b}})\citenamefont {Aprile} \emph
  {et~al.}}]{XENON:2019gfna}%
  \BibitemOpen
  \bibfield  {author} {\bibinfo {author} {\bibfnamefont {E.}~\bibnamefont
  {Aprile}} \emph {et~al.} (\bibinfo {collaboration} {XENON}),\ }\href
  {\doibase 10.1103/PhysRevLett.123.251801} {\bibfield  {journal} {\bibinfo
  {journal} {Phys. Rev. Lett.}\ }\textbf {\bibinfo {volume} {123}},\ \bibinfo
  {pages} {251801} (\bibinfo {year} {2019}{\natexlab{b}})},\ \Eprint
  {http://arxiv.org/abs/1907.11485} {arXiv:1907.11485 [hep-ex]} \BibitemShut
  {NoStop}%
\bibitem [{\citenamefont {Bloch}\ \emph {et~al.}(2017)\citenamefont {Bloch},
  \citenamefont {Essig}, \citenamefont {Tobioka}, \citenamefont {Volansky},\
  and\ \citenamefont {Yu}}]{Bloch:2016sjj}%
  \BibitemOpen
  \bibfield  {author} {\bibinfo {author} {\bibfnamefont {I.~M.}\ \bibnamefont
  {Bloch}}, \bibinfo {author} {\bibfnamefont {R.}~\bibnamefont {Essig}},
  \bibinfo {author} {\bibfnamefont {K.}~\bibnamefont {Tobioka}}, \bibinfo
  {author} {\bibfnamefont {T.}~\bibnamefont {Volansky}}, \ and\ \bibinfo
  {author} {\bibfnamefont {T.-T.}\ \bibnamefont {Yu}},\ }\href {\doibase
  10.1007/JHEP06(2017)087} {\bibfield  {journal} {\bibinfo  {journal} {JHEP}\
  }\textbf {\bibinfo {volume} {06}},\ \bibinfo {pages} {087} (\bibinfo {year}
  {2017})},\ \Eprint {http://arxiv.org/abs/1608.02123} {arXiv:1608.02123
  [hep-ph]} \BibitemShut {NoStop}%
\bibitem [{\citenamefont {Graham}\ \emph {et~al.}(2016)\citenamefont {Graham},
  \citenamefont {Mardon},\ and\ \citenamefont {Rajendran}}]{Graham:2015rva}%
  \BibitemOpen
  \bibfield  {author} {\bibinfo {author} {\bibfnamefont {P.~W.}\ \bibnamefont
  {Graham}}, \bibinfo {author} {\bibfnamefont {J.}~\bibnamefont {Mardon}}, \
  and\ \bibinfo {author} {\bibfnamefont {S.}~\bibnamefont {Rajendran}},\ }\href
  {\doibase 10.1103/PhysRevD.93.103520} {\bibfield  {journal} {\bibinfo
  {journal} {Phys. Rev. D}\ }\textbf {\bibinfo {volume} {93}},\ \bibinfo
  {pages} {103520} (\bibinfo {year} {2016})},\ \Eprint
  {http://arxiv.org/abs/1504.02102} {arXiv:1504.02102 [hep-ph]} \BibitemShut
  {NoStop}%
\bibitem [{\citenamefont {Dror}\ \emph {et~al.}(2019)\citenamefont {Dror},
  \citenamefont {Harigaya},\ and\ \citenamefont {Narayan}}]{Dror:2018pdh}%
  \BibitemOpen
  \bibfield  {author} {\bibinfo {author} {\bibfnamefont {J.~A.}\ \bibnamefont
  {Dror}}, \bibinfo {author} {\bibfnamefont {K.}~\bibnamefont {Harigaya}}, \
  and\ \bibinfo {author} {\bibfnamefont {V.}~\bibnamefont {Narayan}},\ }\href
  {\doibase 10.1103/PhysRevD.99.035036} {\bibfield  {journal} {\bibinfo
  {journal} {Phys. Rev. D}\ }\textbf {\bibinfo {volume} {99}},\ \bibinfo
  {pages} {035036} (\bibinfo {year} {2019})},\ \Eprint
  {http://arxiv.org/abs/1810.07195} {arXiv:1810.07195 [hep-ph]} \BibitemShut
  {NoStop}%
\bibitem [{\citenamefont {Adshead}\ \emph {et~al.}(2023)\citenamefont
  {Adshead}, \citenamefont {Lozanov},\ and\ \citenamefont
  {Weiner}}]{Adshead:2023qiw}%
  \BibitemOpen
  \bibfield  {author} {\bibinfo {author} {\bibfnamefont {P.}~\bibnamefont
  {Adshead}}, \bibinfo {author} {\bibfnamefont {K.~D.}\ \bibnamefont
  {Lozanov}}, \ and\ \bibinfo {author} {\bibfnamefont {Z.~J.}\ \bibnamefont
  {Weiner}},\ }\href {\doibase 10.1103/PhysRevD.107.083519} {\bibfield
  {journal} {\bibinfo  {journal} {Phys. Rev. D}\ }\textbf {\bibinfo {volume}
  {107}},\ \bibinfo {pages} {083519} (\bibinfo {year} {2023})},\ \Eprint
  {http://arxiv.org/abs/2301.07718} {arXiv:2301.07718 [hep-ph]} \BibitemShut
  {NoStop}%
\bibitem [{\citenamefont {Arias}\ \emph {et~al.}(2012)\citenamefont {Arias},
  \citenamefont {Cadamuro}, \citenamefont {Goodsell}, \citenamefont {Jaeckel},
  \citenamefont {Redondo},\ and\ \citenamefont {Ringwald}}]{Arias:2012az}%
  \BibitemOpen
  \bibfield  {author} {\bibinfo {author} {\bibfnamefont {P.}~\bibnamefont
  {Arias}}, \bibinfo {author} {\bibfnamefont {D.}~\bibnamefont {Cadamuro}},
  \bibinfo {author} {\bibfnamefont {M.}~\bibnamefont {Goodsell}}, \bibinfo
  {author} {\bibfnamefont {J.}~\bibnamefont {Jaeckel}}, \bibinfo {author}
  {\bibfnamefont {J.}~\bibnamefont {Redondo}}, \ and\ \bibinfo {author}
  {\bibfnamefont {A.}~\bibnamefont {Ringwald}},\ }\href {\doibase
  10.1088/1475-7516/2012/06/013} {\bibfield  {journal} {\bibinfo  {journal}
  {JCAP}\ }\textbf {\bibinfo {volume} {06}},\ \bibinfo {pages} {013} (\bibinfo
  {year} {2012})},\ \Eprint {http://arxiv.org/abs/1201.5902} {arXiv:1201.5902
  [hep-ph]} \BibitemShut {NoStop}%
\bibitem [{\citenamefont {Nelson}\ and\ \citenamefont
  {Scholtz}(2011)}]{Nelson:2011sf}%
  \BibitemOpen
  \bibfield  {author} {\bibinfo {author} {\bibfnamefont {A.~E.}\ \bibnamefont
  {Nelson}}\ and\ \bibinfo {author} {\bibfnamefont {J.}~\bibnamefont
  {Scholtz}},\ }\href {\doibase 10.1103/PhysRevD.84.103501} {\bibfield
  {journal} {\bibinfo  {journal} {Phys. Rev. D}\ }\textbf {\bibinfo {volume}
  {84}},\ \bibinfo {pages} {103501} (\bibinfo {year} {2011})},\ \Eprint
  {http://arxiv.org/abs/1105.2812} {arXiv:1105.2812 [hep-ph]} \BibitemShut
  {NoStop}%
\bibitem [{\citenamefont {Dimopoulos}(2006)}]{Dimopoulos:2006ms}%
  \BibitemOpen
  \bibfield  {author} {\bibinfo {author} {\bibfnamefont {K.}~\bibnamefont
  {Dimopoulos}},\ }\href {\doibase 10.1103/PhysRevD.74.083502} {\bibfield
  {journal} {\bibinfo  {journal} {Phys. Rev. D}\ }\textbf {\bibinfo {volume}
  {74}},\ \bibinfo {pages} {083502} (\bibinfo {year} {2006})},\ \Eprint
  {http://arxiv.org/abs/hep-ph/0607229} {arXiv:hep-ph/0607229} \BibitemShut
  {NoStop}%
\bibitem [{\citenamefont {Hochberg}\ \emph
  {et~al.}(2016{\natexlab{a}})\citenamefont {Hochberg}, \citenamefont {Pyle},
  \citenamefont {Zhao},\ and\ \citenamefont {Zurek}}]{Hochberg:2015fth}%
  \BibitemOpen
  \bibfield  {author} {\bibinfo {author} {\bibfnamefont {Y.}~\bibnamefont
  {Hochberg}}, \bibinfo {author} {\bibfnamefont {M.}~\bibnamefont {Pyle}},
  \bibinfo {author} {\bibfnamefont {Y.}~\bibnamefont {Zhao}}, \ and\ \bibinfo
  {author} {\bibfnamefont {K.~M.}\ \bibnamefont {Zurek}},\ }\href {\doibase
  10.1007/JHEP08(2016)057} {\bibfield  {journal} {\bibinfo  {journal} {JHEP}\
  }\textbf {\bibinfo {volume} {08}},\ \bibinfo {pages} {057} (\bibinfo {year}
  {2016}{\natexlab{a}})},\ \Eprint {http://arxiv.org/abs/1512.04533}
  {arXiv:1512.04533 [hep-ph]} \BibitemShut {NoStop}%
\bibitem [{\citenamefont {Hochberg}\ \emph
  {et~al.}(2016{\natexlab{b}})\citenamefont {Hochberg}, \citenamefont {Zhao},\
  and\ \citenamefont {Zurek}}]{Hochberg:2015pha}%
  \BibitemOpen
  \bibfield  {author} {\bibinfo {author} {\bibfnamefont {Y.}~\bibnamefont
  {Hochberg}}, \bibinfo {author} {\bibfnamefont {Y.}~\bibnamefont {Zhao}}, \
  and\ \bibinfo {author} {\bibfnamefont {K.~M.}\ \bibnamefont {Zurek}},\ }\href
  {\doibase 10.1103/PhysRevLett.116.011301} {\bibfield  {journal} {\bibinfo
  {journal} {Phys. Rev. Lett.}\ }\textbf {\bibinfo {volume} {116}},\ \bibinfo
  {pages} {011301} (\bibinfo {year} {2016}{\natexlab{b}})},\ \Eprint
  {http://arxiv.org/abs/1504.07237} {arXiv:1504.07237 [hep-ph]} \BibitemShut
  {NoStop}%
\bibitem [{\citenamefont {Hochberg}\ \emph
  {et~al.}(2016{\natexlab{c}})\citenamefont {Hochberg}, \citenamefont {Lin},\
  and\ \citenamefont {Zurek}}]{Hochberg:2016ajh}%
  \BibitemOpen
  \bibfield  {author} {\bibinfo {author} {\bibfnamefont {Y.}~\bibnamefont
  {Hochberg}}, \bibinfo {author} {\bibfnamefont {T.}~\bibnamefont {Lin}}, \
  and\ \bibinfo {author} {\bibfnamefont {K.~M.}\ \bibnamefont {Zurek}},\ }\href
  {\doibase 10.1103/PhysRevD.94.015019} {\bibfield  {journal} {\bibinfo
  {journal} {Phys. Rev. D}\ }\textbf {\bibinfo {volume} {94}},\ \bibinfo
  {pages} {015019} (\bibinfo {year} {2016}{\natexlab{c}})},\ \Eprint
  {http://arxiv.org/abs/1604.06800} {arXiv:1604.06800 [hep-ph]} \BibitemShut
  {NoStop}%
\bibitem [{\citenamefont {Hochberg}\ \emph {et~al.}(2019)\citenamefont
  {Hochberg}, \citenamefont {Charaev}, \citenamefont {Nam}, \citenamefont
  {Verma}, \citenamefont {Colangelo},\ and\ \citenamefont
  {Berggren}}]{Hochberg:2019cyy}%
  \BibitemOpen
  \bibfield  {author} {\bibinfo {author} {\bibfnamefont {Y.}~\bibnamefont
  {Hochberg}}, \bibinfo {author} {\bibfnamefont {I.}~\bibnamefont {Charaev}},
  \bibinfo {author} {\bibfnamefont {S.-W.}\ \bibnamefont {Nam}}, \bibinfo
  {author} {\bibfnamefont {V.}~\bibnamefont {Verma}}, \bibinfo {author}
  {\bibfnamefont {M.}~\bibnamefont {Colangelo}}, \ and\ \bibinfo {author}
  {\bibfnamefont {K.~K.}\ \bibnamefont {Berggren}},\ }\href {\doibase
  10.1103/PhysRevLett.123.151802} {\bibfield  {journal} {\bibinfo  {journal}
  {Phys. Rev. Lett.}\ }\textbf {\bibinfo {volume} {123}},\ \bibinfo {pages}
  {151802} (\bibinfo {year} {2019})},\ \Eprint
  {http://arxiv.org/abs/1903.05101} {arXiv:1903.05101 [hep-ph]} \BibitemShut
  {NoStop}%
\bibitem [{\citenamefont {Hochberg}\ \emph {et~al.}(2023)\citenamefont
  {Hochberg}, \citenamefont {Kramer}, \citenamefont {Kurinsky},\ and\
  \citenamefont {Lehmann}}]{Hochberg:2021ymx}%
  \BibitemOpen
  \bibfield  {author} {\bibinfo {author} {\bibfnamefont {Y.}~\bibnamefont
  {Hochberg}}, \bibinfo {author} {\bibfnamefont {E.~D.}\ \bibnamefont
  {Kramer}}, \bibinfo {author} {\bibfnamefont {N.}~\bibnamefont {Kurinsky}}, \
  and\ \bibinfo {author} {\bibfnamefont {B.~V.}\ \bibnamefont {Lehmann}},\
  }\href {\doibase 10.1103/PhysRevD.107.076015} {\bibfield  {journal} {\bibinfo
   {journal} {Phys. Rev. D}\ }\textbf {\bibinfo {volume} {107}},\ \bibinfo
  {pages} {076015} (\bibinfo {year} {2023})},\ \Eprint
  {http://arxiv.org/abs/2109.04473} {arXiv:2109.04473 [hep-ph]} \BibitemShut
  {NoStop}%
\bibitem [{\citenamefont {Hochberg}\ \emph {et~al.}(2022)\citenamefont
  {Hochberg}, \citenamefont {Lehmann}, \citenamefont {Charaev}, \citenamefont
  {Chiles}, \citenamefont {Colangelo}, \citenamefont {Nam},\ and\ \citenamefont
  {Berggren}}]{Hochberg:2021yud}%
  \BibitemOpen
  \bibfield  {author} {\bibinfo {author} {\bibfnamefont {Y.}~\bibnamefont
  {Hochberg}}, \bibinfo {author} {\bibfnamefont {B.~V.}\ \bibnamefont
  {Lehmann}}, \bibinfo {author} {\bibfnamefont {I.}~\bibnamefont {Charaev}},
  \bibinfo {author} {\bibfnamefont {J.}~\bibnamefont {Chiles}}, \bibinfo
  {author} {\bibfnamefont {M.}~\bibnamefont {Colangelo}}, \bibinfo {author}
  {\bibfnamefont {S.~W.}\ \bibnamefont {Nam}}, \ and\ \bibinfo {author}
  {\bibfnamefont {K.~K.}\ \bibnamefont {Berggren}},\ }\href {\doibase
  10.1103/PhysRevD.106.112005} {\bibfield  {journal} {\bibinfo  {journal}
  {Phys. Rev. D}\ }\textbf {\bibinfo {volume} {106}},\ \bibinfo {pages}
  {112005} (\bibinfo {year} {2022})},\ \Eprint
  {http://arxiv.org/abs/2110.01586} {arXiv:2110.01586 [hep-ph]} \BibitemShut
  {NoStop}%
\bibitem [{\citenamefont {Gelmini}\ \emph {et~al.}(2020)\citenamefont
  {Gelmini}, \citenamefont {Takhistov},\ and\ \citenamefont
  {Vitagliano}}]{Gelmini:2020xir}%
  \BibitemOpen
  \bibfield  {author} {\bibinfo {author} {\bibfnamefont {G.~B.}\ \bibnamefont
  {Gelmini}}, \bibinfo {author} {\bibfnamefont {V.}~\bibnamefont {Takhistov}},
  \ and\ \bibinfo {author} {\bibfnamefont {E.}~\bibnamefont {Vitagliano}},\
  }\href {\doibase 10.1016/j.physletb.2020.135779} {\bibfield  {journal}
  {\bibinfo  {journal} {Phys. Lett. B}\ }\textbf {\bibinfo {volume} {809}},\
  \bibinfo {pages} {135779} (\bibinfo {year} {2020})},\ \Eprint
  {http://arxiv.org/abs/2006.13909} {arXiv:2006.13909 [hep-ph]} \BibitemShut
  {NoStop}%
\bibitem [{\citenamefont {Mitridate}\ \emph
  {et~al.}(2021{\natexlab{a}})\citenamefont {Mitridate}, \citenamefont
  {Trickle}, \citenamefont {Zhang},\ and\ \citenamefont
  {Zurek}}]{Mitridate:2021ctr}%
  \BibitemOpen
  \bibfield  {author} {\bibinfo {author} {\bibfnamefont {A.}~\bibnamefont
  {Mitridate}}, \bibinfo {author} {\bibfnamefont {T.}~\bibnamefont {Trickle}},
  \bibinfo {author} {\bibfnamefont {Z.}~\bibnamefont {Zhang}}, \ and\ \bibinfo
  {author} {\bibfnamefont {K.~M.}\ \bibnamefont {Zurek}},\ }\href {\doibase
  10.1007/JHEP09(2021)123} {\bibfield  {journal} {\bibinfo  {journal} {JHEP}\
  }\textbf {\bibinfo {volume} {09}},\ \bibinfo {pages} {123} (\bibinfo {year}
  {2021}{\natexlab{a}})},\ \Eprint {http://arxiv.org/abs/2106.12586}
  {arXiv:2106.12586 [hep-ph]} \BibitemShut {NoStop}%
\bibitem [{\citenamefont {Hochberg}\ \emph {et~al.}(2018)\citenamefont
  {Hochberg}, \citenamefont {Kahn}, \citenamefont {Lisanti}, \citenamefont
  {Zurek}, \citenamefont {Grushin}, \citenamefont {Ilan}, \citenamefont
  {Griffin}, \citenamefont {Liu}, \citenamefont {Weber},\ and\ \citenamefont
  {Neaton}}]{Hochberg:2017wce}%
  \BibitemOpen
  \bibfield  {author} {\bibinfo {author} {\bibfnamefont {Y.}~\bibnamefont
  {Hochberg}}, \bibinfo {author} {\bibfnamefont {Y.}~\bibnamefont {Kahn}},
  \bibinfo {author} {\bibfnamefont {M.}~\bibnamefont {Lisanti}}, \bibinfo
  {author} {\bibfnamefont {K.~M.}\ \bibnamefont {Zurek}}, \bibinfo {author}
  {\bibfnamefont {A.~G.}\ \bibnamefont {Grushin}}, \bibinfo {author}
  {\bibfnamefont {R.}~\bibnamefont {Ilan}}, \bibinfo {author} {\bibfnamefont
  {S.~M.}\ \bibnamefont {Griffin}}, \bibinfo {author} {\bibfnamefont {Z.-F.}\
  \bibnamefont {Liu}}, \bibinfo {author} {\bibfnamefont {S.~F.}\ \bibnamefont
  {Weber}}, \ and\ \bibinfo {author} {\bibfnamefont {J.~B.}\ \bibnamefont
  {Neaton}},\ }\href {\doibase 10.1103/PhysRevD.97.015004} {\bibfield
  {journal} {\bibinfo  {journal} {Phys. Rev. D}\ }\textbf {\bibinfo {volume}
  {97}},\ \bibinfo {pages} {015004} (\bibinfo {year} {2018})},\ \Eprint
  {http://arxiv.org/abs/1708.08929} {arXiv:1708.08929 [hep-ph]} \BibitemShut
  {NoStop}%
\bibitem [{\citenamefont {Coskuner}\ \emph {et~al.}(2021)\citenamefont
  {Coskuner}, \citenamefont {Mitridate}, \citenamefont {Olivares},\ and\
  \citenamefont {Zurek}}]{Coskuner:2019odd}%
  \BibitemOpen
  \bibfield  {author} {\bibinfo {author} {\bibfnamefont {A.}~\bibnamefont
  {Coskuner}}, \bibinfo {author} {\bibfnamefont {A.}~\bibnamefont {Mitridate}},
  \bibinfo {author} {\bibfnamefont {A.}~\bibnamefont {Olivares}}, \ and\
  \bibinfo {author} {\bibfnamefont {K.~M.}\ \bibnamefont {Zurek}},\ }\href
  {\doibase 10.1103/PhysRevD.103.016006} {\bibfield  {journal} {\bibinfo
  {journal} {Phys. Rev. D}\ }\textbf {\bibinfo {volume} {103}},\ \bibinfo
  {pages} {016006} (\bibinfo {year} {2021})},\ \Eprint
  {http://arxiv.org/abs/1909.09170} {arXiv:1909.09170 [hep-ph]} \BibitemShut
  {NoStop}%
\bibitem [{\citenamefont {Geilhufe}\ \emph {et~al.}(2020)\citenamefont
  {Geilhufe}, \citenamefont {Kahlhoefer},\ and\ \citenamefont
  {Winkler}}]{Geilhufe:2019ndy}%
  \BibitemOpen
  \bibfield  {author} {\bibinfo {author} {\bibfnamefont {R.~M.}\ \bibnamefont
  {Geilhufe}}, \bibinfo {author} {\bibfnamefont {F.}~\bibnamefont
  {Kahlhoefer}}, \ and\ \bibinfo {author} {\bibfnamefont {M.~W.}\ \bibnamefont
  {Winkler}},\ }\href {\doibase 10.1103/PhysRevD.101.055005} {\bibfield
  {journal} {\bibinfo  {journal} {Phys. Rev. D}\ }\textbf {\bibinfo {volume}
  {101}},\ \bibinfo {pages} {055005} (\bibinfo {year} {2020})},\ \Eprint
  {http://arxiv.org/abs/1910.02091} {arXiv:1910.02091 [hep-ph]} \BibitemShut
  {NoStop}%
\bibitem [{\citenamefont {Du}\ \emph {et~al.}(2022)\citenamefont {Du},
  \citenamefont {Ega\~na Ugrinovic}, \citenamefont {Essig},\ and\ \citenamefont
  {Sholapurkar}}]{Du:2022dxf}%
  \BibitemOpen
  \bibfield  {author} {\bibinfo {author} {\bibfnamefont {P.}~\bibnamefont
  {Du}}, \bibinfo {author} {\bibfnamefont {D.}~\bibnamefont {Ega\~na
  Ugrinovic}}, \bibinfo {author} {\bibfnamefont {R.}~\bibnamefont {Essig}}, \
  and\ \bibinfo {author} {\bibfnamefont {M.}~\bibnamefont {Sholapurkar}},\
  }\href@noop {} {\  (\bibinfo {year} {2022})},\ \Eprint
  {http://arxiv.org/abs/2212.04504} {arXiv:2212.04504 [hep-ph]} \BibitemShut
  {NoStop}%
\bibitem [{\citenamefont {Hochberg}\ \emph {et~al.}(2017)\citenamefont
  {Hochberg}, \citenamefont {Kahn}, \citenamefont {Lisanti}, \citenamefont
  {Tully},\ and\ \citenamefont {Zurek}}]{Hochberg:2016ntt}%
  \BibitemOpen
  \bibfield  {author} {\bibinfo {author} {\bibfnamefont {Y.}~\bibnamefont
  {Hochberg}}, \bibinfo {author} {\bibfnamefont {Y.}~\bibnamefont {Kahn}},
  \bibinfo {author} {\bibfnamefont {M.}~\bibnamefont {Lisanti}}, \bibinfo
  {author} {\bibfnamefont {C.~G.}\ \bibnamefont {Tully}}, \ and\ \bibinfo
  {author} {\bibfnamefont {K.~M.}\ \bibnamefont {Zurek}},\ }\href {\doibase
  10.1016/j.physletb.2017.06.051} {\bibfield  {journal} {\bibinfo  {journal}
  {Phys. Lett. B}\ }\textbf {\bibinfo {volume} {772}},\ \bibinfo {pages} {239}
  (\bibinfo {year} {2017})},\ \Eprint {http://arxiv.org/abs/1606.08849}
  {arXiv:1606.08849 [hep-ph]} \BibitemShut {NoStop}%
\bibitem [{\citenamefont {Catena}\ \emph
  {et~al.}(2023{\natexlab{a}})\citenamefont {Catena}, \citenamefont {Emken},
  \citenamefont {Matas}, \citenamefont {Spaldin},\ and\ \citenamefont
  {Urdshals}}]{Catena:2023qkj}%
  \BibitemOpen
  \bibfield  {author} {\bibinfo {author} {\bibfnamefont {R.}~\bibnamefont
  {Catena}}, \bibinfo {author} {\bibfnamefont {T.}~\bibnamefont {Emken}},
  \bibinfo {author} {\bibfnamefont {M.}~\bibnamefont {Matas}}, \bibinfo
  {author} {\bibfnamefont {N.~A.}\ \bibnamefont {Spaldin}}, \ and\ \bibinfo
  {author} {\bibfnamefont {E.}~\bibnamefont {Urdshals}},\ }\href@noop {} {\
  (\bibinfo {year} {2023}{\natexlab{a}})},\ \Eprint
  {http://arxiv.org/abs/2303.15497} {arXiv:2303.15497 [hep-ph]} \BibitemShut
  {NoStop}%
\bibitem [{\citenamefont {Catena}\ \emph
  {et~al.}(2023{\natexlab{b}})\citenamefont {Catena}, \citenamefont {Emken},
  \citenamefont {Matas}, \citenamefont {Spaldin},\ and\ \citenamefont
  {Urdshals}}]{Catena:2023awl}%
  \BibitemOpen
  \bibfield  {author} {\bibinfo {author} {\bibfnamefont {R.}~\bibnamefont
  {Catena}}, \bibinfo {author} {\bibfnamefont {T.}~\bibnamefont {Emken}},
  \bibinfo {author} {\bibfnamefont {M.}~\bibnamefont {Matas}}, \bibinfo
  {author} {\bibfnamefont {N.~A.}\ \bibnamefont {Spaldin}}, \ and\ \bibinfo
  {author} {\bibfnamefont {E.}~\bibnamefont {Urdshals}},\ }\href@noop {} {\
  (\bibinfo {year} {2023}{\natexlab{b}})},\ \Eprint
  {http://arxiv.org/abs/2303.15509} {arXiv:2303.15509 [hep-ph]} \BibitemShut
  {NoStop}%
\bibitem [{\citenamefont {Albakry}\ \emph {et~al.}(2022)\citenamefont {Albakry}
  \emph {et~al.}}]{SuperCDMS:2022kse}%
  \BibitemOpen
  \bibfield  {author} {\bibinfo {author} {\bibfnamefont {M.~F.}\ \bibnamefont
  {Albakry}} \emph {et~al.} (\bibinfo {collaboration} {SuperCDMS}),\ }in\
  \href@noop {} {\emph {\bibinfo {booktitle} {{Snowmass 2021}}}}\ (\bibinfo
  {year} {2022})\ \Eprint {http://arxiv.org/abs/2203.08463} {arXiv:2203.08463
  [physics.ins-det]} \BibitemShut {NoStop}%
\bibitem [{\citenamefont {Inzani}\ \emph {et~al.}(2021)\citenamefont {Inzani},
  \citenamefont {Faghaninia},\ and\ \citenamefont {Griffin}}]{Inzani:2020szg}%
  \BibitemOpen
  \bibfield  {author} {\bibinfo {author} {\bibfnamefont {K.}~\bibnamefont
  {Inzani}}, \bibinfo {author} {\bibfnamefont {A.}~\bibnamefont {Faghaninia}},
  \ and\ \bibinfo {author} {\bibfnamefont {S.~M.}\ \bibnamefont {Griffin}},\
  }\href {\doibase 10.1103/PhysRevResearch.3.013069} {\bibfield  {journal}
  {\bibinfo  {journal} {Phys. Rev. Res.}\ }\textbf {\bibinfo {volume} {3}},\
  \bibinfo {pages} {013069} (\bibinfo {year} {2021})},\ \Eprint
  {http://arxiv.org/abs/2008.05062} {arXiv:2008.05062 [cond-mat.mtrl-sci]}
  \BibitemShut {NoStop}%
\bibitem [{\citenamefont {Chen}\ \emph {et~al.}(2022)\citenamefont {Chen},
  \citenamefont {Mitridate}, \citenamefont {Trickle}, \citenamefont {Zhang},
  \citenamefont {Bernardi},\ and\ \citenamefont {Zurek}}]{Chen:2022pyd}%
  \BibitemOpen
  \bibfield  {author} {\bibinfo {author} {\bibfnamefont {H.-Y.}\ \bibnamefont
  {Chen}}, \bibinfo {author} {\bibfnamefont {A.}~\bibnamefont {Mitridate}},
  \bibinfo {author} {\bibfnamefont {T.}~\bibnamefont {Trickle}}, \bibinfo
  {author} {\bibfnamefont {Z.}~\bibnamefont {Zhang}}, \bibinfo {author}
  {\bibfnamefont {M.}~\bibnamefont {Bernardi}}, \ and\ \bibinfo {author}
  {\bibfnamefont {K.~M.}\ \bibnamefont {Zurek}},\ }\href {\doibase
  10.1103/PhysRevD.106.015024} {\bibfield  {journal} {\bibinfo  {journal}
  {Phys. Rev. D}\ }\textbf {\bibinfo {volume} {106}},\ \bibinfo {pages}
  {015024} (\bibinfo {year} {2022})},\ \Eprint
  {http://arxiv.org/abs/2202.11716} {arXiv:2202.11716 [hep-ph]} \BibitemShut
  {NoStop}%
\bibitem [{\citenamefont {Schutz}\ and\ \citenamefont
  {Zurek}(2016)}]{Schutz:2016tid}%
  \BibitemOpen
  \bibfield  {author} {\bibinfo {author} {\bibfnamefont {K.}~\bibnamefont
  {Schutz}}\ and\ \bibinfo {author} {\bibfnamefont {K.~M.}\ \bibnamefont
  {Zurek}},\ }\href {\doibase 10.1103/PhysRevLett.117.121302} {\bibfield
  {journal} {\bibinfo  {journal} {Phys. Rev. Lett.}\ }\textbf {\bibinfo
  {volume} {117}},\ \bibinfo {pages} {121302} (\bibinfo {year} {2016})},\
  \Eprint {http://arxiv.org/abs/1604.08206} {arXiv:1604.08206 [hep-ph]}
  \BibitemShut {NoStop}%
\bibitem [{\citenamefont {Knapen}\ \emph
  {et~al.}(2017{\natexlab{a}})\citenamefont {Knapen}, \citenamefont {Lin},\
  and\ \citenamefont {Zurek}}]{Knapen:2016cue}%
  \BibitemOpen
  \bibfield  {author} {\bibinfo {author} {\bibfnamefont {S.}~\bibnamefont
  {Knapen}}, \bibinfo {author} {\bibfnamefont {T.}~\bibnamefont {Lin}}, \ and\
  \bibinfo {author} {\bibfnamefont {K.~M.}\ \bibnamefont {Zurek}},\ }\href
  {\doibase 10.1103/PhysRevD.95.056019} {\bibfield  {journal} {\bibinfo
  {journal} {Phys. Rev. D}\ }\textbf {\bibinfo {volume} {95}},\ \bibinfo
  {pages} {056019} (\bibinfo {year} {2017}{\natexlab{a}})},\ \Eprint
  {http://arxiv.org/abs/1611.06228} {arXiv:1611.06228 [hep-ph]} \BibitemShut
  {NoStop}%
\bibitem [{\citenamefont {Knapen}\ \emph {et~al.}(2018)\citenamefont {Knapen},
  \citenamefont {Lin}, \citenamefont {Pyle},\ and\ \citenamefont
  {Zurek}}]{Knapen:2017ekk}%
  \BibitemOpen
  \bibfield  {author} {\bibinfo {author} {\bibfnamefont {S.}~\bibnamefont
  {Knapen}}, \bibinfo {author} {\bibfnamefont {T.}~\bibnamefont {Lin}},
  \bibinfo {author} {\bibfnamefont {M.}~\bibnamefont {Pyle}}, \ and\ \bibinfo
  {author} {\bibfnamefont {K.~M.}\ \bibnamefont {Zurek}},\ }\href {\doibase
  10.1016/j.physletb.2018.08.064} {\bibfield  {journal} {\bibinfo  {journal}
  {Phys. Lett. B}\ }\textbf {\bibinfo {volume} {785}},\ \bibinfo {pages} {386}
  (\bibinfo {year} {2018})},\ \Eprint {http://arxiv.org/abs/1712.06598}
  {arXiv:1712.06598 [hep-ph]} \BibitemShut {NoStop}%
\bibitem [{\citenamefont {Griffin}\ \emph {et~al.}(2018)\citenamefont
  {Griffin}, \citenamefont {Knapen}, \citenamefont {Lin},\ and\ \citenamefont
  {Zurek}}]{Griffin:2018bjn}%
  \BibitemOpen
  \bibfield  {author} {\bibinfo {author} {\bibfnamefont {S.}~\bibnamefont
  {Griffin}}, \bibinfo {author} {\bibfnamefont {S.}~\bibnamefont {Knapen}},
  \bibinfo {author} {\bibfnamefont {T.}~\bibnamefont {Lin}}, \ and\ \bibinfo
  {author} {\bibfnamefont {K.~M.}\ \bibnamefont {Zurek}},\ }\href {\doibase
  10.1103/PhysRevD.98.115034} {\bibfield  {journal} {\bibinfo  {journal} {Phys.
  Rev. D}\ }\textbf {\bibinfo {volume} {98}},\ \bibinfo {pages} {115034}
  (\bibinfo {year} {2018})},\ \Eprint {http://arxiv.org/abs/1807.10291}
  {arXiv:1807.10291 [hep-ph]} \BibitemShut {NoStop}%
\bibitem [{\citenamefont {Trickle}\ \emph
  {et~al.}(2020{\natexlab{a}})\citenamefont {Trickle}, \citenamefont {Zhang},
  \citenamefont {Zurek}, \citenamefont {Inzani},\ and\ \citenamefont
  {Griffin}}]{Trickle:2019nya}%
  \BibitemOpen
  \bibfield  {author} {\bibinfo {author} {\bibfnamefont {T.}~\bibnamefont
  {Trickle}}, \bibinfo {author} {\bibfnamefont {Z.}~\bibnamefont {Zhang}},
  \bibinfo {author} {\bibfnamefont {K.~M.}\ \bibnamefont {Zurek}}, \bibinfo
  {author} {\bibfnamefont {K.}~\bibnamefont {Inzani}}, \ and\ \bibinfo {author}
  {\bibfnamefont {S.~M.}\ \bibnamefont {Griffin}},\ }\href {\doibase
  10.1007/JHEP03(2020)036} {\bibfield  {journal} {\bibinfo  {journal} {JHEP}\
  }\textbf {\bibinfo {volume} {03}},\ \bibinfo {pages} {036} (\bibinfo {year}
  {2020}{\natexlab{a}})},\ \Eprint {http://arxiv.org/abs/1910.08092}
  {arXiv:1910.08092 [hep-ph]} \BibitemShut {NoStop}%
\bibitem [{\citenamefont {Barbieri}\ \emph {et~al.}(1989)\citenamefont
  {Barbieri}, \citenamefont {Cerdonio}, \citenamefont {Fiorentini},\ and\
  \citenamefont {Vitale}}]{Barbieri:1985cp}%
  \BibitemOpen
  \bibfield  {author} {\bibinfo {author} {\bibfnamefont {R.}~\bibnamefont
  {Barbieri}}, \bibinfo {author} {\bibfnamefont {M.}~\bibnamefont {Cerdonio}},
  \bibinfo {author} {\bibfnamefont {G.}~\bibnamefont {Fiorentini}}, \ and\
  \bibinfo {author} {\bibfnamefont {S.}~\bibnamefont {Vitale}},\ }\href
  {\doibase 10.1016/0370-2693(89)91209-4} {\bibfield  {journal} {\bibinfo
  {journal} {Phys. Lett. B}\ }\textbf {\bibinfo {volume} {226}},\ \bibinfo
  {pages} {357} (\bibinfo {year} {1989})}\BibitemShut {NoStop}%
\bibitem [{\citenamefont {Trickle}\ \emph
  {et~al.}(2020{\natexlab{b}})\citenamefont {Trickle}, \citenamefont {Zhang},\
  and\ \citenamefont {Zurek}}]{Trickle:2019ovy}%
  \BibitemOpen
  \bibfield  {author} {\bibinfo {author} {\bibfnamefont {T.}~\bibnamefont
  {Trickle}}, \bibinfo {author} {\bibfnamefont {Z.}~\bibnamefont {Zhang}}, \
  and\ \bibinfo {author} {\bibfnamefont {K.~M.}\ \bibnamefont {Zurek}},\ }\href
  {\doibase 10.1103/PhysRevLett.124.201801} {\bibfield  {journal} {\bibinfo
  {journal} {Phys. Rev. Lett.}\ }\textbf {\bibinfo {volume} {124}},\ \bibinfo
  {pages} {201801} (\bibinfo {year} {2020}{\natexlab{b}})},\ \Eprint
  {http://arxiv.org/abs/1905.13744} {arXiv:1905.13744 [hep-ph]} \BibitemShut
  {NoStop}%
\bibitem [{\citenamefont {Chigusa}\ \emph {et~al.}(2020)\citenamefont
  {Chigusa}, \citenamefont {Moroi},\ and\ \citenamefont
  {Nakayama}}]{Chigusa:2020gfs}%
  \BibitemOpen
  \bibfield  {author} {\bibinfo {author} {\bibfnamefont {S.}~\bibnamefont
  {Chigusa}}, \bibinfo {author} {\bibfnamefont {T.}~\bibnamefont {Moroi}}, \
  and\ \bibinfo {author} {\bibfnamefont {K.}~\bibnamefont {Nakayama}},\ }\href
  {\doibase 10.1103/PhysRevD.101.096013} {\bibfield  {journal} {\bibinfo
  {journal} {Phys. Rev. D}\ }\textbf {\bibinfo {volume} {101}},\ \bibinfo
  {pages} {096013} (\bibinfo {year} {2020})},\ \Eprint
  {http://arxiv.org/abs/2001.10666} {arXiv:2001.10666 [hep-ph]} \BibitemShut
  {NoStop}%
\bibitem [{\citenamefont {Mitridate}\ \emph {et~al.}(2020)\citenamefont
  {Mitridate}, \citenamefont {Trickle}, \citenamefont {Zhang},\ and\
  \citenamefont {Zurek}}]{Mitridate:2020kly}%
  \BibitemOpen
  \bibfield  {author} {\bibinfo {author} {\bibfnamefont {A.}~\bibnamefont
  {Mitridate}}, \bibinfo {author} {\bibfnamefont {T.}~\bibnamefont {Trickle}},
  \bibinfo {author} {\bibfnamefont {Z.}~\bibnamefont {Zhang}}, \ and\ \bibinfo
  {author} {\bibfnamefont {K.~M.}\ \bibnamefont {Zurek}},\ }\href {\doibase
  10.1103/PhysRevD.102.095005} {\bibfield  {journal} {\bibinfo  {journal}
  {Phys. Rev. D}\ }\textbf {\bibinfo {volume} {102}},\ \bibinfo {pages}
  {095005} (\bibinfo {year} {2020})},\ \Eprint
  {http://arxiv.org/abs/2005.10256} {arXiv:2005.10256 [hep-ph]} \BibitemShut
  {NoStop}%
\bibitem [{\citenamefont {Trickle}\ \emph {et~al.}(2022)\citenamefont
  {Trickle}, \citenamefont {Zhang},\ and\ \citenamefont
  {Zurek}}]{Trickle:2020oki}%
  \BibitemOpen
  \bibfield  {author} {\bibinfo {author} {\bibfnamefont {T.}~\bibnamefont
  {Trickle}}, \bibinfo {author} {\bibfnamefont {Z.}~\bibnamefont {Zhang}}, \
  and\ \bibinfo {author} {\bibfnamefont {K.~M.}\ \bibnamefont {Zurek}},\ }\href
  {\doibase 10.1103/PhysRevD.105.015001} {\bibfield  {journal} {\bibinfo
  {journal} {Phys. Rev. D}\ }\textbf {\bibinfo {volume} {105}},\ \bibinfo
  {pages} {015001} (\bibinfo {year} {2022})},\ \Eprint
  {http://arxiv.org/abs/2009.13534} {arXiv:2009.13534 [hep-ph]} \BibitemShut
  {NoStop}%
\bibitem [{\citenamefont {Chang}\ \emph {et~al.}(2020)\citenamefont {Chang}
  \emph {et~al.}}]{Chang2020}%
  \BibitemOpen
  \bibfield  {author} {\bibinfo {author} {\bibfnamefont {C.}~\bibnamefont
  {Chang}} \emph {et~al.},\ }\href
  {https://www.snowmass21.org/docs/files/summaries/CF/SNOWMASS21-CF1_CF2-IF1_IF8-120.pdf}
  {\enquote {\bibinfo {title} {Snowmass 2021 letter of interest: The tessaract
  dark matter project},}\ } (\bibinfo {year} {2020})\BibitemShut {NoStop}%
\bibitem [{\citenamefont {Caputo}\ \emph {et~al.}(2019)\citenamefont {Caputo},
  \citenamefont {Esposito},\ and\ \citenamefont {Polosa}}]{Caputo:2019cyg}%
  \BibitemOpen
  \bibfield  {author} {\bibinfo {author} {\bibfnamefont {A.}~\bibnamefont
  {Caputo}}, \bibinfo {author} {\bibfnamefont {A.}~\bibnamefont {Esposito}}, \
  and\ \bibinfo {author} {\bibfnamefont {A.~D.}\ \bibnamefont {Polosa}},\
  }\href {\doibase 10.1103/PhysRevD.100.116007} {\bibfield  {journal} {\bibinfo
   {journal} {Phys. Rev. D}\ }\textbf {\bibinfo {volume} {100}},\ \bibinfo
  {pages} {116007} (\bibinfo {year} {2019})},\ \Eprint
  {http://arxiv.org/abs/1907.10635} {arXiv:1907.10635 [hep-ph]} \BibitemShut
  {NoStop}%
\bibitem [{\citenamefont {Cox}\ \emph {et~al.}(2019)\citenamefont {Cox},
  \citenamefont {Melia},\ and\ \citenamefont {Rajendran}}]{Cox:2019cod}%
  \BibitemOpen
  \bibfield  {author} {\bibinfo {author} {\bibfnamefont {P.}~\bibnamefont
  {Cox}}, \bibinfo {author} {\bibfnamefont {T.}~\bibnamefont {Melia}}, \ and\
  \bibinfo {author} {\bibfnamefont {S.}~\bibnamefont {Rajendran}},\ }\href
  {\doibase 10.1103/PhysRevD.100.055011} {\bibfield  {journal} {\bibinfo
  {journal} {Phys. Rev. D}\ }\textbf {\bibinfo {volume} {100}},\ \bibinfo
  {pages} {055011} (\bibinfo {year} {2019})},\ \Eprint
  {http://arxiv.org/abs/1905.05575} {arXiv:1905.05575 [hep-ph]} \BibitemShut
  {NoStop}%
\bibitem [{\citenamefont {Griffin}\ \emph {et~al.}(2020)\citenamefont
  {Griffin}, \citenamefont {Inzani}, \citenamefont {Trickle}, \citenamefont
  {Zhang},\ and\ \citenamefont {Zurek}}]{Griffin:2019mvc}%
  \BibitemOpen
  \bibfield  {author} {\bibinfo {author} {\bibfnamefont {S.~M.}\ \bibnamefont
  {Griffin}}, \bibinfo {author} {\bibfnamefont {K.}~\bibnamefont {Inzani}},
  \bibinfo {author} {\bibfnamefont {T.}~\bibnamefont {Trickle}}, \bibinfo
  {author} {\bibfnamefont {Z.}~\bibnamefont {Zhang}}, \ and\ \bibinfo {author}
  {\bibfnamefont {K.~M.}\ \bibnamefont {Zurek}},\ }\href {\doibase
  10.1103/PhysRevD.101.055004} {\bibfield  {journal} {\bibinfo  {journal}
  {Phys. Rev. D}\ }\textbf {\bibinfo {volume} {101}},\ \bibinfo {pages}
  {055004} (\bibinfo {year} {2020})},\ \Eprint
  {http://arxiv.org/abs/1910.10716} {arXiv:1910.10716 [hep-ph]} \BibitemShut
  {NoStop}%
\bibitem [{\citenamefont {Lasenby}\ and\ \citenamefont
  {Prabhu}(2022)}]{Lasenby:2021wsc}%
  \BibitemOpen
  \bibfield  {author} {\bibinfo {author} {\bibfnamefont {R.}~\bibnamefont
  {Lasenby}}\ and\ \bibinfo {author} {\bibfnamefont {A.}~\bibnamefont
  {Prabhu}},\ }\href {\doibase 10.1103/PhysRevD.105.095009} {\bibfield
  {journal} {\bibinfo  {journal} {Phys. Rev. D}\ }\textbf {\bibinfo {volume}
  {105}},\ \bibinfo {pages} {095009} (\bibinfo {year} {2022})},\ \Eprint
  {http://arxiv.org/abs/2110.01587} {arXiv:2110.01587 [hep-ph]} \BibitemShut
  {NoStop}%
\bibitem [{\citenamefont {Fitzpatrick}\ \emph {et~al.}(2013)\citenamefont
  {Fitzpatrick}, \citenamefont {Haxton}, \citenamefont {Katz}, \citenamefont
  {Lubbers},\ and\ \citenamefont {Xu}}]{Fitzpatrick:2012ix}%
  \BibitemOpen
  \bibfield  {author} {\bibinfo {author} {\bibfnamefont {A.~L.}\ \bibnamefont
  {Fitzpatrick}}, \bibinfo {author} {\bibfnamefont {W.}~\bibnamefont {Haxton}},
  \bibinfo {author} {\bibfnamefont {E.}~\bibnamefont {Katz}}, \bibinfo {author}
  {\bibfnamefont {N.}~\bibnamefont {Lubbers}}, \ and\ \bibinfo {author}
  {\bibfnamefont {Y.}~\bibnamefont {Xu}},\ }\href {\doibase
  10.1088/1475-7516/2013/02/004} {\bibfield  {journal} {\bibinfo  {journal}
  {JCAP}\ }\textbf {\bibinfo {volume} {02}},\ \bibinfo {pages} {004} (\bibinfo
  {year} {2013})},\ \Eprint {http://arxiv.org/abs/1203.3542} {arXiv:1203.3542
  [hep-ph]} \BibitemShut {NoStop}%
\bibitem [{\citenamefont {Cirelli}\ \emph {et~al.}(2013)\citenamefont
  {Cirelli}, \citenamefont {Del~Nobile},\ and\ \citenamefont
  {Panci}}]{Cirelli:2013ufw}%
  \BibitemOpen
  \bibfield  {author} {\bibinfo {author} {\bibfnamefont {M.}~\bibnamefont
  {Cirelli}}, \bibinfo {author} {\bibfnamefont {E.}~\bibnamefont {Del~Nobile}},
  \ and\ \bibinfo {author} {\bibfnamefont {P.}~\bibnamefont {Panci}},\ }\href
  {\doibase 10.1088/1475-7516/2013/10/019} {\bibfield  {journal} {\bibinfo
  {journal} {JCAP}\ }\textbf {\bibinfo {volume} {10}},\ \bibinfo {pages} {019}
  (\bibinfo {year} {2013})},\ \Eprint {http://arxiv.org/abs/1307.5955}
  {arXiv:1307.5955 [hep-ph]} \BibitemShut {NoStop}%
\bibitem [{\citenamefont {Anand}\ \emph {et~al.}(2014)\citenamefont {Anand},
  \citenamefont {Fitzpatrick},\ and\ \citenamefont {Haxton}}]{Anand:2013yka}%
  \BibitemOpen
  \bibfield  {author} {\bibinfo {author} {\bibfnamefont {N.}~\bibnamefont
  {Anand}}, \bibinfo {author} {\bibfnamefont {A.~L.}\ \bibnamefont
  {Fitzpatrick}}, \ and\ \bibinfo {author} {\bibfnamefont {W.~C.}\ \bibnamefont
  {Haxton}},\ }\href {\doibase 10.1103/PhysRevC.89.065501} {\bibfield
  {journal} {\bibinfo  {journal} {Phys. Rev. C}\ }\textbf {\bibinfo {volume}
  {89}},\ \bibinfo {pages} {065501} (\bibinfo {year} {2014})},\ \Eprint
  {http://arxiv.org/abs/1308.6288} {arXiv:1308.6288 [hep-ph]} \BibitemShut
  {NoStop}%
\bibitem [{\citenamefont {Gresham}\ and\ \citenamefont
  {Zurek}(2014)}]{Gresham:2014vja}%
  \BibitemOpen
  \bibfield  {author} {\bibinfo {author} {\bibfnamefont {M.~I.}\ \bibnamefont
  {Gresham}}\ and\ \bibinfo {author} {\bibfnamefont {K.~M.}\ \bibnamefont
  {Zurek}},\ }\href {\doibase 10.1103/PhysRevD.89.123521} {\bibfield  {journal}
  {\bibinfo  {journal} {Phys. Rev. D}\ }\textbf {\bibinfo {volume} {89}},\
  \bibinfo {pages} {123521} (\bibinfo {year} {2014})},\ \Eprint
  {http://arxiv.org/abs/1401.3739} {arXiv:1401.3739 [hep-ph]} \BibitemShut
  {NoStop}%
\bibitem [{\citenamefont {Anand}\ \emph {et~al.}(2015)\citenamefont {Anand},
  \citenamefont {Fitzpatrick},\ and\ \citenamefont {Haxton}}]{Anand:2014kea}%
  \BibitemOpen
  \bibfield  {author} {\bibinfo {author} {\bibfnamefont {N.}~\bibnamefont
  {Anand}}, \bibinfo {author} {\bibfnamefont {A.~L.}\ \bibnamefont
  {Fitzpatrick}}, \ and\ \bibinfo {author} {\bibfnamefont {W.~C.}\ \bibnamefont
  {Haxton}},\ }\href {\doibase 10.1016/j.phpro.2014.12.017} {\bibfield
  {journal} {\bibinfo  {journal} {Phys. Procedia}\ }\textbf {\bibinfo {volume}
  {61}},\ \bibinfo {pages} {97} (\bibinfo {year} {2015})},\ \Eprint
  {http://arxiv.org/abs/1405.6690} {arXiv:1405.6690 [nucl-th]} \BibitemShut
  {NoStop}%
\bibitem [{\citenamefont {Del~Nobile}(2018)}]{DelNobile:2018dfg}%
  \BibitemOpen
  \bibfield  {author} {\bibinfo {author} {\bibfnamefont {E.}~\bibnamefont
  {Del~Nobile}},\ }\href {\doibase 10.1103/PhysRevD.98.123003} {\bibfield
  {journal} {\bibinfo  {journal} {Phys. Rev. D}\ }\textbf {\bibinfo {volume}
  {98}},\ \bibinfo {pages} {123003} (\bibinfo {year} {2018})},\ \Eprint
  {http://arxiv.org/abs/1806.01291} {arXiv:1806.01291 [hep-ph]} \BibitemShut
  {NoStop}%
\bibitem [{\citenamefont {Knapen}\ \emph {et~al.}(2022)\citenamefont {Knapen},
  \citenamefont {Kozaczuk},\ and\ \citenamefont {Lin}}]{Knapen:2021bwg}%
  \BibitemOpen
  \bibfield  {author} {\bibinfo {author} {\bibfnamefont {S.}~\bibnamefont
  {Knapen}}, \bibinfo {author} {\bibfnamefont {J.}~\bibnamefont {Kozaczuk}}, \
  and\ \bibinfo {author} {\bibfnamefont {T.}~\bibnamefont {Lin}},\ }\href
  {\doibase 10.1103/PhysRevD.105.015014} {\bibfield  {journal} {\bibinfo
  {journal} {Phys. Rev. D}\ }\textbf {\bibinfo {volume} {105}},\ \bibinfo
  {pages} {015014} (\bibinfo {year} {2022})},\ \Eprint
  {http://arxiv.org/abs/2104.12786} {arXiv:2104.12786 [hep-ph]} \BibitemShut
  {NoStop}%
\bibitem [{\citenamefont {Krnjaic}\ and\ \citenamefont
  {Trickle}(2023)}]{Krnjaic:2023nxe}%
  \BibitemOpen
  \bibfield  {author} {\bibinfo {author} {\bibfnamefont {G.}~\bibnamefont
  {Krnjaic}}\ and\ \bibinfo {author} {\bibfnamefont {T.}~\bibnamefont
  {Trickle}},\ }\href {\doibase 10.1103/PhysRevD.108.015024} {\bibfield
  {journal} {\bibinfo  {journal} {Phys. Rev. D}\ }\textbf {\bibinfo {volume}
  {108}},\ \bibinfo {pages} {015024} (\bibinfo {year} {2023})},\ \Eprint
  {http://arxiv.org/abs/2303.11344} {arXiv:2303.11344 [hep-ph]} \BibitemShut
  {NoStop}%
\bibitem [{\citenamefont {Mitridate}\ \emph
  {et~al.}(2021{\natexlab{b}})\citenamefont {Mitridate}, \citenamefont
  {Trickle}, \citenamefont {Zhang},\ and\ \citenamefont
  {Zurek}}]{Mitridate:2021ctra}%
  \BibitemOpen
  \bibfield  {author} {\bibinfo {author} {\bibfnamefont {A.}~\bibnamefont
  {Mitridate}}, \bibinfo {author} {\bibfnamefont {T.}~\bibnamefont {Trickle}},
  \bibinfo {author} {\bibfnamefont {Z.}~\bibnamefont {Zhang}}, \ and\ \bibinfo
  {author} {\bibfnamefont {K.~M.}\ \bibnamefont {Zurek}},\ }\href {\doibase
  10.1007/JHEP09(2021)123} {\bibfield  {journal} {\bibinfo  {journal} {JHEP}\
  }\textbf {\bibinfo {volume} {09}},\ \bibinfo {pages} {123} (\bibinfo {year}
  {2021}{\natexlab{b}})},\ \Eprint {http://arxiv.org/abs/2106.12586}
  {arXiv:2106.12586 [hep-ph]} \BibitemShut {NoStop}%
\bibitem [{\citenamefont {Hardy}\ and\ \citenamefont
  {Lasenby}(2017)}]{Hardy:2016kme}%
  \BibitemOpen
  \bibfield  {author} {\bibinfo {author} {\bibfnamefont {E.}~\bibnamefont
  {Hardy}}\ and\ \bibinfo {author} {\bibfnamefont {R.}~\bibnamefont
  {Lasenby}},\ }\href {\doibase 10.1007/JHEP02(2017)033} {\bibfield  {journal}
  {\bibinfo  {journal} {JHEP}\ }\textbf {\bibinfo {volume} {02}},\ \bibinfo
  {pages} {033} (\bibinfo {year} {2017})},\ \Eprint
  {http://arxiv.org/abs/1611.05852} {arXiv:1611.05852 [hep-ph]} \BibitemShut
  {NoStop}%
\bibitem [{\citenamefont {Trickle}\ and\ \citenamefont
  {Zhang}(2023)}]{PhonoDark}%
  \BibitemOpen
  \bibfield  {author} {\bibinfo {author} {\bibfnamefont {T.}~\bibnamefont
  {Trickle}}\ and\ \bibinfo {author} {\bibfnamefont {Z.}~\bibnamefont
  {Zhang}},\ }\href {https://github.com/tanner-trickle/PhonoDark} {\enquote
  {\bibinfo {title} {https://github.com/tanner-trickle/phonodark},}\ }
  (\bibinfo {year} {2023})\BibitemShut {NoStop}%
\bibitem [{\citenamefont {Mahan}(2000)}]{mah00}%
  \BibitemOpen
  \bibfield  {author} {\bibinfo {author} {\bibfnamefont {G.~D.}\ \bibnamefont
  {Mahan}},\ }\href@noop {} {\emph {\bibinfo {title} {Many Particle Physics,
  Third Edition}}}\ (\bibinfo  {publisher} {Plenum},\ \bibinfo {address} {New
  York},\ \bibinfo {year} {2000})\BibitemShut {NoStop}%
\bibitem [{\citenamefont {Zatorski}\ and\ \citenamefont
  {Pachucki}(2010)}]{PhysRevA.82.052520}%
  \BibitemOpen
  \bibfield  {author} {\bibinfo {author} {\bibfnamefont {J.}~\bibnamefont
  {Zatorski}}\ and\ \bibinfo {author} {\bibfnamefont {K.}~\bibnamefont
  {Pachucki}},\ }\href {\doibase 10.1103/PhysRevA.82.052520} {\bibfield
  {journal} {\bibinfo  {journal} {Phys. Rev. A}\ }\textbf {\bibinfo {volume}
  {82}},\ \bibinfo {pages} {052520} (\bibinfo {year} {2010})}\BibitemShut
  {NoStop}%
\bibitem [{\citenamefont {Balk}\ \emph
  {et~al.}(1994{\natexlab{a}})\citenamefont {Balk}, \citenamefont {Korner},\
  and\ \citenamefont {Pirjol}}]{Balk:1993eva}%
  \BibitemOpen
  \bibfield  {author} {\bibinfo {author} {\bibfnamefont {S.}~\bibnamefont
  {Balk}}, \bibinfo {author} {\bibfnamefont {J.~G.}\ \bibnamefont {Korner}}, \
  and\ \bibinfo {author} {\bibfnamefont {D.}~\bibnamefont {Pirjol}},\ }\href
  {\doibase 10.1016/0550-3213(94)90211-9} {\bibfield  {journal} {\bibinfo
  {journal} {Nucl. Phys. B}\ }\textbf {\bibinfo {volume} {428}},\ \bibinfo
  {pages} {499} (\bibinfo {year} {1994}{\natexlab{a}})},\ \Eprint
  {http://arxiv.org/abs/hep-ph/9307230} {arXiv:hep-ph/9307230} \BibitemShut
  {NoStop}%
\bibitem [{\citenamefont {Gardestig}\ \emph
  {et~al.}(2007{\natexlab{a}})\citenamefont {Gardestig}, \citenamefont
  {Kubodera},\ and\ \citenamefont {Myhrer}}]{Gardestig:2007mka}%
  \BibitemOpen
  \bibfield  {author} {\bibinfo {author} {\bibfnamefont {A.}~\bibnamefont
  {Gardestig}}, \bibinfo {author} {\bibfnamefont {K.}~\bibnamefont {Kubodera}},
  \ and\ \bibinfo {author} {\bibfnamefont {F.}~\bibnamefont {Myhrer}},\ }\href
  {\doibase 10.1103/PhysRevC.76.014005} {\bibfield  {journal} {\bibinfo
  {journal} {Phys. Rev. C}\ }\textbf {\bibinfo {volume} {76}},\ \bibinfo
  {pages} {014005} (\bibinfo {year} {2007}{\natexlab{a}})},\ \Eprint
  {http://arxiv.org/abs/0705.2885} {arXiv:0705.2885 [nucl-th]} \BibitemShut
  {NoStop}%
\bibitem [{\citenamefont {Bjorken}(1965)}]{bjorken}%
  \BibitemOpen
  \bibfield  {author} {\bibinfo {author} {\bibfnamefont {J.~D.}\ \bibnamefont
  {Bjorken}},\ }\href@noop {} {\emph {\bibinfo {title} {Relativistic quantum
  mechanics}}}\ (\bibinfo  {publisher} {McGraw-Hill},\ \bibinfo {year}
  {1965})\BibitemShut {NoStop}%
\bibitem [{\citenamefont {Gardestig}\ \emph
  {et~al.}(2007{\natexlab{b}})\citenamefont {Gardestig}, \citenamefont
  {Kubodera},\ and\ \citenamefont {Myhrer}}]{Gardestig:2007mk}%
  \BibitemOpen
  \bibfield  {author} {\bibinfo {author} {\bibfnamefont {A.}~\bibnamefont
  {Gardestig}}, \bibinfo {author} {\bibfnamefont {K.}~\bibnamefont {Kubodera}},
  \ and\ \bibinfo {author} {\bibfnamefont {F.}~\bibnamefont {Myhrer}},\ }\href
  {\doibase 10.1103/PhysRevC.76.014005} {\bibfield  {journal} {\bibinfo
  {journal} {Phys. Rev. C}\ }\textbf {\bibinfo {volume} {76}},\ \bibinfo
  {pages} {014005} (\bibinfo {year} {2007}{\natexlab{b}})},\ \Eprint
  {http://arxiv.org/abs/0705.2885} {arXiv:0705.2885 [nucl-th]} \BibitemShut
  {NoStop}%
\bibitem [{\citenamefont {Foldy}\ and\ \citenamefont
  {Wouthuysen}(1950)}]{Foldy:1949wa}%
  \BibitemOpen
  \bibfield  {author} {\bibinfo {author} {\bibfnamefont {L.~L.}\ \bibnamefont
  {Foldy}}\ and\ \bibinfo {author} {\bibfnamefont {S.~A.}\ \bibnamefont
  {Wouthuysen}},\ }\href {\doibase 10.1103/PhysRev.78.29} {\bibfield  {journal}
  {\bibinfo  {journal} {Phys. Rev.}\ }\textbf {\bibinfo {volume} {78}},\
  \bibinfo {pages} {29} (\bibinfo {year} {1950})}\BibitemShut {NoStop}%
\bibitem [{\citenamefont {Foldy}(1952)}]{Foldy_1952}%
  \BibitemOpen
  \bibfield  {author} {\bibinfo {author} {\bibfnamefont {L.~L.}\ \bibnamefont
  {Foldy}},\ }\href {\doibase 10.1103/physrev.87.688} {\bibfield  {journal}
  {\bibinfo  {journal} {Physical Review}\ }\textbf {\bibinfo {volume} {87}},\
  \bibinfo {pages} {688} (\bibinfo {year} {1952})}\BibitemShut {NoStop}%
\bibitem [{\citenamefont {Balk}\ \emph
  {et~al.}(1994{\natexlab{b}})\citenamefont {Balk}, \citenamefont {Korner},\
  and\ \citenamefont {Pirjol}}]{Balk:1993ev}%
  \BibitemOpen
  \bibfield  {author} {\bibinfo {author} {\bibfnamefont {S.}~\bibnamefont
  {Balk}}, \bibinfo {author} {\bibfnamefont {J.~G.}\ \bibnamefont {Korner}}, \
  and\ \bibinfo {author} {\bibfnamefont {D.}~\bibnamefont {Pirjol}},\ }\href
  {\doibase 10.1016/0550-3213(94)90211-9} {\bibfield  {journal} {\bibinfo
  {journal} {Nucl. Phys. B}\ }\textbf {\bibinfo {volume} {428}},\ \bibinfo
  {pages} {499} (\bibinfo {year} {1994}{\natexlab{b}})},\ \Eprint
  {http://arxiv.org/abs/hep-ph/9307230} {arXiv:hep-ph/9307230} \BibitemShut
  {NoStop}%
\bibitem [{\citenamefont {Smith}(2023)}]{Smith:2023htu}%
  \BibitemOpen
  \bibfield  {author} {\bibinfo {author} {\bibfnamefont {C.}~\bibnamefont
  {Smith}},\ }\href@noop {} {\  (\bibinfo {year} {2023})},\ \Eprint
  {http://arxiv.org/abs/2302.01142} {arXiv:2302.01142 [hep-ph]} \BibitemShut
  {NoStop}%
\bibitem [{\citenamefont {Martin}(2004)}]{Martin_2004}%
  \BibitemOpen
  \bibfield  {author} {\bibinfo {author} {\bibfnamefont {R.~M.}\ \bibnamefont
  {Martin}},\ }\href {\doibase 10.1017/cbo9780511805769} {\emph {\bibinfo
  {title} {Electronic Structure}}}\ (\bibinfo  {publisher} {Cambridge
  University Press},\ \bibinfo {year} {2004})\BibitemShut {NoStop}%
\bibitem [{\citenamefont {Kresse}\ and\ \citenamefont {Hafner}(1993)}]{VASP_1}%
  \BibitemOpen
  \bibfield  {author} {\bibinfo {author} {\bibfnamefont {G.}~\bibnamefont
  {Kresse}}\ and\ \bibinfo {author} {\bibfnamefont {J.}~\bibnamefont
  {Hafner}},\ }\href {\doibase 10.1103/PhysRevB.47.558} {\bibfield  {journal}
  {\bibinfo  {journal} {Phys. Rev. B}\ }\textbf {\bibinfo {volume} {47}},\
  \bibinfo {pages} {558} (\bibinfo {year} {1993})}\BibitemShut {NoStop}%
\bibitem [{\citenamefont {Kresse}\ and\ \citenamefont {Hafner}(1994)}]{VASP_2}%
  \BibitemOpen
  \bibfield  {author} {\bibinfo {author} {\bibfnamefont {G.}~\bibnamefont
  {Kresse}}\ and\ \bibinfo {author} {\bibfnamefont {J.}~\bibnamefont
  {Hafner}},\ }\href {\doibase 10.1103/PhysRevB.49.14251} {\bibfield  {journal}
  {\bibinfo  {journal} {Phys. Rev. B}\ }\textbf {\bibinfo {volume} {49}},\
  \bibinfo {pages} {14251} (\bibinfo {year} {1994})}\BibitemShut {NoStop}%
\bibitem [{\citenamefont {Kresse}\ and\ \citenamefont
  {Furthmüller}(1996)}]{VASP_3}%
  \BibitemOpen
  \bibfield  {author} {\bibinfo {author} {\bibfnamefont {G.}~\bibnamefont
  {Kresse}}\ and\ \bibinfo {author} {\bibfnamefont {J.}~\bibnamefont
  {Furthmüller}},\ }\href {\doibase
  https://doi.org/10.1016/0927-0256(96)00008-0} {\bibfield  {journal} {\bibinfo
   {journal} {Computational Materials Science}\ }\textbf {\bibinfo {volume}
  {6}},\ \bibinfo {pages} {15} (\bibinfo {year} {1996})}\BibitemShut {NoStop}%
\bibitem [{\citenamefont {Bl\"ochl}(1994)}]{VASP_4}%
  \BibitemOpen
  \bibfield  {author} {\bibinfo {author} {\bibfnamefont {P.~E.}\ \bibnamefont
  {Bl\"ochl}},\ }\href {\doibase 10.1103/PhysRevB.50.17953} {\bibfield
  {journal} {\bibinfo  {journal} {Phys. Rev. B}\ }\textbf {\bibinfo {volume}
  {50}},\ \bibinfo {pages} {17953} (\bibinfo {year} {1994})}\BibitemShut
  {NoStop}%
\bibitem [{\citenamefont {Kresse}\ and\ \citenamefont
  {Joubert}(1999)}]{VASP_5}%
  \BibitemOpen
  \bibfield  {author} {\bibinfo {author} {\bibfnamefont {G.}~\bibnamefont
  {Kresse}}\ and\ \bibinfo {author} {\bibfnamefont {D.}~\bibnamefont
  {Joubert}},\ }\href {\doibase 10.1103/PhysRevB.59.1758} {\bibfield  {journal}
  {\bibinfo  {journal} {Phys. Rev. B}\ }\textbf {\bibinfo {volume} {59}},\
  \bibinfo {pages} {1758} (\bibinfo {year} {1999})}\BibitemShut {NoStop}%
\bibitem [{\citenamefont {Togo}\ and\ \citenamefont {Tanaka}(2015)}]{phonopy}%
  \BibitemOpen
  \bibfield  {author} {\bibinfo {author} {\bibfnamefont {A.}~\bibnamefont
  {Togo}}\ and\ \bibinfo {author} {\bibfnamefont {I.}~\bibnamefont {Tanaka}},\
  }\href {\doibase https://doi.org/10.1016/j.scriptamat.2015.07.021} {\bibfield
   {journal} {\bibinfo  {journal} {Scripta Materialia}\ }\textbf {\bibinfo
  {volume} {108}},\ \bibinfo {pages} {1} (\bibinfo {year} {2015})}\BibitemShut
  {NoStop}%
\bibitem [{\citenamefont {Gonze}\ and\ \citenamefont {Lee}(1997)}]{Gonze_1997}%
  \BibitemOpen
  \bibfield  {author} {\bibinfo {author} {\bibfnamefont {X.}~\bibnamefont
  {Gonze}}\ and\ \bibinfo {author} {\bibfnamefont {C.}~\bibnamefont {Lee}},\
  }\href {\doibase 10.1103/PhysRevB.55.10355} {\bibfield  {journal} {\bibinfo
  {journal} {Phys. Rev. B}\ }\textbf {\bibinfo {volume} {55}},\ \bibinfo
  {pages} {10355} (\bibinfo {year} {1997})}\BibitemShut {NoStop}%
\bibitem [{\citenamefont {Coskuner}\ \emph {et~al.}(2022)\citenamefont
  {Coskuner}, \citenamefont {Trickle}, \citenamefont {Zhang},\ and\
  \citenamefont {Zurek}}]{Coskuner:2021qxo}%
  \BibitemOpen
  \bibfield  {author} {\bibinfo {author} {\bibfnamefont {A.}~\bibnamefont
  {Coskuner}}, \bibinfo {author} {\bibfnamefont {T.}~\bibnamefont {Trickle}},
  \bibinfo {author} {\bibfnamefont {Z.}~\bibnamefont {Zhang}}, \ and\ \bibinfo
  {author} {\bibfnamefont {K.~M.}\ \bibnamefont {Zurek}},\ }\href {\doibase
  10.1103/PhysRevD.105.015010} {\bibfield  {journal} {\bibinfo  {journal}
  {Phys. Rev. D}\ }\textbf {\bibinfo {volume} {105}},\ \bibinfo {pages}
  {015010} (\bibinfo {year} {2022})},\ \Eprint
  {http://arxiv.org/abs/2102.09567} {arXiv:2102.09567 [hep-ph]} \BibitemShut
  {NoStop}%
\bibitem [{\citenamefont {Berlin}\ and\ \citenamefont
  {Trickle}(2023)}]{Berlin:2023ppd}%
  \BibitemOpen
  \bibfield  {author} {\bibinfo {author} {\bibfnamefont {A.}~\bibnamefont
  {Berlin}}\ and\ \bibinfo {author} {\bibfnamefont {T.}~\bibnamefont
  {Trickle}},\ }\href@noop {} {\  (\bibinfo {year} {2023})},\ \Eprint
  {http://arxiv.org/abs/2305.05681} {arXiv:2305.05681 [hep-ph]} \BibitemShut
  {NoStop}%
\bibitem [{\citenamefont {Jain}\ \emph {et~al.}(2013)\citenamefont {Jain},
  \citenamefont {Ong}, \citenamefont {Hautier}, \citenamefont {Chen},
  \citenamefont {Richards}, \citenamefont {Dacek}, \citenamefont {Cholia},
  \citenamefont {Gunter}, \citenamefont {Skinner}, \citenamefont {Ceder},\ and\
  \citenamefont {Persson}}]{Jain_2013}%
  \BibitemOpen
  \bibfield  {author} {\bibinfo {author} {\bibfnamefont {A.}~\bibnamefont
  {Jain}}, \bibinfo {author} {\bibfnamefont {S.~P.}\ \bibnamefont {Ong}},
  \bibinfo {author} {\bibfnamefont {G.}~\bibnamefont {Hautier}}, \bibinfo
  {author} {\bibfnamefont {W.}~\bibnamefont {Chen}}, \bibinfo {author}
  {\bibfnamefont {W.~D.}\ \bibnamefont {Richards}}, \bibinfo {author}
  {\bibfnamefont {S.}~\bibnamefont {Dacek}}, \bibinfo {author} {\bibfnamefont
  {S.}~\bibnamefont {Cholia}}, \bibinfo {author} {\bibfnamefont
  {D.}~\bibnamefont {Gunter}}, \bibinfo {author} {\bibfnamefont
  {D.}~\bibnamefont {Skinner}}, \bibinfo {author} {\bibfnamefont
  {G.}~\bibnamefont {Ceder}}, \ and\ \bibinfo {author} {\bibfnamefont {K.~A.}\
  \bibnamefont {Persson}},\ }\href {\doibase 10.1063/1.4812323} {\bibfield
  {journal} {\bibinfo  {journal} {{APL} Materials}\ }\textbf {\bibinfo {volume}
  {1}} (\bibinfo {year} {2013}),\ 10.1063/1.4812323}\BibitemShut {NoStop}%
\bibitem [{\citenamefont {Ong}\ \emph {et~al.}(2013)\citenamefont {Ong},
  \citenamefont {Richards}, \citenamefont {Jain}, \citenamefont {Hautier},
  \citenamefont {Kocher}, \citenamefont {Cholia}, \citenamefont {Gunter},
  \citenamefont {Chevrier}, \citenamefont {Persson},\ and\ \citenamefont
  {Ceder}}]{Ong_2013}%
  \BibitemOpen
  \bibfield  {author} {\bibinfo {author} {\bibfnamefont {S.~P.}\ \bibnamefont
  {Ong}}, \bibinfo {author} {\bibfnamefont {W.~D.}\ \bibnamefont {Richards}},
  \bibinfo {author} {\bibfnamefont {A.}~\bibnamefont {Jain}}, \bibinfo {author}
  {\bibfnamefont {G.}~\bibnamefont {Hautier}}, \bibinfo {author} {\bibfnamefont
  {M.}~\bibnamefont {Kocher}}, \bibinfo {author} {\bibfnamefont
  {S.}~\bibnamefont {Cholia}}, \bibinfo {author} {\bibfnamefont
  {D.}~\bibnamefont {Gunter}}, \bibinfo {author} {\bibfnamefont {V.~L.}\
  \bibnamefont {Chevrier}}, \bibinfo {author} {\bibfnamefont {K.~A.}\
  \bibnamefont {Persson}}, \ and\ \bibinfo {author} {\bibfnamefont
  {G.}~\bibnamefont {Ceder}},\ }\href {\doibase
  10.1016/j.commatsci.2012.10.028} {\bibfield  {journal} {\bibinfo  {journal}
  {Computational Materials Science}\ }\textbf {\bibinfo {volume} {68}},\
  \bibinfo {pages} {314} (\bibinfo {year} {2013})}\BibitemShut {NoStop}%
\bibitem [{\citenamefont {Ong}\ \emph {et~al.}(2015)\citenamefont {Ong},
  \citenamefont {Cholia}, \citenamefont {Jain}, \citenamefont {Brafman},
  \citenamefont {Gunter}, \citenamefont {Ceder},\ and\ \citenamefont
  {Persson}}]{Ong_2015}%
  \BibitemOpen
  \bibfield  {author} {\bibinfo {author} {\bibfnamefont {S.~P.}\ \bibnamefont
  {Ong}}, \bibinfo {author} {\bibfnamefont {S.}~\bibnamefont {Cholia}},
  \bibinfo {author} {\bibfnamefont {A.}~\bibnamefont {Jain}}, \bibinfo {author}
  {\bibfnamefont {M.}~\bibnamefont {Brafman}}, \bibinfo {author} {\bibfnamefont
  {D.}~\bibnamefont {Gunter}}, \bibinfo {author} {\bibfnamefont
  {G.}~\bibnamefont {Ceder}}, \ and\ \bibinfo {author} {\bibfnamefont {K.~A.}\
  \bibnamefont {Persson}},\ }\href {\doibase 10.1016/j.commatsci.2014.10.037}
  {\bibfield  {journal} {\bibinfo  {journal} {Computational Materials Science}\
  }\textbf {\bibinfo {volume} {97}},\ \bibinfo {pages} {209} (\bibinfo {year}
  {2015})}\BibitemShut {NoStop}%
\bibitem [{\citenamefont {Togo}(2023)}]{phonondb}%
  \BibitemOpen
  \bibfield  {author} {\bibinfo {author} {\bibfnamefont {A.}~\bibnamefont
  {Togo}},\ }\href {http://phonondb.mtl.kyoto-u.ac.jp/index.html} {\enquote
  {\bibinfo {title} {http://phonondb.mtl.kyoto-u.ac.jp/index.html},}\ }
  (\bibinfo {year} {2023})\BibitemShut {NoStop}%
\bibitem [{\citenamefont {McGuire}(2017)}]{FeBr2_data}%
  \BibitemOpen
  \bibfield  {author} {\bibinfo {author} {\bibfnamefont {M.}~\bibnamefont
  {McGuire}},\ }\href {\doibase 10.3390/cryst7050121} {\bibfield  {journal}
  {\bibinfo  {journal} {Crystals}\ }\textbf {\bibinfo {volume} {7}},\ \bibinfo
  {pages} {121} (\bibinfo {year} {2017})}\BibitemShut {NoStop}%
\bibitem [{\citenamefont {Wilkinson}\ \emph {et~al.}(1959)\citenamefont
  {Wilkinson}, \citenamefont {Cable}, \citenamefont {Wollan},\ and\
  \citenamefont {Koehler}}]{FeBr2_structure}%
  \BibitemOpen
  \bibfield  {author} {\bibinfo {author} {\bibfnamefont {M.~K.}\ \bibnamefont
  {Wilkinson}}, \bibinfo {author} {\bibfnamefont {J.~W.}\ \bibnamefont
  {Cable}}, \bibinfo {author} {\bibfnamefont {E.~O.}\ \bibnamefont {Wollan}}, \
  and\ \bibinfo {author} {\bibfnamefont {W.~C.}\ \bibnamefont {Koehler}},\
  }\href {\doibase 10.1103/PhysRev.113.497} {\bibfield  {journal} {\bibinfo
  {journal} {Phys. Rev.}\ }\textbf {\bibinfo {volume} {113}},\ \bibinfo {pages}
  {497} (\bibinfo {year} {1959})}\BibitemShut {NoStop}%
\bibitem [{\citenamefont {Peskin}\ and\ \citenamefont
  {Schroeder}(1995)}]{Peskin:1995ev}%
  \BibitemOpen
  \bibfield  {author} {\bibinfo {author} {\bibfnamefont {M.~E.}\ \bibnamefont
  {Peskin}}\ and\ \bibinfo {author} {\bibfnamefont {D.~V.}\ \bibnamefont
  {Schroeder}},\ }\href@noop {} {\emph {\bibinfo {title} {{An Introduction to
  quantum field theory}}}}\ (\bibinfo  {publisher} {Addison-Wesley},\ \bibinfo
  {address} {Reading, USA},\ \bibinfo {year} {1995})\BibitemShut {NoStop}%
\bibitem [{\citenamefont {Adelberger}\ \emph {et~al.}(2003)\citenamefont
  {Adelberger}, \citenamefont {Heckel},\ and\ \citenamefont
  {Nelson}}]{Adelberger:2003zx}%
  \BibitemOpen
  \bibfield  {author} {\bibinfo {author} {\bibfnamefont {E.~G.}\ \bibnamefont
  {Adelberger}}, \bibinfo {author} {\bibfnamefont {B.~R.}\ \bibnamefont
  {Heckel}}, \ and\ \bibinfo {author} {\bibfnamefont {A.~E.}\ \bibnamefont
  {Nelson}},\ }\href {\doibase 10.1146/annurev.nucl.53.041002.110503}
  {\bibfield  {journal} {\bibinfo  {journal} {Ann. Rev. Nucl. Part. Sci.}\
  }\textbf {\bibinfo {volume} {53}},\ \bibinfo {pages} {77} (\bibinfo {year}
  {2003})},\ \Eprint {http://arxiv.org/abs/hep-ph/0307284}
  {arXiv:hep-ph/0307284} \BibitemShut {NoStop}%
\bibitem [{\citenamefont {Konopliv}\ \emph {et~al.}(2011)\citenamefont
  {Konopliv}, \citenamefont {Asmar}, \citenamefont {Folkner}, \citenamefont
  {Özgür Karatekin}, \citenamefont {Nunes}, \citenamefont {Smrekar},
  \citenamefont {Yoder},\ and\ \citenamefont {Zuber}}]{KONOPLIV2011401}%
  \BibitemOpen
  \bibfield  {author} {\bibinfo {author} {\bibfnamefont {A.~S.}\ \bibnamefont
  {Konopliv}}, \bibinfo {author} {\bibfnamefont {S.~W.}\ \bibnamefont {Asmar}},
  \bibinfo {author} {\bibfnamefont {W.~M.}\ \bibnamefont {Folkner}}, \bibinfo
  {author} {\bibnamefont {Özgür Karatekin}}, \bibinfo {author} {\bibfnamefont
  {D.~C.}\ \bibnamefont {Nunes}}, \bibinfo {author} {\bibfnamefont {S.~E.}\
  \bibnamefont {Smrekar}}, \bibinfo {author} {\bibfnamefont {C.~F.}\
  \bibnamefont {Yoder}}, \ and\ \bibinfo {author} {\bibfnamefont {M.~T.}\
  \bibnamefont {Zuber}},\ }\href {\doibase
  https://doi.org/10.1016/j.icarus.2010.10.004} {\bibfield  {journal} {\bibinfo
   {journal} {Icarus}\ }\textbf {\bibinfo {volume} {211}},\ \bibinfo {pages}
  {401} (\bibinfo {year} {2011})}\BibitemShut {NoStop}%
\bibitem [{\citenamefont {Bottaro}\ \emph {et~al.}(2023)\citenamefont
  {Bottaro}, \citenamefont {Caputo}, \citenamefont {Raffelt},\ and\
  \citenamefont {Vitagliano}}]{Bottaro:2023gep}%
  \BibitemOpen
  \bibfield  {author} {\bibinfo {author} {\bibfnamefont {S.}~\bibnamefont
  {Bottaro}}, \bibinfo {author} {\bibfnamefont {A.}~\bibnamefont {Caputo}},
  \bibinfo {author} {\bibfnamefont {G.}~\bibnamefont {Raffelt}}, \ and\
  \bibinfo {author} {\bibfnamefont {E.}~\bibnamefont {Vitagliano}},\
  }\href@noop {} {\  (\bibinfo {year} {2023})},\ \Eprint
  {http://arxiv.org/abs/2303.00778} {arXiv:2303.00778 [hep-ph]} \BibitemShut
  {NoStop}%
\bibitem [{\citenamefont {Damour}\ and\ \citenamefont
  {Donoghue}(2010)}]{Damour:2010rp}%
  \BibitemOpen
  \bibfield  {author} {\bibinfo {author} {\bibfnamefont {T.}~\bibnamefont
  {Damour}}\ and\ \bibinfo {author} {\bibfnamefont {J.~F.}\ \bibnamefont
  {Donoghue}},\ }\href {\doibase 10.1103/PhysRevD.82.084033} {\bibfield
  {journal} {\bibinfo  {journal} {Phys. Rev. D}\ }\textbf {\bibinfo {volume}
  {82}},\ \bibinfo {pages} {084033} (\bibinfo {year} {2010})},\ \Eprint
  {http://arxiv.org/abs/1007.2792} {arXiv:1007.2792 [gr-qc]} \BibitemShut
  {NoStop}%
\bibitem [{\citenamefont {Knapen}\ \emph
  {et~al.}(2017{\natexlab{b}})\citenamefont {Knapen}, \citenamefont {Lin},\
  and\ \citenamefont {Zurek}}]{Knapen:2017xzo}%
  \BibitemOpen
  \bibfield  {author} {\bibinfo {author} {\bibfnamefont {S.}~\bibnamefont
  {Knapen}}, \bibinfo {author} {\bibfnamefont {T.}~\bibnamefont {Lin}}, \ and\
  \bibinfo {author} {\bibfnamefont {K.~M.}\ \bibnamefont {Zurek}},\ }\href
  {\doibase 10.1103/PhysRevD.96.115021} {\bibfield  {journal} {\bibinfo
  {journal} {Phys. Rev. D}\ }\textbf {\bibinfo {volume} {96}},\ \bibinfo
  {pages} {115021} (\bibinfo {year} {2017}{\natexlab{b}})},\ \Eprint
  {http://arxiv.org/abs/1709.07882} {arXiv:1709.07882 [hep-ph]} \BibitemShut
  {NoStop}%
\bibitem [{\citenamefont {Capozzi}\ and\ \citenamefont
  {Raffelt}(2020)}]{Capozzi:2020cbu}%
  \BibitemOpen
  \bibfield  {author} {\bibinfo {author} {\bibfnamefont {F.}~\bibnamefont
  {Capozzi}}\ and\ \bibinfo {author} {\bibfnamefont {G.}~\bibnamefont
  {Raffelt}},\ }\href {\doibase 10.1103/PhysRevD.102.083007} {\bibfield
  {journal} {\bibinfo  {journal} {Phys. Rev. D}\ }\textbf {\bibinfo {volume}
  {102}},\ \bibinfo {pages} {083007} (\bibinfo {year} {2020})},\ \Eprint
  {http://arxiv.org/abs/2007.03694} {arXiv:2007.03694 [astro-ph.SR]}
  \BibitemShut {NoStop}%
\bibitem [{\citenamefont {Giannotti}\ \emph {et~al.}(2017)\citenamefont
  {Giannotti}, \citenamefont {Irastorza}, \citenamefont {Redondo},
  \citenamefont {Ringwald},\ and\ \citenamefont {Saikawa}}]{Giannotti:2017hny}%
  \BibitemOpen
  \bibfield  {author} {\bibinfo {author} {\bibfnamefont {M.}~\bibnamefont
  {Giannotti}}, \bibinfo {author} {\bibfnamefont {I.~G.}\ \bibnamefont
  {Irastorza}}, \bibinfo {author} {\bibfnamefont {J.}~\bibnamefont {Redondo}},
  \bibinfo {author} {\bibfnamefont {A.}~\bibnamefont {Ringwald}}, \ and\
  \bibinfo {author} {\bibfnamefont {K.}~\bibnamefont {Saikawa}},\ }\href
  {\doibase 10.1088/1475-7516/2017/10/010} {\bibfield  {journal} {\bibinfo
  {journal} {JCAP}\ }\textbf {\bibinfo {volume} {10}},\ \bibinfo {pages} {010}
  (\bibinfo {year} {2017})},\ \Eprint {http://arxiv.org/abs/1708.02111}
  {arXiv:1708.02111 [hep-ph]} \BibitemShut {NoStop}%
\bibitem [{\citenamefont {Dennis}\ and\ \citenamefont
  {Sakstein}(2023)}]{Dennis:2023kfe}%
  \BibitemOpen
  \bibfield  {author} {\bibinfo {author} {\bibfnamefont {M.~T.}\ \bibnamefont
  {Dennis}}\ and\ \bibinfo {author} {\bibfnamefont {J.}~\bibnamefont
  {Sakstein}},\ }\href@noop {} {\  (\bibinfo {year} {2023})},\ \Eprint
  {http://arxiv.org/abs/2305.03113} {arXiv:2305.03113 [hep-ph]} \BibitemShut
  {NoStop}%
\bibitem [{\citenamefont {Buschmann}\ \emph
  {et~al.}(2022{\natexlab{a}})\citenamefont {Buschmann}, \citenamefont
  {Dessert}, \citenamefont {Foster}, \citenamefont {Long},\ and\ \citenamefont
  {Safdi}}]{Buschmann:2021juv}%
  \BibitemOpen
  \bibfield  {author} {\bibinfo {author} {\bibfnamefont {M.}~\bibnamefont
  {Buschmann}}, \bibinfo {author} {\bibfnamefont {C.}~\bibnamefont {Dessert}},
  \bibinfo {author} {\bibfnamefont {J.~W.}\ \bibnamefont {Foster}}, \bibinfo
  {author} {\bibfnamefont {A.~J.}\ \bibnamefont {Long}}, \ and\ \bibinfo
  {author} {\bibfnamefont {B.~R.}\ \bibnamefont {Safdi}},\ }\href {\doibase
  10.1103/PhysRevLett.128.091102} {\bibfield  {journal} {\bibinfo  {journal}
  {Phys. Rev. Lett.}\ }\textbf {\bibinfo {volume} {128}},\ \bibinfo {pages}
  {091102} (\bibinfo {year} {2022}{\natexlab{a}})},\ \Eprint
  {http://arxiv.org/abs/2111.09892} {arXiv:2111.09892 [hep-ph]} \BibitemShut
  {NoStop}%
\bibitem [{\citenamefont {Workman}\ \emph {et~al.}(2022)\citenamefont {Workman}
  \emph {et~al.}}]{ParticleDataGroup:2022pth}%
  \BibitemOpen
  \bibfield  {author} {\bibinfo {author} {\bibfnamefont {R.~L.}\ \bibnamefont
  {Workman}} \emph {et~al.} (\bibinfo {collaboration} {Particle Data Group}),\
  }\href {\doibase 10.1093/ptep/ptac097} {\bibfield  {journal} {\bibinfo
  {journal} {PTEP}\ }\textbf {\bibinfo {volume} {2022}},\ \bibinfo {pages}
  {083C01} (\bibinfo {year} {2022})}\BibitemShut {NoStop}%
\bibitem [{\citenamefont {Peccei}\ and\ \citenamefont
  {Quinn}(1977{\natexlab{a}})}]{Peccei:1977ur}%
  \BibitemOpen
  \bibfield  {author} {\bibinfo {author} {\bibfnamefont {R.~D.}\ \bibnamefont
  {Peccei}}\ and\ \bibinfo {author} {\bibfnamefont {H.~R.}\ \bibnamefont
  {Quinn}},\ }\href {\doibase 10.1103/PhysRevD.16.1791} {\bibfield  {journal}
  {\bibinfo  {journal} {Phys. Rev. D}\ }\textbf {\bibinfo {volume} {16}},\
  \bibinfo {pages} {1791} (\bibinfo {year} {1977}{\natexlab{a}})}\BibitemShut
  {NoStop}%
\bibitem [{\citenamefont {Peccei}\ and\ \citenamefont
  {Quinn}(1977{\natexlab{b}})}]{Peccei:1977hh}%
  \BibitemOpen
  \bibfield  {author} {\bibinfo {author} {\bibfnamefont {R.~D.}\ \bibnamefont
  {Peccei}}\ and\ \bibinfo {author} {\bibfnamefont {H.~R.}\ \bibnamefont
  {Quinn}},\ }\href {\doibase 10.1103/PhysRevLett.38.1440} {\bibfield
  {journal} {\bibinfo  {journal} {Phys. Rev. Lett.}\ }\textbf {\bibinfo
  {volume} {38}},\ \bibinfo {pages} {1440} (\bibinfo {year}
  {1977}{\natexlab{b}})}\BibitemShut {NoStop}%
\bibitem [{\citenamefont {Wilczek}(1978)}]{Wilczek:1977pj}%
  \BibitemOpen
  \bibfield  {author} {\bibinfo {author} {\bibfnamefont {F.}~\bibnamefont
  {Wilczek}},\ }\href {\doibase 10.1103/PhysRevLett.40.279} {\bibfield
  {journal} {\bibinfo  {journal} {Phys. Rev. Lett.}\ }\textbf {\bibinfo
  {volume} {40}},\ \bibinfo {pages} {279} (\bibinfo {year} {1978})}\BibitemShut
  {NoStop}%
\bibitem [{\citenamefont {Weinberg}(1975)}]{Weinberg:1975ui}%
  \BibitemOpen
  \bibfield  {author} {\bibinfo {author} {\bibfnamefont {S.}~\bibnamefont
  {Weinberg}},\ }\href {\doibase 10.1103/PhysRevD.11.3583} {\bibfield
  {journal} {\bibinfo  {journal} {Phys. Rev. D}\ }\textbf {\bibinfo {volume}
  {11}},\ \bibinfo {pages} {3583} (\bibinfo {year} {1975})}\BibitemShut
  {NoStop}%
\bibitem [{\citenamefont {Abbott}\ and\ \citenamefont
  {Sikivie}(1983)}]{Abbott:1982af}%
  \BibitemOpen
  \bibfield  {author} {\bibinfo {author} {\bibfnamefont {L.~F.}\ \bibnamefont
  {Abbott}}\ and\ \bibinfo {author} {\bibfnamefont {P.}~\bibnamefont
  {Sikivie}},\ }\href {\doibase 10.1016/0370-2693(83)90638-X} {\bibfield
  {journal} {\bibinfo  {journal} {Phys. Lett. B}\ }\textbf {\bibinfo {volume}
  {120}},\ \bibinfo {pages} {133} (\bibinfo {year} {1983})}\BibitemShut
  {NoStop}%
\bibitem [{\citenamefont {Turner}(1983)}]{Turner:1983he}%
  \BibitemOpen
  \bibfield  {author} {\bibinfo {author} {\bibfnamefont {M.~S.}\ \bibnamefont
  {Turner}},\ }\href {\doibase 10.1103/PhysRevD.28.1243} {\bibfield  {journal}
  {\bibinfo  {journal} {Phys. Rev. D}\ }\textbf {\bibinfo {volume} {28}},\
  \bibinfo {pages} {1243} (\bibinfo {year} {1983})}\BibitemShut {NoStop}%
\bibitem [{\citenamefont {Turner}(1986)}]{Turner:1985si}%
  \BibitemOpen
  \bibfield  {author} {\bibinfo {author} {\bibfnamefont {M.~S.}\ \bibnamefont
  {Turner}},\ }\href {\doibase 10.1103/PhysRevD.33.889} {\bibfield  {journal}
  {\bibinfo  {journal} {Phys. Rev. D}\ }\textbf {\bibinfo {volume} {33}},\
  \bibinfo {pages} {889} (\bibinfo {year} {1986})}\BibitemShut {NoStop}%
\bibitem [{\citenamefont {Preskill}\ \emph {et~al.}(1983)\citenamefont
  {Preskill}, \citenamefont {Wise},\ and\ \citenamefont
  {Wilczek}}]{Preskill:1982cy}%
  \BibitemOpen
  \bibfield  {author} {\bibinfo {author} {\bibfnamefont {J.}~\bibnamefont
  {Preskill}}, \bibinfo {author} {\bibfnamefont {M.~B.}\ \bibnamefont {Wise}},
  \ and\ \bibinfo {author} {\bibfnamefont {F.}~\bibnamefont {Wilczek}},\ }\href
  {\doibase 10.1016/0370-2693(83)90637-8} {\bibfield  {journal} {\bibinfo
  {journal} {Phys. Lett. B}\ }\textbf {\bibinfo {volume} {120}},\ \bibinfo
  {pages} {127} (\bibinfo {year} {1983})}\BibitemShut {NoStop}%
\bibitem [{\citenamefont {Dine}\ and\ \citenamefont
  {Fischler}(1983)}]{Dine:1982ah}%
  \BibitemOpen
  \bibfield  {author} {\bibinfo {author} {\bibfnamefont {M.}~\bibnamefont
  {Dine}}\ and\ \bibinfo {author} {\bibfnamefont {W.}~\bibnamefont
  {Fischler}},\ }\href {\doibase 10.1016/0370-2693(83)90639-1} {\bibfield
  {journal} {\bibinfo  {journal} {Phys. Lett. B}\ }\textbf {\bibinfo {volume}
  {120}},\ \bibinfo {pages} {137} (\bibinfo {year} {1983})}\BibitemShut
  {NoStop}%
\bibitem [{\citenamefont {Chang}\ and\ \citenamefont
  {Cui}(2020)}]{Chang:2019tvx}%
  \BibitemOpen
  \bibfield  {author} {\bibinfo {author} {\bibfnamefont {C.-F.}\ \bibnamefont
  {Chang}}\ and\ \bibinfo {author} {\bibfnamefont {Y.}~\bibnamefont {Cui}},\
  }\href {\doibase 10.1103/PhysRevD.102.015003} {\bibfield  {journal} {\bibinfo
   {journal} {Phys. Rev. D}\ }\textbf {\bibinfo {volume} {102}},\ \bibinfo
  {pages} {015003} (\bibinfo {year} {2020})},\ \Eprint
  {http://arxiv.org/abs/1911.11885} {arXiv:1911.11885 [hep-ph]} \BibitemShut
  {NoStop}%
\bibitem [{\citenamefont {Co}\ \emph {et~al.}(2020{\natexlab{a}})\citenamefont
  {Co}, \citenamefont {Hall},\ and\ \citenamefont {Harigaya}}]{Co:2019jts}%
  \BibitemOpen
  \bibfield  {author} {\bibinfo {author} {\bibfnamefont {R.~T.}\ \bibnamefont
  {Co}}, \bibinfo {author} {\bibfnamefont {L.~J.}\ \bibnamefont {Hall}}, \ and\
  \bibinfo {author} {\bibfnamefont {K.}~\bibnamefont {Harigaya}},\ }\href
  {\doibase 10.1103/PhysRevLett.124.251802} {\bibfield  {journal} {\bibinfo
  {journal} {Phys. Rev. Lett.}\ }\textbf {\bibinfo {volume} {124}},\ \bibinfo
  {pages} {251802} (\bibinfo {year} {2020}{\natexlab{a}})},\ \Eprint
  {http://arxiv.org/abs/1910.14152} {arXiv:1910.14152 [hep-ph]} \BibitemShut
  {NoStop}%
\bibitem [{\citenamefont {Co}\ and\ \citenamefont
  {Harigaya}(2020)}]{Co:2019wyp}%
  \BibitemOpen
  \bibfield  {author} {\bibinfo {author} {\bibfnamefont {R.~T.}\ \bibnamefont
  {Co}}\ and\ \bibinfo {author} {\bibfnamefont {K.}~\bibnamefont {Harigaya}},\
  }\href {\doibase 10.1103/PhysRevLett.124.111602} {\bibfield  {journal}
  {\bibinfo  {journal} {Phys. Rev. Lett.}\ }\textbf {\bibinfo {volume} {124}},\
  \bibinfo {pages} {111602} (\bibinfo {year} {2020})},\ \Eprint
  {http://arxiv.org/abs/1910.02080} {arXiv:1910.02080 [hep-ph]} \BibitemShut
  {NoStop}%
\bibitem [{\citenamefont {Co}\ \emph {et~al.}(2018)\citenamefont {Co},
  \citenamefont {Hall},\ and\ \citenamefont {Harigaya}}]{Co:2017mop}%
  \BibitemOpen
  \bibfield  {author} {\bibinfo {author} {\bibfnamefont {R.~T.}\ \bibnamefont
  {Co}}, \bibinfo {author} {\bibfnamefont {L.~J.}\ \bibnamefont {Hall}}, \ and\
  \bibinfo {author} {\bibfnamefont {K.}~\bibnamefont {Harigaya}},\ }\href
  {\doibase 10.1103/PhysRevLett.120.211602} {\bibfield  {journal} {\bibinfo
  {journal} {Phys. Rev. Lett.}\ }\textbf {\bibinfo {volume} {120}},\ \bibinfo
  {pages} {211602} (\bibinfo {year} {2018})},\ \Eprint
  {http://arxiv.org/abs/1711.10486} {arXiv:1711.10486 [hep-ph]} \BibitemShut
  {NoStop}%
\bibitem [{\citenamefont {Harigaya}\ and\ \citenamefont
  {Leedom}(2020)}]{Harigaya:2019qnl}%
  \BibitemOpen
  \bibfield  {author} {\bibinfo {author} {\bibfnamefont {K.}~\bibnamefont
  {Harigaya}}\ and\ \bibinfo {author} {\bibfnamefont {J.~M.}\ \bibnamefont
  {Leedom}},\ }\href {\doibase 10.1007/JHEP06(2020)034} {\bibfield  {journal}
  {\bibinfo  {journal} {JHEP}\ }\textbf {\bibinfo {volume} {06}},\ \bibinfo
  {pages} {034} (\bibinfo {year} {2020})},\ \Eprint
  {http://arxiv.org/abs/1910.04163} {arXiv:1910.04163 [hep-ph]} \BibitemShut
  {NoStop}%
\bibitem [{\citenamefont {Co}\ \emph {et~al.}(2020{\natexlab{b}})\citenamefont
  {Co}, \citenamefont {Hall}, \citenamefont {Harigaya}, \citenamefont {Olive},\
  and\ \citenamefont {Verner}}]{Co:2020dya}%
  \BibitemOpen
  \bibfield  {author} {\bibinfo {author} {\bibfnamefont {R.~T.}\ \bibnamefont
  {Co}}, \bibinfo {author} {\bibfnamefont {L.~J.}\ \bibnamefont {Hall}},
  \bibinfo {author} {\bibfnamefont {K.}~\bibnamefont {Harigaya}}, \bibinfo
  {author} {\bibfnamefont {K.~A.}\ \bibnamefont {Olive}}, \ and\ \bibinfo
  {author} {\bibfnamefont {S.}~\bibnamefont {Verner}},\ }\href {\doibase
  10.1088/1475-7516/2020/08/036} {\bibfield  {journal} {\bibinfo  {journal}
  {JCAP}\ }\textbf {\bibinfo {volume} {08}},\ \bibinfo {pages} {036} (\bibinfo
  {year} {2020}{\natexlab{b}})},\ \Eprint {http://arxiv.org/abs/2004.00629}
  {arXiv:2004.00629 [hep-ph]} \BibitemShut {NoStop}%
\bibitem [{\citenamefont {Daido}\ \emph {et~al.}(2017)\citenamefont {Daido},
  \citenamefont {Takahashi},\ and\ \citenamefont {Yin}}]{Daido:2017wwb}%
  \BibitemOpen
  \bibfield  {author} {\bibinfo {author} {\bibfnamefont {R.}~\bibnamefont
  {Daido}}, \bibinfo {author} {\bibfnamefont {F.}~\bibnamefont {Takahashi}}, \
  and\ \bibinfo {author} {\bibfnamefont {W.}~\bibnamefont {Yin}},\ }\href
  {\doibase 10.1088/1475-7516/2017/05/044} {\bibfield  {journal} {\bibinfo
  {journal} {JCAP}\ }\textbf {\bibinfo {volume} {05}},\ \bibinfo {pages} {044}
  (\bibinfo {year} {2017})},\ \Eprint {http://arxiv.org/abs/1702.03284}
  {arXiv:1702.03284 [hep-ph]} \BibitemShut {NoStop}%
\bibitem [{\citenamefont {Daido}\ \emph {et~al.}(2018)\citenamefont {Daido},
  \citenamefont {Takahashi},\ and\ \citenamefont {Yin}}]{Daido:2017tbr}%
  \BibitemOpen
  \bibfield  {author} {\bibinfo {author} {\bibfnamefont {R.}~\bibnamefont
  {Daido}}, \bibinfo {author} {\bibfnamefont {F.}~\bibnamefont {Takahashi}}, \
  and\ \bibinfo {author} {\bibfnamefont {W.}~\bibnamefont {Yin}},\ }\href
  {\doibase 10.1007/JHEP02(2018)104} {\bibfield  {journal} {\bibinfo  {journal}
  {JHEP}\ }\textbf {\bibinfo {volume} {02}},\ \bibinfo {pages} {104} (\bibinfo
  {year} {2018})},\ \Eprint {http://arxiv.org/abs/1710.11107} {arXiv:1710.11107
  [hep-ph]} \BibitemShut {NoStop}%
\bibitem [{\citenamefont {Battye}\ and\ \citenamefont
  {Shellard}(1994)}]{Battye:1994au}%
  \BibitemOpen
  \bibfield  {author} {\bibinfo {author} {\bibfnamefont {R.~A.}\ \bibnamefont
  {Battye}}\ and\ \bibinfo {author} {\bibfnamefont {E.~P.~S.}\ \bibnamefont
  {Shellard}},\ }\href {\doibase 10.1103/PhysRevLett.73.2954} {\bibfield
  {journal} {\bibinfo  {journal} {Phys. Rev. Lett.}\ }\textbf {\bibinfo
  {volume} {73}},\ \bibinfo {pages} {2954} (\bibinfo {year} {1994})},\ \bibinfo
  {note} {[Erratum: Phys.Rev.Lett. 76, 2203--2204 (1996)]},\ \Eprint
  {http://arxiv.org/abs/astro-ph/9403018} {arXiv:astro-ph/9403018} \BibitemShut
  {NoStop}%
\bibitem [{\citenamefont {Hagmann}\ \emph {et~al.}(1999)\citenamefont
  {Hagmann}, \citenamefont {Chang},\ and\ \citenamefont
  {Sikivie}}]{Hagmann:1998me}%
  \BibitemOpen
  \bibfield  {author} {\bibinfo {author} {\bibfnamefont {C.}~\bibnamefont
  {Hagmann}}, \bibinfo {author} {\bibfnamefont {S.}~\bibnamefont {Chang}}, \
  and\ \bibinfo {author} {\bibfnamefont {P.}~\bibnamefont {Sikivie}},\ }\href
  {\doibase 10.1016/S0920-5632(98)00506-4} {\bibfield  {journal} {\bibinfo
  {journal} {Nucl. Phys. B Proc. Suppl.}\ }\textbf {\bibinfo {volume} {72}},\
  \bibinfo {pages} {81} (\bibinfo {year} {1999})},\ \Eprint
  {http://arxiv.org/abs/hep-ph/9807428} {arXiv:hep-ph/9807428} \BibitemShut
  {NoStop}%
\bibitem [{\citenamefont {Hiramatsu}\ \emph {et~al.}(2011)\citenamefont
  {Hiramatsu}, \citenamefont {Kawasaki}, \citenamefont {Sekiguchi},
  \citenamefont {Yamaguchi},\ and\ \citenamefont
  {Yokoyama}}]{Hiramatsu:2010yu}%
  \BibitemOpen
  \bibfield  {author} {\bibinfo {author} {\bibfnamefont {T.}~\bibnamefont
  {Hiramatsu}}, \bibinfo {author} {\bibfnamefont {M.}~\bibnamefont {Kawasaki}},
  \bibinfo {author} {\bibfnamefont {T.}~\bibnamefont {Sekiguchi}}, \bibinfo
  {author} {\bibfnamefont {M.}~\bibnamefont {Yamaguchi}}, \ and\ \bibinfo
  {author} {\bibfnamefont {J.}~\bibnamefont {Yokoyama}},\ }\href {\doibase
  10.1103/PhysRevD.83.123531} {\bibfield  {journal} {\bibinfo  {journal} {Phys.
  Rev. D}\ }\textbf {\bibinfo {volume} {83}},\ \bibinfo {pages} {123531}
  (\bibinfo {year} {2011})},\ \Eprint {http://arxiv.org/abs/1012.5502}
  {arXiv:1012.5502 [hep-ph]} \BibitemShut {NoStop}%
\bibitem [{\citenamefont {Klaer}\ and\ \citenamefont
  {Moore}(2017{\natexlab{a}})}]{Klaer:2017qhr}%
  \BibitemOpen
  \bibfield  {author} {\bibinfo {author} {\bibfnamefont {V.~B.}\ \bibnamefont
  {Klaer}}\ and\ \bibinfo {author} {\bibfnamefont {G.~D.}\ \bibnamefont
  {Moore}},\ }\href {\doibase 10.1088/1475-7516/2017/10/043} {\bibfield
  {journal} {\bibinfo  {journal} {JCAP}\ }\textbf {\bibinfo {volume} {10}},\
  \bibinfo {pages} {043} (\bibinfo {year} {2017}{\natexlab{a}})},\ \Eprint
  {http://arxiv.org/abs/1707.05566} {arXiv:1707.05566 [hep-ph]} \BibitemShut
  {NoStop}%
\bibitem [{\citenamefont {Klaer}\ and\ \citenamefont
  {Moore}(2017{\natexlab{b}})}]{Klaer:2017ond}%
  \BibitemOpen
  \bibfield  {author} {\bibinfo {author} {\bibfnamefont {V.~B.~.}\ \bibnamefont
  {Klaer}}\ and\ \bibinfo {author} {\bibfnamefont {G.~D.}\ \bibnamefont
  {Moore}},\ }\href {\doibase 10.1088/1475-7516/2017/11/049} {\bibfield
  {journal} {\bibinfo  {journal} {JCAP}\ }\textbf {\bibinfo {volume} {11}},\
  \bibinfo {pages} {049} (\bibinfo {year} {2017}{\natexlab{b}})},\ \Eprint
  {http://arxiv.org/abs/1708.07521} {arXiv:1708.07521 [hep-ph]} \BibitemShut
  {NoStop}%
\bibitem [{\citenamefont {Gorghetto}\ \emph {et~al.}(2018)\citenamefont
  {Gorghetto}, \citenamefont {Hardy},\ and\ \citenamefont
  {Villadoro}}]{Gorghetto:2018myk}%
  \BibitemOpen
  \bibfield  {author} {\bibinfo {author} {\bibfnamefont {M.}~\bibnamefont
  {Gorghetto}}, \bibinfo {author} {\bibfnamefont {E.}~\bibnamefont {Hardy}}, \
  and\ \bibinfo {author} {\bibfnamefont {G.}~\bibnamefont {Villadoro}},\ }\href
  {\doibase 10.1007/JHEP07(2018)151} {\bibfield  {journal} {\bibinfo  {journal}
  {JHEP}\ }\textbf {\bibinfo {volume} {07}},\ \bibinfo {pages} {151} (\bibinfo
  {year} {2018})},\ \Eprint {http://arxiv.org/abs/1806.04677} {arXiv:1806.04677
  [hep-ph]} \BibitemShut {NoStop}%
\bibitem [{\citenamefont {Vaquero}\ \emph {et~al.}(2019)\citenamefont
  {Vaquero}, \citenamefont {Redondo},\ and\ \citenamefont
  {Stadler}}]{Vaquero:2018tib}%
  \BibitemOpen
  \bibfield  {author} {\bibinfo {author} {\bibfnamefont {A.}~\bibnamefont
  {Vaquero}}, \bibinfo {author} {\bibfnamefont {J.}~\bibnamefont {Redondo}}, \
  and\ \bibinfo {author} {\bibfnamefont {J.}~\bibnamefont {Stadler}},\ }\href
  {\doibase 10.1088/1475-7516/2019/04/012} {\bibfield  {journal} {\bibinfo
  {journal} {JCAP}\ }\textbf {\bibinfo {volume} {04}},\ \bibinfo {pages} {012}
  (\bibinfo {year} {2019})},\ \Eprint {http://arxiv.org/abs/1809.09241}
  {arXiv:1809.09241 [astro-ph.CO]} \BibitemShut {NoStop}%
\bibitem [{\citenamefont {Buschmann}\ \emph {et~al.}(2020)\citenamefont
  {Buschmann}, \citenamefont {Foster},\ and\ \citenamefont
  {Safdi}}]{Buschmann:2019icd}%
  \BibitemOpen
  \bibfield  {author} {\bibinfo {author} {\bibfnamefont {M.}~\bibnamefont
  {Buschmann}}, \bibinfo {author} {\bibfnamefont {J.~W.}\ \bibnamefont
  {Foster}}, \ and\ \bibinfo {author} {\bibfnamefont {B.~R.}\ \bibnamefont
  {Safdi}},\ }\href {\doibase 10.1103/PhysRevLett.124.161103} {\bibfield
  {journal} {\bibinfo  {journal} {Phys. Rev. Lett.}\ }\textbf {\bibinfo
  {volume} {124}},\ \bibinfo {pages} {161103} (\bibinfo {year} {2020})},\
  \Eprint {http://arxiv.org/abs/1906.00967} {arXiv:1906.00967 [astro-ph.CO]}
  \BibitemShut {NoStop}%
\bibitem [{\citenamefont {Hindmarsh}\ \emph {et~al.}(2020)\citenamefont
  {Hindmarsh}, \citenamefont {Lizarraga}, \citenamefont {Lopez-Eiguren},\ and\
  \citenamefont {Urrestilla}}]{Hindmarsh:2019csc}%
  \BibitemOpen
  \bibfield  {author} {\bibinfo {author} {\bibfnamefont {M.}~\bibnamefont
  {Hindmarsh}}, \bibinfo {author} {\bibfnamefont {J.}~\bibnamefont
  {Lizarraga}}, \bibinfo {author} {\bibfnamefont {A.}~\bibnamefont
  {Lopez-Eiguren}}, \ and\ \bibinfo {author} {\bibfnamefont {J.}~\bibnamefont
  {Urrestilla}},\ }\href {\doibase 10.1103/PhysRevLett.124.021301} {\bibfield
  {journal} {\bibinfo  {journal} {Phys. Rev. Lett.}\ }\textbf {\bibinfo
  {volume} {124}},\ \bibinfo {pages} {021301} (\bibinfo {year} {2020})},\
  \Eprint {http://arxiv.org/abs/1908.03522} {arXiv:1908.03522 [astro-ph.CO]}
  \BibitemShut {NoStop}%
\bibitem [{\citenamefont {Gorghetto}\ \emph {et~al.}(2021)\citenamefont
  {Gorghetto}, \citenamefont {Hardy},\ and\ \citenamefont
  {Villadoro}}]{Gorghetto:2020qws}%
  \BibitemOpen
  \bibfield  {author} {\bibinfo {author} {\bibfnamefont {M.}~\bibnamefont
  {Gorghetto}}, \bibinfo {author} {\bibfnamefont {E.}~\bibnamefont {Hardy}}, \
  and\ \bibinfo {author} {\bibfnamefont {G.}~\bibnamefont {Villadoro}},\ }\href
  {\doibase 10.21468/SciPostPhys.10.2.050} {\bibfield  {journal} {\bibinfo
  {journal} {SciPost Phys.}\ }\textbf {\bibinfo {volume} {10}},\ \bibinfo
  {pages} {050} (\bibinfo {year} {2021})},\ \Eprint
  {http://arxiv.org/abs/2007.04990} {arXiv:2007.04990 [hep-ph]} \BibitemShut
  {NoStop}%
\bibitem [{\citenamefont {Dine}\ \emph {et~al.}(2021)\citenamefont {Dine},
  \citenamefont {Fernandez}, \citenamefont {Ghalsasi},\ and\ \citenamefont
  {Patel}}]{Dine:2020pds}%
  \BibitemOpen
  \bibfield  {author} {\bibinfo {author} {\bibfnamefont {M.}~\bibnamefont
  {Dine}}, \bibinfo {author} {\bibfnamefont {N.}~\bibnamefont {Fernandez}},
  \bibinfo {author} {\bibfnamefont {A.}~\bibnamefont {Ghalsasi}}, \ and\
  \bibinfo {author} {\bibfnamefont {H.~H.}\ \bibnamefont {Patel}},\ }\href
  {\doibase 10.1088/1475-7516/2021/11/041} {\bibfield  {journal} {\bibinfo
  {journal} {JCAP}\ }\textbf {\bibinfo {volume} {11}},\ \bibinfo {pages} {041}
  (\bibinfo {year} {2021})},\ \Eprint {http://arxiv.org/abs/2012.13065}
  {arXiv:2012.13065 [hep-ph]} \BibitemShut {NoStop}%
\bibitem [{\citenamefont {Buschmann}\ \emph
  {et~al.}(2022{\natexlab{b}})\citenamefont {Buschmann}, \citenamefont
  {Foster}, \citenamefont {Hook}, \citenamefont {Peterson}, \citenamefont
  {Willcox}, \citenamefont {Zhang},\ and\ \citenamefont
  {Safdi}}]{Buschmann:2021sdq}%
  \BibitemOpen
  \bibfield  {author} {\bibinfo {author} {\bibfnamefont {M.}~\bibnamefont
  {Buschmann}}, \bibinfo {author} {\bibfnamefont {J.~W.}\ \bibnamefont
  {Foster}}, \bibinfo {author} {\bibfnamefont {A.}~\bibnamefont {Hook}},
  \bibinfo {author} {\bibfnamefont {A.}~\bibnamefont {Peterson}}, \bibinfo
  {author} {\bibfnamefont {D.~E.}\ \bibnamefont {Willcox}}, \bibinfo {author}
  {\bibfnamefont {W.}~\bibnamefont {Zhang}}, \ and\ \bibinfo {author}
  {\bibfnamefont {B.~R.}\ \bibnamefont {Safdi}},\ }\href {\doibase
  10.1038/s41467-022-28669-y} {\bibfield  {journal} {\bibinfo  {journal}
  {Nature Commun.}\ }\textbf {\bibinfo {volume} {13}},\ \bibinfo {pages} {1049}
  (\bibinfo {year} {2022}{\natexlab{b}})},\ \Eprint
  {http://arxiv.org/abs/2108.05368} {arXiv:2108.05368 [hep-ph]} \BibitemShut
  {NoStop}%
\bibitem [{\citenamefont {Hindmarsh}\ \emph {et~al.}(2021)\citenamefont
  {Hindmarsh}, \citenamefont {Lizarraga}, \citenamefont {Lopez-Eiguren},\ and\
  \citenamefont {Urrestilla}}]{Hindmarsh:2021zkt}%
  \BibitemOpen
  \bibfield  {author} {\bibinfo {author} {\bibfnamefont {M.}~\bibnamefont
  {Hindmarsh}}, \bibinfo {author} {\bibfnamefont {J.}~\bibnamefont
  {Lizarraga}}, \bibinfo {author} {\bibfnamefont {A.}~\bibnamefont
  {Lopez-Eiguren}}, \ and\ \bibinfo {author} {\bibfnamefont {J.}~\bibnamefont
  {Urrestilla}},\ }\href@noop {} {\  (\bibinfo {year} {2021})},\ \Eprint
  {http://arxiv.org/abs/2109.09679} {arXiv:2109.09679 [astro-ph.CO]}
  \BibitemShut {NoStop}%
\bibitem [{\citenamefont {Berlin}\ and\ \citenamefont
  {Zhou}(2022)}]{Berlin:2022mia}%
  \BibitemOpen
  \bibfield  {author} {\bibinfo {author} {\bibfnamefont {A.}~\bibnamefont
  {Berlin}}\ and\ \bibinfo {author} {\bibfnamefont {K.}~\bibnamefont {Zhou}},\
  }\href@noop {} {\  (\bibinfo {year} {2022})},\ \Eprint
  {http://arxiv.org/abs/2209.12901} {arXiv:2209.12901 [hep-ph]} \BibitemShut
  {NoStop}%
\bibitem [{\citenamefont {Adelberger}\ \emph {et~al.}(2009)\citenamefont
  {Adelberger}, \citenamefont {Gundlach}, \citenamefont {Heckel}, \citenamefont
  {Hoedl},\ and\ \citenamefont {Schlamminger}}]{ADELBERGER2009102}%
  \BibitemOpen
  \bibfield  {author} {\bibinfo {author} {\bibfnamefont {E.}~\bibnamefont
  {Adelberger}}, \bibinfo {author} {\bibfnamefont {J.}~\bibnamefont
  {Gundlach}}, \bibinfo {author} {\bibfnamefont {B.}~\bibnamefont {Heckel}},
  \bibinfo {author} {\bibfnamefont {S.}~\bibnamefont {Hoedl}}, \ and\ \bibinfo
  {author} {\bibfnamefont {S.}~\bibnamefont {Schlamminger}},\ }\href {\doibase
  https://doi.org/10.1016/j.ppnp.2008.08.002} {\bibfield  {journal} {\bibinfo
  {journal} {Progress in Particle and Nuclear Physics}\ }\textbf {\bibinfo
  {volume} {62}},\ \bibinfo {pages} {102} (\bibinfo {year} {2009})}\BibitemShut
  {NoStop}%
\bibitem [{\citenamefont {Fabbrichesi}\ \emph {et~al.}(2020)\citenamefont
  {Fabbrichesi}, \citenamefont {Gabrielli},\ and\ \citenamefont
  {Lanfranchi}}]{Fabbrichesi:2020wbt}%
  \BibitemOpen
  \bibfield  {author} {\bibinfo {author} {\bibfnamefont {M.}~\bibnamefont
  {Fabbrichesi}}, \bibinfo {author} {\bibfnamefont {E.}~\bibnamefont
  {Gabrielli}}, \ and\ \bibinfo {author} {\bibfnamefont {G.}~\bibnamefont
  {Lanfranchi}},\ }\href {\doibase 10.1007/978-3-030-62519-1} {\  (\bibinfo
  {year} {2020}),\ 10.1007/978-3-030-62519-1},\ \Eprint
  {http://arxiv.org/abs/2005.01515} {arXiv:2005.01515 [hep-ph]} \BibitemShut
  {NoStop}%
\bibitem [{\citenamefont {Krnjaic}\ \emph {et~al.}(2022)\citenamefont
  {Krnjaic}, \citenamefont {Rocha},\ and\ \citenamefont
  {Sokolenko}}]{Krnjaic:2022wor}%
  \BibitemOpen
  \bibfield  {author} {\bibinfo {author} {\bibfnamefont {G.}~\bibnamefont
  {Krnjaic}}, \bibinfo {author} {\bibfnamefont {D.}~\bibnamefont {Rocha}}, \
  and\ \bibinfo {author} {\bibfnamefont {A.}~\bibnamefont {Sokolenko}},\
  }\href@noop {} {\  (\bibinfo {year} {2022})},\ \Eprint
  {http://arxiv.org/abs/2210.06487} {arXiv:2210.06487 [hep-ph]} \BibitemShut
  {NoStop}%
\bibitem [{\citenamefont {Dobrescu}(2005)}]{Dobrescu:2004wz}%
  \BibitemOpen
  \bibfield  {author} {\bibinfo {author} {\bibfnamefont {B.~A.}\ \bibnamefont
  {Dobrescu}},\ }\href {\doibase 10.1103/PhysRevLett.94.151802} {\bibfield
  {journal} {\bibinfo  {journal} {Phys. Rev. Lett.}\ }\textbf {\bibinfo
  {volume} {94}},\ \bibinfo {pages} {151802} (\bibinfo {year} {2005})},\
  \Eprint {http://arxiv.org/abs/hep-ph/0411004} {arXiv:hep-ph/0411004}
  \BibitemShut {NoStop}%
\bibitem [{\citenamefont {Taufertsh\"ofer}\ \emph {et~al.}(2023)\citenamefont
  {Taufertsh\"ofer}, \citenamefont {Garcia-Sciveres},\ and\ \citenamefont
  {Griffin}}]{Taufertshofer:2023rgq}%
  \BibitemOpen
  \bibfield  {author} {\bibinfo {author} {\bibfnamefont {N.}~\bibnamefont
  {Taufertsh\"ofer}}, \bibinfo {author} {\bibfnamefont {M.}~\bibnamefont
  {Garcia-Sciveres}}, \ and\ \bibinfo {author} {\bibfnamefont {S.~M.}\
  \bibnamefont {Griffin}},\ }\href@noop {} {\  (\bibinfo {year} {2023})},\
  \Eprint {http://arxiv.org/abs/2301.04778} {arXiv:2301.04778 [hep-ph]}
  \BibitemShut {NoStop}%
\bibitem [{\citenamefont {Romao}\ \emph {et~al.}(2023)\citenamefont {Romao},
  \citenamefont {Catena}, \citenamefont {Spaldin},\ and\ \citenamefont
  {Matas}}]{Romao:2023zqf}%
  \BibitemOpen
  \bibfield  {author} {\bibinfo {author} {\bibfnamefont {C.~P.}\ \bibnamefont
  {Romao}}, \bibinfo {author} {\bibfnamefont {R.}~\bibnamefont {Catena}},
  \bibinfo {author} {\bibfnamefont {N.~A.}\ \bibnamefont {Spaldin}}, \ and\
  \bibinfo {author} {\bibfnamefont {M.}~\bibnamefont {Matas}},\ }\href@noop {}
  {\  (\bibinfo {year} {2023})},\ \Eprint {http://arxiv.org/abs/2301.07617}
  {arXiv:2301.07617 [hep-ph]} \BibitemShut {NoStop}%
\bibitem [{\citenamefont {Esposito}\ and\ \citenamefont
  {Pavaskar}(2023)}]{Esposito:2022bnu}%
  \BibitemOpen
  \bibfield  {author} {\bibinfo {author} {\bibfnamefont {A.}~\bibnamefont
  {Esposito}}\ and\ \bibinfo {author} {\bibfnamefont {S.}~\bibnamefont
  {Pavaskar}},\ }\href {\doibase 10.1103/PhysRevD.108.L011901} {\bibfield
  {journal} {\bibinfo  {journal} {Phys. Rev. D}\ }\textbf {\bibinfo {volume}
  {108}},\ \bibinfo {pages} {L011901} (\bibinfo {year} {2023})},\ \Eprint
  {http://arxiv.org/abs/2210.13516} {arXiv:2210.13516 [hep-ph]} \BibitemShut
  {NoStop}%
\end{thebibliography}%

\end{document}